\documentclass[twocolumn,showpacs,preprintnumbers,amsmath,amssymb]{revtex4}
\usepackage{epsfig}

\usepackage[table]{xcolor}
\usepackage{graphicx}% Include figure files
\usepackage{dcolumn}% Align table columns on decimal point
\usepackage{bm}% bold math
\usepackage{esvect}

\usepackage{hyperref}
\hypersetup{colorlinks=true,linkcolor=magenta,anchorcolor=green,citecolor=cyan,filecolor=black,menucolor=black,urlcolor=brown}
\usepackage{slashed}
\newcommand{\be}{\begin{equation}}
\newcommand{\ee}{\end{equation}}
\newcommand{\ba}{\begin{eqnarray}}
\newcommand{\ea}{\end{eqnarray}}

\begin{document}

\title{Projection of the gravitational dynamics on a subspace of probability distributions: 
curl-free Gaussian ansatz}

\author{Patrick Valageas}
\affiliation{Institut de Physique Th\'eorique, Universit\'e  Paris-Saclay, CEA, CNRS, F-91191 Gif-sur-Yvette Cedex, France}

\begin{abstract}

We present a new approach to model the gravitational dynamics of large-scale structures.
Instead of solving the equations of motion up to a finite perturbative order or 
building phenomenological models, we follow the evolution of the probability
distribution of the displacement and velocity fields within an approximation subspace.
Keeping the exact equations of motion with their full nonlinearity, this provides
a nonperturbative scheme that goes beyond shell crossing. 
Focusing on the simplest case of a curl-free Gaussian ansatz
for the displacement and velocity fields, we find that truncations of the power spectra
on nonlinear scales directly arise from the equations of motion. This leads to a truncated
Zeldovich approximation for the density power spectrum, but with a truncation that
is not set a priori and with different power spectra for the displacement and velocity fields.
The positivity of their auto power spectra also follows from the equations of motion.
Although the density power spectrum is only recovered up to a smooth drift on BAO scales,
the predicted density correlation function agrees with numerical simulations
within $2\%$ from BAO scales down to $7 h^{-1} {\rm Mpc}$ at $z \geq 0.35$, 
without any free parameter.

\end{abstract}

\date{\today}% It is always \today, today,

\maketitle

%\tableofcontents

\section{Introduction}
\label{sec:introduction}

The large-scale structures that we observe in the current Universe and at low
redshifts, i.e. the cosmic web, its filaments, galaxies and clusters of galaxies,
and Lyman-$\alpha$ absorption clouds, have emerged from the amplification by
gravitational instability of small primordial perturbations. 
In the standard inflation scenario, these were generated by quantum fluctuations during 
the inflationary epoch. Next, once these density perturbations reach the nonlinear regime
and form astrophysical objects such as galaxies or X-ray clusters, baryonic physics
comes into play through heating and cooling processes, star formation, feedback
from active galactic nuclei (AGN), ... Therefore, the measurements of these large-scale structures
provide key probes of the primordial mechanisms generating the initial seeds, 
of the underlying cosmological model (e.g., the amount of dark matter and dark energy)
that affects the growth rate of the density fluctuations at all redshifts, and of the astrophysical
processes associated with various objects (e.g., the bias of the tracers of the matter density
field).
More specifically, the baryon acoustic oscillations (BAO) of the matter or galaxy power
spectra, which correspond to a peak at about $110 h^{-1} {\rm Mpc}$ in the correlation
functions, are a robust signature of the acoustic oscillations in the baryon-photon
fluid before recombination that are also seen in the cosmic microwave background (CMB)
\cite{Eisenstein:2006nj}.
This provides a standard ruler that is able to constrain the low-redshift expansion of the
Universe and the standard $\Lambda$-CDM cosmological scenario \cite{Eisenstein:1998tu}.
Distortions of the images of background galaxies by the fluctuations of the gravitational 
potential along the lines of sight (weak gravitational lensing) also probe the total
matter density fluctuations and provide direct constraints on the cosmological
scenarios \cite{Munshi:2006fn}.
This has provided the motivation for several galaxy surveys in the last decades or the near
future, such as the Baryon Oscillation Spectroscopic Survey \cite{Ross:2016gvb},
the WiggleZ Dark Energy Survey \cite{Blake:2011en}, 
the Dark Energy Spectroscopic Instrument \cite{Martini:2018kdj},
Euclid \cite{Laureijs:2011gra} or the Large Synoptic Survey Telescope \cite{Abell:2009aa}.

The formation of these large-scale structures is often studied with numerical
simulations, which can tackle highly nonlinear scales and also include various
baryonic effects, such as star formation and feedback from AGN, if they include
an hydrodynamic description for the gas in addition to the N-body codes that are
adequate for cold dark matter (CDM).
However, it remains desirable to develop analytic or semi-analytic methods.
On large scales they provide efficient tools that are more practical than numerical
simulations to explore a large parameter space. On a qualitative level, they also help 
to understand how different parameters or alternative theories (e.g., models of
dark matter and dark energy, or modified gravity scenarios) affect the
cosmological structures.

The standard analytical approach to study gravitational clustering in the late
Universe is the standard perturbation theory (SPT) \cite{Goroff:1986ep,Bernardeau:2001qr}.
There, one writes the equations of motion in Eulerian space for the matter
density and velocity fields, $\rho({\bf x},t)$ and ${\bf v}({\bf x},t)$, that is, 
the continuity and Euler equations, supplemented
by the Poisson equation for the gravitational force. These equations being nonlinear
(quadratic), one writes a perturbative expansion in powers of the primordial fluctuations
and solves for the density and velocity fields up to some finite order.
One can also employ various partial resummation schemes
\cite{Crocce:2005xy,Valageas:2006bi,Bernardeau:2008fa,Taruya:2012ut}.
Finally, assuming Gaussian initial conditions, one takes the Gaussian average
of products of these fields to compute the density and velocity polyspectra or
$n$-point correlations.
Going to second or third order in the linear power spectrum $P_L$ improves the
agreement with numerical simulations on large scales, as compared with the linear
theory. However, the accuracy does not keep improving at higher orders and this scheme
cannot reach nonlinear scales, even if all perturbative diagrams were resummed
\cite{Carlson:2009it,Valageas:2010rx,Blas:2013aba,Valageas:2013hxa}.
Indeed, the Euler equation itself is only an approximation that neglects shell crossing,
where different streams coexist at a given location and give rise to nonzero velocity
dispersion and vorticity \cite{Pichon:1999tk,Pueblas:2008uv,Valageas:2010rx}.

A method to handle this problem is to explicitly consider coarsed-grained
equations of motion \cite{Pietroni:2011iz}.
Another recent approach is the effective field theory (EFT) of large-scale
structures \cite{Baumann:2010tm,Carrasco:2012cv}. 
Following methods devised in other fields where the equations of motion,
or the Lagrangians, are not exactly known, one derives low-energy effective actions
that are based on the symmetries of the problem, by taking into account all possible
operators up to some order in a derivative expansion (for instance). For the cosmological
dynamics, one considers a large-scale effective theory, taking into account all operators
up to some order over the wave number $k$. In practice, this adds new counterterms to the
SPT diagrams, which should capture the impact on large scales of small-scale nonperturbative 
processes, like shell crossing. The coefficients of these new terms cannot be derived
and need to be fitted to numerical simulations. However, once these parameters have been
set by fitting a few quantities, such as the power spectrum at a given scale, once can 
compute other statistical quantities. Thus, this framework remains predictive
\cite{DAmico:2019fhj}.
An advantage of this approach is that it can also handle baryons and biased tracers, 
such as galaxies, where indeed the equations of motion are not explicitly known or too 
complex to be of any use (e.g., one cannot include all astrophysical processes associated 
with star formation)
\cite{Senatore:2014eva,Perko:2016puo,Lewandowski:2014rca}.
Then, an effective approach is unavoidable.
In practice, EFT schemes usually assume a curl-free velocity field and neglect the
generation of vorticity by small-scale nonlinearities, so that they do not include all possible
nonlinear effects. But this is expected to be a small effect on large scales and could
be added to the formalism.

On the other hand, if we only consider dark matter, that is, if we neglect baryonic physics, 
the equations of motion are exactly known and given by Newton's (or Einstein's) gravity.
Then, the traditional approach to handle shell crossing is to work in Lagrangian space,
where we follow the trajectories of particles 
\cite{Zeldovich:1969sb,Buchert:1992ya,Bouchet:1992uh,Buchert:1993xz,Bouchet:1994xp,Matsubara:2007wj,Vlah:2014nta,Matsubara:2015ipa,Taruya:2017ohk,McDonald:2017ths}.
Then, the fundamental object is the
displacement field ${\bf \Psi}({\bf q},t)$ and nothing peculiar appears at shell crossing.
A disadvantage of this method is that one eventually needs to compute the statistics
of the density field from the displacement field. This is a highly nonlinear transformation
that leads to practical difficulties for many-point polyspectra or correlation functions. 
An alternative to this Lagrangian route is to go from the hydrodynamical equations,
associated with the density and velocity fields, to the Vlasov equation, associated with
the phase-space distribution $f({\bf x},{\bf v},t)$.
This provides an exact Eulerian-space description of the gravitational dynamics
\cite{Valageas:2003gm,Tassev:2010us}.
However, this leads to 7-dimensional fields, which makes computations very heavy and
time consuming.
Another recent alternative is to replace the hydrodynamical equations by the
Schrodinger equation, which in some regime can provide an approximation to the
Vlasov equation \cite{Widrow:1993qq,Uhlemann:2014npa}.

Another level of distinction between the different analytical approaches is whether 
they work with the equations of motion or directly with statistical quantities.
The popular methods above work at the level of the equations of motion.
There, one computes an approximation for  $\rho({\bf x},t)$ in terms
of the initial condition $\delta_{L0}({\bf x})$, determined by the growing mode of the
linear density contrast. Next, statistical quantities such as the power spectrum
are obtained by taking the Gaussian average over products of such nonlinear functionals.
Another approach is to first write the evolution equations satisfied by those statistical quantities
and next solve them with some approximation scheme 
\cite{Taruya:2007xy,Pietroni:2008jx,Anselmi:2012cn}.
This typically leads to infinite series of equations that relate $n$- and $(n+1)$-point 
correlation functions or polyspectra, as in the BBGKY hierarchy \cite{Davis_1977}.
Alternatively, one can work with the
probability distribution functional of the fields \cite{Blas:2015qsi}
or the generating functional of the correlation
functions \cite{Valageas:2006bi,Valageas:2007su,McDonald:2017ths}.
An advantage of this approach is that such statistical quantities satisfy
symmetries (e.g., translation invariance) that are not obeyed by individual realizations of
the random fields, which can simplify some expressions. However, computations 
become cumbersome when going beyond three-point correlations. 

In this paper, we present a new approach to follow the gravitational dynamics
of large-scale structures. In contrast with most previous schemes, 
we wish to build a scheme that is meaningful from large to small scales, 
hence goes beyond perturbative treatments, and does not introduce free parameters
that require fitting to numerical simulations. To handle shell crossings, we adopt
a Lagrangian framework (in principle we could also opt for the  phase-space distribution 
$f({\bf x},{\bf v},t)$). We also work at the level of the probability distribution 
${\cal P}({\bf \Psi},{\bf v},t)$ of the displacement and velocity fields, instead of trying to solve 
the dynamics of individual realizations.
Then, we propose to follow the progress of gravitational clustering by 
``projecting'' the dynamics onto a subspace of trial distributions ${\cal P}$.
This idea is an extension of the standard procedure to estimate the minimum of a nonlinear
cost functional $S[\varphi(x)]$. There, one can expand $\varphi$ over a finite
basis, $\varphi=\sum_i a_i \psi_i$, which defines a subspace of possible functions $\varphi$,
and look for the minimum of $S(\{a_i\})$.
However, instead of a minimization problem, we use the equations of motion to follow
the evolution of ${\cal P}({\bf \Psi},{\bf v},t)$ within the lower-dimensional subspace
of trial distributions $\{{\cal P}\}_{\rm trial}$.

In this article, we consider the simplest case of Gaussian distributions $\{{\cal P}\}_{\rm trial}$,
which are fully defined by the displacement and velocity power spectra.
Then, the evolution of ${\cal P}$ is determined by the equations of motion for these 
power spectra. As for the Eulerian-space BBGKY hierarchy \cite{Davis_1977}, this is not a closed
system, because these equations involve correlations with the gravitational force,
which is a nonlinear functional of the displacement field.
However, within the Gaussian ansatz for ${\cal P}$ (or more generally, given the
form of ${\cal P}$ within the approximate subspace), we can exactly compute
such correlations and close the system.
In other words, in contrast with most approaches, we keep the exact equations
of motion and only perform the truncation, or approximation, at the level
of the distribution ${\cal P}$.
This provides a nonperturbative scheme that can handle shell crossing and does not
require parameters to be fitted by numerical simulations. The equations of motion
themselves determine the parameters that enter the probability distribution ${\cal P}$,
here the displacement and velocity power spectra.
For the density field, this leads to a prediction that coincides with the truncated Zeldovich
approximation. However, the truncation is not introduced by hand but arises from the
equations of motion. Also, in contrast with the truncated Zeldovich approximation,
the displacement and velocity power spectra are different.

This paper is organized as follows. 
In Sec.~\ref{sec:eom} we recall the equations of motion for the displacement field
and its probability distribution, as well as the expression of the gravitational force.
In Sec.~\ref{sec:constraints}, we give the evolution equations of 
the displacement and velocity power spectra, which provide constraints on the
evolution of the distribution ${\cal P}$.
Then, in Sec.~\ref{sec:curl-free-Gaussian}, we present the simplest Gaussian ansatz for
the distribution ${\cal P}$ and derive its closure of the system of equations of motion.
We briefly compare our approach with other analytical schemes in 
Sec.~\ref{sec:comparison-theory}.
Then, in Sec.~\ref{sec:self-similar}, we first present our numerical computations for the 
case of self-similar dynamics, with power-law linear power spectra in the 
Einstein-de Sitter cosmology.
We turn to the realistic $\Lambda$-CDM cosmology in Sec.~\ref{sec:LCDM}
and conclude in Sec.~\ref{sec:Conclusion}.

\section{Equations of motion}
\label{sec:eom}

\subsection{Lagrangian displacement field}
\label{sec:displacement}

In a Lagrangian framework, we describe the dynamics by following the comoving trajectories
${\bf x}({\bf q},t)$ of the particles, labeled by their initial comoving position ${\bf q}$.
As usual, it is convenient to introduce the displacement ${\bf \Psi}({\bf q},t)$ so that
the positions at time $t$ read
\be
{\bf x}({\bf q},t) = {\bf q} + {\bf \Psi}({\bf q},t) .
\ee
Then, standard Lagrangian perturbation theory aims at computing the trajectories
${\bf x}({\bf q},t)$ as a perturbative expansion over powers of the displacement ${\bf \Psi}$
\cite{Buchert:1992ya,Buchert:1993xz,Bouchet:1994xp,Matsubara:2007wj}. 
In the expanding Universe, the equation of motion of the gravitational dynamics reads
\be
\ddot{\bf \Psi} + 2 H \dot{\bf \Psi} = - \frac{\nabla_{\bf x} \Phi}{a^2} ,
\label{eq:Psi-eom}
\ee
where the dot denotes the partial derivative with respect to time, $a(t)$ is the scale factor,
$H=\dot{a}/a$ the Hubble expansion rate, and $\Phi$ the gravitational potential. 
Because we work with comoving coordinates, the background expansion has been 
subtracted and $\Phi$ is only sourced by the density perturbations $\rho-\bar\rho$,
where $\bar\rho$ is the background density. Thus, $\Phi$ is given by the Poisson equation
\be
\nabla^2_{\bf x} \Phi = a^2 4\pi{\cal G} ( \rho-\bar\rho) ,
\label{eq:Poisson}
\ee
where ${\cal G}$ is Newton's constant, and its explicit expression is often written as
\be
\Phi({\bf x},t) = - a^2 {\cal G} \int d{\bf x}' \, \frac{\rho({\bf x}')-\bar\rho}{|{\bf x}'-{\bf x}|} .
\ee
The background counterterm also corresponds to the well-known Jeans ``swindle'', 
which regularizes the infrared divergence of the gravitational force due to an infinite
homogeneous background.
As pointed out in \cite{Kiessling:1999eq,2009PhRvE..80d1108G,Gabrielli:2010uk},
a more satisfactory expression is obtained by introducing a screening of the gravitational
interaction with distance,
\be
\Phi({\bf x},t) = - a^2 {\cal G} \int d{\bf x}' \, \rho({\bf x}') 
\frac{e^{-\mu |{\bf x}'-{\bf x}|}}{|{\bf x}'-{\bf x}|} ,
\label{eq:Phi-mu}
\ee
and taking the limit $\mu\to 0$ at the end of the computations. 
Indeed, an homogeneous background gives a finite constant contribution to the potential $\Phi$,
which does not contribute to the gravitational force ${\bf F}=-\nabla_{\bf x}\Phi$.
This corresponds to the screened Poisson equation,
\be
\nabla^2_{\bf x} \Phi - \mu^2 \Phi = a^2 4\pi{\cal G} \rho .
\ee
Solving this equation in Fourier space, we obtain at once
\be
\Phi({\bf x},t) = - a^2 4\pi {\cal G} \int \frac{d{\bf x}'d{\bf k}}{(2\pi)^3} \,
 e^{i {\bf k} \cdot ( {\bf x} - {\bf x}') } \frac{1}{k^2+\mu^2} \rho({\bf x}') .
\label{eq:Phi-mu-k}
\ee
This is simply Eq.(\ref{eq:Phi-mu}) with the Fourier representation of the screened
gravitational interaction.

A difficulty that one often encounters in Lagrangian perturbation theory is that the
gravitational force $\nabla_{\bf x}\Phi$ in the equation of motion (\ref{eq:Psi-eom}) 
is naturally written in Eulerian space ${\bf x}$, as in the Poisson equation (\ref{eq:Poisson})
or the expression (\ref{eq:Phi-mu}). Then, in the course of the perturbative computation,
one may switch back and forth from Eulerian to Lagrangian space.
In the standard Lagrangian perturbation theory
\cite{Buchert:1992ya,Buchert:1993xz,Bouchet:1994xp,Matsubara:2007wj},
one takes the divergence of the equation of motion (\ref{eq:Psi-eom}) with respect
to ${\bf x}$, so as to use the Poisson equation to eliminate the gravitational potential
in favor of the density. The latter is obtained from the conservation of matter
as $\rho/\bar\rho = | \det( \partial{\bf q}/\partial{\bf x} ) |$.
This leads to a nonlinear equation in ${\bf \Psi}$, of cubic order in 3 dimensions
\cite{Bernardeau:2008ss}.
It is often supplemented by the requirement of a curl-free Eulerian velocity field.
This latter step is valid at all orders of Eulerian perturbation theory but it is not exact 
because shell crossing generates a nonzero vorticity \cite{Pichon:1999tk,Pueblas:2008uv}.

In this paper we follow a different approach, as we do not take the divergence of
the equation of motion (\ref{eq:Psi-eom}). Instead, as in \cite{McDonald:2017ths}
we directly obtain the gravitational
force from the explicit expression (\ref{eq:Phi-mu-k}) of the gravitational potential
in terms of the density field. 
Indeed, mass conservation allows us to derive a simple expression that only involves the 
Lagrangian trajectories.
Before shell crossing we have $\rho({\bf x}) d{\bf x} = \bar\rho d{\bf q}$, while after
shell crossing we need to sum over all streams. In both cases, the gravitational potential
(\ref{eq:Phi-mu-k}) simply writes as
\be
\Phi({\bf x},t) = - a^2 4\pi {\cal G} \bar\rho \int \frac{d{\bf q}'d{\bf k}}{(2\pi)^3} \,
 e^{i {\bf k} \cdot [ {\bf x} - {\bf x}({\bf q}',t) ]} \frac{1}{k^2+\mu^2} .
\label{eq:Phi-mu-k-qp}
\ee
Thus, instead of counting the mass in Eulerian space with the density field, we simply
count the particles, labeled by the initial position ${\bf q}'$.

In the linear regime over the displacement field ${\bf \Psi}$ or the density perturbation
$\delta = (\rho-\bar\rho)/\bar\rho$, denoted by the subscript ``L'', the linear growing mode 
is the curl-free displacement given in Fourier space by
\be
{\bf \Psi}_L({\bf k},t) = \frac{i {\bf k}}{k^2} \delta_L({\bf k},t) , \;\;\;
\delta_L({\bf k},t) = D_+(t) \delta_{L0}({\bf k}) .
\label{eq:PsiL-deltaL}
\ee
The linear growing mode $D_+(t)$ is given by
\be
\ddot{D}_+  + 2 H \dot{D}_+ = 4 \pi {\cal G} \bar\rho D_+ ,
\ee
and at early times in the matter era we have $D_+(t) \propto a(t) \propto t^{2/3}$.
It is convenient to use $\eta=\ln D_+(t)$ as the time coordinate. Then, the equation of
motion (\ref{eq:Psi-eom}) reads
\be
\frac{\partial^2 {\bf \Psi}}{\partial\eta^2} + \left( \frac{3\Omega_{\rm m}}{2f^2} - 1 \right) 
\frac{\partial {\bf \Psi}}{\partial\eta} = \frac{3\Omega_{\rm m}}{2f^2} {\bf F} ,
\label{eq:eom-k-F}
\ee
where we introduced the linear growth rate $f(t)$,
\be
f(t) = \frac{d\ln D_+}{d\ln a} = \frac{\dot{D}_+}{H D_+} ,
\ee
and the gravitational force ${\bf F}({\bf q},\eta)$ on the particle ${\bf q}$ reads
\be
{\bf F}({\bf q},\eta) =  \int \frac{d{\bf q}'d{\bf k}}{(2\pi)^3} \,
 e^{i {\bf k} \cdot [ {\bf x}({\bf q}) - {\bf x}({\bf q}') ]} \frac{i {\bf k}}{k^2+\mu^2} .
\label{eq:Fq}
\ee
As compared with the expression in \cite{McDonald:2017ths}, we have added 
the regularization factor $\mu^2$.
In this fashion, the equation of motion (\ref{eq:eom-k-F}) is fully written in terms of the
Lagrangian-space displacement field, at the price of a strong nonlinearity as the exponential
generates terms at all orders in powers of ${\bf \Psi}$.
All the cosmological dependence is captured by the factors $\Omega_{\rm m}/f^2$.
This factor remains close to unity at all redshifts and it is often approximated by unity
in perturbative computations.
This approximate symmetry can actually be used to derive approximate consistency relations
that go beyond low-order perturbation theory \cite{Valageas:2013zda,Kehagias:2013paa}.

\subsection{Linear displacement field}
\label{sec:linear-displacement}

We can check that the linear growing mode (\ref{eq:PsiL-deltaL}) is a solution of the
linearized equation derived from Eq.(\ref{eq:eom-k-F}).
At linear order, the force reads
\be
{\bf F}_L({\bf q}) =  \int \frac{d{\bf q}'d{\bf k}}{(2\pi)^3} \,
 e^{i {\bf k} \cdot ( {\bf q} - {\bf q}' ) } \left[ 1 + i {\bf k} \cdot ( {\bf\Psi}_L - {\bf\Psi}'_L ) \right] 
 \frac{i {\bf k}}{k^2+\mu^2} .
\ee
The terms $1+ i {\bf k} \cdot {\bf\Psi}_L$, which do not depend on ${\bf q}'$, give a vanishing
contribution as the integral over ${\bf q}'$ gives a Dirac factor $\delta_D({\bf k})$.
(Thus, we explicitly see how the background contribution vanishes thanks to the
screening beyond distance $1/\mu$.)
Substituting the linear expression (\ref{eq:PsiL-deltaL}) gives
\be
{\bf F}_L({\bf q}) =  \int \frac{d{\bf q}'d{\bf k}d{\bf k}'}{(2\pi)^3} \,
 e^{i {\bf k} \cdot ( {\bf q} - {\bf q}' ) + i {\bf k}'\cdot {\bf q}' } \frac{{\bf k} \cdot {\bf k}' }{k'^2} 
\delta_L({\bf k}') \frac{i {\bf k}}{k^2+\mu^2} .
\ee
The integration over ${\bf q}'$ gives the Dirac factor $\delta_D({\bf k}'-{\bf k})$,
and the integration over ${\bf k}'$ gives
\be
{\bf F}_L({\bf q}) =  \int d{\bf k} \, e^{i {\bf k} \cdot {\bf q} } \frac{i {\bf k}}{k^2+\mu^2}
\delta_L({\bf k}) . 
\label{eq:F_L}
\ee
Then, for $\mu\to 0$ we obtain ${\bf F}_L({\bf q}) \to {\bf \Psi}_L({\bf q})$ and we can
see that ${\bf \Psi}_L$ is solution of Eq.(\ref{eq:eom-k-F}), as 
${\bf \Psi}_L({\bf q},\eta) = e^{\eta} {\bf \Psi}_{L0}({\bf q})$ and
$\partial{\bf\Psi}_L/\partial\eta = {\bf \Psi}_L$.

\subsection{Probability distribution}
\label{sec:probability-distribution}

The second-order differential equation of motion (\ref{eq:eom-k-F}) can be written
as a system of two first-order differential equations if we introduce the velocity
field ${\bf v}$,
\be
{\bf v}({\bf q},\eta) \equiv \frac{\partial{\bf\Psi}}{\partial\eta} .
\ee
This gives the coupled first-order system
\ba
&& \frac{\partial{\bf\Psi}}{\partial\eta} = {\bf v} , 
\label{eq:dPsi-deta} \\
&& \frac{\partial{\bf v}}{\partial\eta} + \left( \frac{3\Omega_{\rm m}}{2f^2} - 1 \right) {\bf v} 
= \frac{3\Omega_{\rm m}}{2f^2} {\bf F}[{\bf \Psi}] ,
\label{eq:dv-deta}
\ea
where we made explicit that ${\bf F}$ is a functional of ${\bf \Psi}$.

The probability distribution functional ${\cal P}({\bf \Psi},{\bf v};\eta)$ of the
displacements and velocities obeys the continuity equation
(see also \cite{Blas:2015qsi} for the Eulerian-space probability distribution)
\be
\frac{\partial{\cal P}}{\partial\eta} + \int d{\bf q} \left[ \frac{\delta}{\delta{\bf\Psi}({\bf q})} 
\left(  \frac{\partial{\bf\Psi}}{\partial\eta} {\cal P} \right) + \frac{\delta}{\delta{\bf v}({\bf q})} 
\left( \frac{\partial{\bf v}}{\partial\eta} {\cal P} \right) \right] = 0 .
\label{eq:P-Liouville}
\ee
As for the usual Liouville equation, it describes the conservation of probability
in phase space, here the functional space $\{ {\bf \Psi},{\bf v} \}$.
Substituting the dynamical equations (\ref{eq:dPsi-deta})-(\ref{eq:dv-deta}), we obtain 
a closed evolution equation for ${\cal P}({\bf \Psi},{\bf v};\eta)$.
The advantage of the evolution equation (\ref{eq:P-Liouville}) is that it does not
require keeping track of past history.
In contrast, perturbative approaches based on the equation of motion (\ref{eq:Psi-eom}),
or its Eulerian counterparts for the density and velocity fields, generate an increasingly large
number of integrations over past times as one goes to higher orders.
Indeed, each new order involves one more integration over the time-dependent Green function 
associated with the linearized equation of motion.
To bypass this complication, we can attempt to solve directly the equation (\ref{eq:P-Liouville})
for the distribution ${\cal P}$: from the (approximate) knowledge of ${\cal P}$ at a given time
$\eta$ we can derive the distribution at the next time step $\eta+\Delta\eta$, without
needing the cross-correlations with earlier times. This is actually what numerical simulations
do, advancing particles over one time step from their current positions and velocities.

In practice, we do not expect to find the exact solution of the nonlinear
functional equation (\ref{eq:P-Liouville}).
One possibility is to look for a perturbative expansion of ${\cal P}$ around the Gaussian,
which describes the linear regime. This is the method investigated in \cite{Blas:2015qsi}
for the probablity distribution of the density and velocity fields in the Eulerian framework.
In contrast, the main idea of this paper is to apply a nonperturbative method, by considering
trial distributions and using the dynamical equation (\ref{eq:P-Liouville}) to derive constraints
that fully determine the free parameters of such ansatze.
The hope is that by considering a sequence of increasingly detailed and versatile
ansatze, each one satisfying the equation of motion (\ref{eq:P-Liouville}) to the ``best
possible accuracy'' within its class, we converge to the true distribution ${\cal P}$.
This is similar to a standard minimization problem, where we look for the absolute
minimum of a nonlinear cost functional $S[\varphi(x)]$. One method is to expand the function
$\varphi(x)$ over a basis of orthonormal functions $\psi_i$, 
$\varphi = \sum_i a_i \psi_i$, which is truncated at some order $N$, and to minimize
the associated cost function $S(\{a_i\})$. If the basis $\{\psi_i\}$ is well chosen, in favorable
cases the sequence of approximations $\{\varphi_N\}$ will converge to the exact minimum.

However, our problem is more complex than this minimization problem, as we do not have
a uniquely defined cost functional $S$. Thus, within a given class of trial distributions ${\cal P}$,
it is not obvious how we select the ``best'' choice. Our approach will be to use the evolution
equation (\ref{eq:P-Liouville}) to derive a set of constraints satisfied by ${\cal P}$,
choosing the simplest ones that we can build. Then, we determine ${\cal P}$ from
a self-consistency condition, by requiring it satisfies this set of constraints.
As we increase the complexity and versatility of ${\cal P}$, hence its number of free parameters,
we can take into account an increasing number of constraints.
For instance, if we intend to characterize the probability distribution by its moments,
we can obtain from the evolution equation (\ref{eq:P-Liouville}) an expression for the
time derivative of each moment. Then, truncating at a finite order $N$,
as in the Edgeworth expansion of a probability distribution around the Gaussian,
we can determine the moments or cumulants up to order $N$ from these $N$ constraint
equations.
Here, we can see the ambiguity associated with this method. Although it is more natural
to use the constraints derived from the time derivative of the moments of order one to
$N$, to determine a distribution parameterized by its $N$ first moments, in principle
we could have chosen the constraints derived from the time derivative of the moments 
of order $p$ to $p-1+N$, for any $p$, or any other set of $N$ constraints. 
The true distribution satisfies an infinite number of constraints, e.g. for all higher-order
cumulants, and we can expect to improve the accuracy of our trial distributions by
including an increasing number of constraints.

\subsection{Density power spectrum}
\label{sec:density-power}

Assuming we have obtained the statistics of the displacement field $\Psi$,
we can obtain the statistics of the density field, as for the well-known Zeldovich
approximation \cite{Zeldovich:1969sb}. Indeed, as for Eq.(\ref{eq:Phi-mu-k-qp}), 
integrals over the density field in Eulerian space can be written as integrals over 
Lagrangian space, and we have
\ba
\delta({\bf k}) & = & \int \frac{d{\bf x}}{(2\pi)^3} e^{-i{\bf k}\cdot{\bf x}} \delta({\bf x}) \\ 
&=&  \int \frac{d{\bf q}}{(2\pi)^3} \left( e^{-i{\bf k}\cdot{\bf x}({\bf q})} - e^{-i{\bf k}\cdot{\bf q}} \right) .
\ea
Defining the density power spectrum as
\be
\langle \delta({\bf k}_1) \delta({\bf k}_2) \rangle = \delta_D({\bf k}_1+{\bf k}_2) P(k_1) ,
\ee
this gives for $k>0$ \cite{1995MNRAS.273..475S,Taylor:1996ne}
\be
P(k) = \int \frac{d{\bf q}}{(2\pi)^3} \langle e^{i {\bf k}\cdot[{\bf x}({\bf q}) - {\bf x}(0)]} \rangle .
\label{eq:P-deltadelta-def}
\ee
This expression is exact, so that in principles no further approximation is needed to go 
from the displacement field $\Psi$ to the density field.
However, if $\Psi$ is not Gaussian the average in Eq.(\ref{eq:P-deltadelta-def})
may be difficult to compute. In particular, it involves the moments of $\Psi$ at all orders.

\section{Constraint equations}
\label{sec:constraints}

As explained in the previous section, because we cannot fully solve Eq.(\ref{eq:P-Liouville})
for the evolution of the probability distribution ${\cal P}({\bf \Psi},{\bf v};\eta)$, the approach
we propose in this paper is to use the more limited information associated with 
constraint equations that are consequences of this evolution equation.
The hope is that this reduction can make the problem tractable while retaining enough
information to strongly constrain the final approximation.
As we shall see in the next section, because we consider in this paper a Gaussian ansatz
for the probability distribution ${\cal P}({\bf \Psi},{\bf v};\eta)$, constraints associated
with the displacement and velocity power spectra will be sufficient for our purpose.
More precisely, let us define the divergences in Lagrangian space $\chi$ and $\theta$
of the displacement and velocity fields,
\be
\chi({\bf q},\eta) = - \nabla_{\bf q} \cdot{\bf\Psi} , \;\;\;
\theta({\bf q},\eta) = - \nabla_{\bf q} \cdot{\bf v} .
\label{eq:chi-theta-def} 
\ee
Then, taking the divergence of the equations of motion (\ref{eq:dPsi-deta})-(\ref{eq:dv-deta}),
we obtain
\ba
&& \frac{\partial \chi}{\partial\eta} = \theta , 
\label{eq:dchi-deta} \\
&& \frac{\partial\theta}{\partial\eta} + \left( \frac{3\Omega_{\rm m}}{2f^2} - 1 \right) \theta 
= \frac{3\Omega_{\rm m}}{2f^2} \zeta ,
\label{eq:dtheta-deta}
\ea
where we introduced the divergence of the force in Lagrangian space,
\be
\zeta({\bf q},\eta) = - \nabla_{\bf q} \cdot{\bf F}  .
\label{eq:zeta-def}
\ee
From the equation of motion (\ref{eq:dchi-deta}), we obtain for the time derivative
of the equal-times product
\be
\frac{\partial}{\partial\eta} \langle \chi_1 \chi_2 \rangle =
\langle  \theta_1 \chi_2 + \chi_1 \theta_2 \rangle ,
\ee
where $\chi_i=\chi({\bf q}_i,\eta)$. 
This gives for the equal-times power spectrum $P_{\chi\chi}(k,\eta)$
\be
\frac{\partial P_{\chi\chi}}{\partial\eta} = 2 P_{\chi\theta} .
\label{eq:dPchichi-deta}
\ee
In the same fashion, from Eqs.(\ref{eq:dchi-deta})-(\ref{eq:dtheta-deta}) we obtain
\be
\frac{\partial P_{\chi\theta}}{\partial\eta} = P_{\theta\theta} 
+ \left( 1 - \frac{3\Omega_{\rm m}}{2f^2} \right) P_{\chi\theta} + \frac{3\Omega_{\rm m}}{2f^2}
P_{\chi\zeta} 
\label{eq:dPchitheta-deta}
\ee
and
\be
\frac{\partial P_{\theta\theta}}{\partial\eta} = \left( 2 - \frac{3\Omega_{\rm m}}{f^2} \right) 
P_{\theta\theta} + \frac{3\Omega_{\rm m}}{f^2} P_{\theta\zeta} .
\label{eq:dPthetatheta-deta}
\ee
All equations written so far are exact. Of course, the problem is that the system 
(\ref{eq:dPchichi-deta})-(\ref{eq:dPthetatheta-deta}) is not closed, as it involves
the force cross power spectra $P_{\chi\zeta}$ and $P_{\theta\zeta}$.

In the linear regime, we have seen from Eq.(\ref{eq:F_L}) that ${\bf F}_L = {\bf\Psi}_L$.
Therefore, $\zeta_L=\chi_L$ and we have
\be
P_{L \chi\chi} \!=\! P_{L \chi\theta} \!=\! P_{L \chi\zeta} \!=\! P_{L\theta\theta} 
\!=\! P_{L \theta\zeta} \!=\! P_{L \zeta\zeta} \!=\!  e^{2\eta} P_{L0}(k) ,
\label{eq:PL-cross}
\ee
and we can check that this is a solution of the system 
(\ref{eq:dPchichi-deta})-(\ref{eq:dPthetatheta-deta}).

In the nonlinear regime, to be able to use the system 
(\ref{eq:dPchichi-deta})-(\ref{eq:dPthetatheta-deta})
we also need to express $P_{\chi\zeta}$ and $P_{\theta\zeta}$ in terms
of $\{ P_{\chi\chi}, P_{\chi\theta}, P_{\theta\theta} \}$.
This is the point where our approximation scheme enters, as described in 
Sec.~\ref{sec:curl-free-Gaussian} below for the case of a curl-free Gaussian ansatz.

As noticed in the previous section, there is some freedom in the choice of the constraint 
equations, and instead of considering these two-point statistics we could have chosen
the constraints associated with the time derivatives of higher-order moments 
$\langle \chi^n \theta^m \rangle$, or more intricate nonlinear functionals.
The constraints (\ref{eq:dPchichi-deta})-(\ref{eq:dPthetatheta-deta}) have the advantage
of simplicity and seem more natural to constrain a Gaussian ansatz, such as the one
presented in Sec.~\ref{sec:curl-free-Gaussian} below.

\section{Curl-free Gaussian ansatz}
\label{sec:curl-free-Gaussian}

\subsection{Definition of the ansatz}
\label{sec:ansatz-curl-free}

To illustrate the method proposed in this paper, we consider the simplest ansatz for the
probability distribution ${\cal P}$: the curl-free Gaussian displacement field.
Thus, we generalize the linear solution (\ref{eq:PsiL-deltaL}) by writing
\be
{\bf \Psi}({\bf k}) = \frac{i {\bf k}}{k^2} \chi({\bf k}) , \;\;\;
{\bf v}({\bf k}) = \frac{i {\bf k}}{k^2} \theta({\bf k}) ,
\label{eq:chi-theta-Fourier}
\ee
where $\chi$ and $\theta$ are the displacement and velocity divergences defined
in Eq.(\ref{eq:chi-theta-def}), and we take $\chi$ and $\theta$ to be
Gaussian scalar fields with zero mean.

Then, the power spectra $P_{\chi\chi}(k,\eta)$, $P_{\chi\theta}(k,\eta)$ and
$P_{\theta\theta}(k,\eta)$ fully define the Gaussian probability distribution 
${\cal P}({\bf \Psi},{\bf v};\eta)$.
This ansatz goes beyond the linear regime in two manners; firstly, the power spectrum
$P_{\chi\chi}$ can be different from the linear density power spectrum, secondly,
the spectra $P_{\chi\chi}(k,\eta)$, $P_{\chi\theta}(k,\eta)$ and $P_{\theta\theta}(k,\eta)$
can be different from one another.
This means that this ansatz is also more general than the Zeldovich approximation.

Clearly, this Gaussian ansatz allows us to close  the system 
(\ref{eq:dPchichi-deta})-(\ref{eq:dPthetatheta-deta}), because by its definition
all statistical properties of the fields ${\bf\Psi}$ and ${\bf v}$ are determined by the power 
spectra $\{ P_{\chi\chi}, P_{\chi\theta}, P_{\theta\theta} \}$. Then, the correlations
$\langle \chi\zeta\rangle$ and $\langle \theta\zeta\rangle$ are also fully determined
by these three power spectra, because the gravitational force is fully determined by
the positions of the particles.
Note that the divergence of the force $\zeta$ is not Gaussian, as it is a nonlinear functional
of the displacement, but within our Gaussian ansatz we only need the two-point
spectra $P_{\chi\zeta}$ and $P_{\theta\zeta}$ to close the system.
Then, once we have expressed $P_{\chi\zeta}$ and $P_{\theta\zeta}$ in terms
of $\{ P_{\chi\chi}, P_{\chi\theta}, P_{\theta\theta} \}$, we obtain a closed system that
fully determines the evolution of the power spectra 
$\{ P_{\chi\chi}, P_{\chi\theta}, P_{\theta\theta} \}$,
given their initial conditions set by the linear regime (\ref{eq:PsiL-deltaL}),
\be
\eta \to -\infty : \;\;\; P_{\chi\chi} = P_{\chi\theta} = P_{\theta\theta} =  e^{2\eta} P_{L0}(k) .
\ee

\subsection{Force-displacement and force-velocity cross power spectra}
\label{sec:force-cross-spectra}

\subsubsection{Damping factor $\lambda(k)$}
\label{sec:damping}

To close the system (\ref{eq:dPchichi-deta})-(\ref{eq:dPthetatheta-deta}) and compute the
time evolution of the power spectra, hence of the Gaussian distribution ${\cal P}$,
we need to compute the force-displacement and force-velocity spectra
$P_{\chi\zeta}$ and $P_{\theta\zeta}$.
To avoid the problems associated with the homogeneous background and to focus on the 
divergence $\zeta$ of the force, it is convenient to consider the quantity
\be
C_{\chi\zeta}(Q) = - \langle \chi({\bf q}_1) \int_{|{\bf q}_2-{\bf q}_1|=Q} dS_2
\; {\bf n}_2 \cdot {\bf F}({\bf q}_2) \rangle .
\label{eq:C-chizeta-def}
\ee
The integral is the flux of the force through the sphere of radius $Q$, around
the point ${\bf q}_1$ in Lagrangian space. Here ${\bf n}_2$ is the outward
normal vector to the sphere $S_2$.
Using the divergence theorem and Eq.(\ref{eq:zeta-def}), it only depends on the
divergence $\zeta$ of the force and it writes
\be
C_{\chi\zeta}(Q) = \int_{|{\bf q}_2-{\bf q}_1| \leq Q} d{\bf q}_2 \langle \chi({\bf q}_1)  
\zeta({\bf q}_2) \rangle .
\label{eq:C-chizeta-1}
\ee
Going to Fourier space, we obtain
\be
C_{\chi\zeta}(Q) = (4\pi)^2 Q^2 \int_0^{\infty} dk \, k P_{\chi\zeta}(k) j_1(kQ) ,
\label{eq:C-chizeta-Hankel}
\ee
which is a simple Hankel transform of the power spectrum $P_{\chi\zeta}$.
Then, to derive $P_{\chi\zeta}$ we only need to compute $C_{\chi\zeta}(Q)$ from
its definition (\ref{eq:C-chizeta-def}), using the explicit expression (\ref{eq:Fq})
of the gravitational force. This reads
\ba
C_{\chi\zeta}(Q) & = & - i Q^2 \int d{\bf \Omega}_2 \int 
\frac{d{\bf q}' d{\bf k}'}{(2\pi)^3} \frac{{\bf n}_2 \cdot {\bf k}'}{k'^2+\mu^2} 
\nonumber \\
&& \times \langle \chi({\bf q}_1) 
e^{i {\bf k}' \cdot [ {\bf x}({\bf q}_2) - {\bf x}({\bf q}') ]} \rangle ,
\ea
where ${\bf q}_2={\bf q}_1+Q{\bf n}_2$.
Using the general property that if $\varphi$ and $\Phi$ are Gaussian fields of zero mean
we have
\be
\langle \varphi \, e^{\Phi} \rangle = \langle \varphi \Phi \rangle \, 
e^{\langle \Phi^2\rangle/2} ,
\ee
we obtain
\ba
&& \hspace{-0.5cm} C_{\chi\zeta}(Q) = Q^2 \int d{\bf \Omega}_2 
\int \frac{d{\bf q}' d{\bf k}'}{(2\pi)^3} \frac{{\bf n}_2 \cdot {\bf k}'}{k'^2+\mu^2} 
e^{i {\bf k}' \cdot ({\bf q}_2-{\bf q}')} \nonumber \\
&& \hspace{-0.5cm} \times \langle \chi({\bf q}_1) [ {\bf k}' \cdot 
({\bf\Psi}({\bf q}_2) - {\bf\Psi}({\bf q}')) ] \rangle 
e^{-\langle [{\bf k}' \cdot ({\bf\Psi}({\bf q}_2) - {\bf\Psi}({\bf q}'))]^2 \rangle/2} 
\nonumber \\
&&
\ea
Going to Fourier space, we obtain from Eq.(\ref{eq:chi-theta-Fourier})
\be
\langle \chi({\bf q}_1) {\bf\Psi}({\bf q}_2) \rangle = -i \int d{\bf k}_1 
\frac{{\bf k}_1}{k_1^2} e^{i {\bf k}_1 \cdot ( {\bf q}_1 - {\bf q}_2 ) }
P_{\chi\chi}(k_1) ,
\ee
and
\ba
&& \frac{1}{2} \langle [ {\bf k}' \cdot ( {\bf\Psi}({\bf q}_2) - {\bf\Psi}({\bf q}_1) ) ]^2 
\rangle = \int d{\bf k}'' \, P_{\chi\chi}(k'') \nonumber \\
&& \hspace{0.5cm} \times \left( \frac{ {\bf k}' \cdot {\bf k}'' }{k''^2} \right)^2 
\left[1-\cos({\bf k''}\cdot({\bf q}_2-{\bf q}_1))\right] .
\label{eq:Psi-dq-variance}
\ea
With the change of variable ${\bf q}'={\bf q}_2-{\bf q}$, this gives
\ba 
&& C_{\chi\zeta}(Q) = i Q^2 \int d{\bf \Omega}_2 
\int \frac{d{\bf q} d{\bf k}' d{\bf k}_1}{(2\pi)^3} 
\frac{{\bf n}_2 \cdot {\bf k}'}{k'^2+\mu^2} \frac{{\bf k}' \cdot {\bf k}_1}{k_1^2} 
\nonumber \\
&& \times P_{\chi\chi}(k_1) e^{-\int d{\bf k}'' \, P_{\chi\chi}(k'') 
[ 1 - \cos( {\bf k''} \cdot {\bf q} ) ] ({\bf k}' \cdot {\bf k}'')^2/k''^4 } \nonumber \\
&& \times e^{i {\bf k}'\cdot {\bf q} - i Q {\bf n}_2 \cdot {\bf k}_1} 
\left( e^{i {\bf k}_1 \cdot {\bf q}} -1 \right) .
\ea
Using the property
\be
\int d{\bf\Omega} \; ({\bf n}\cdot{\bf k}_2) \; e^{i Q {\bf n} \cdot {\bf k}_1} =
i 4\pi \frac{{\bf k}_2\cdot{\bf k}_1}{k_1} j_1(k_1 Q) ,
\ee
where $j_1(z)$ is the first-order spherical Bessel function, the integration over
${\bf\Omega}_2$ gives
\ba 
&& C_{\chi\zeta}(Q) = 4\pi Q^2 \int \frac{d{\bf q} d{\bf k}' d{\bf k}_1}{(2\pi)^3} 
\frac{({\bf k}'\cdot{\bf k}_1)^2}{(k'^2+\mu^2)k_1^3} P_{\chi\chi}(k_1) \nonumber \\
&& \times j_1(k_1 Q) \, e^{-\int d{\bf k}'' \, P_{\chi\chi}(k'') 
[1-\cos({\bf k''}\cdot{\bf q})] ({\bf k}' \cdot {\bf k}'')^2/k''^4 } \nonumber \\
&& \times e^{i{\bf k}'\cdot{\bf q}} \left( e^{i{\bf k}_1 \cdot{\bf q}} -1 \right) .
\ea
The integral over the angles of ${\bf q}$ and ${\bf k}'$ leaves a quantity that no
longer depends on the direction of ${\bf k}_1$. 
Therefore, the comparison with Eq.(\ref{eq:C-chizeta-Hankel}) directly gives
\be
P_{\chi\zeta}(k) =  P_{\chi\chi}(k) \lambda(k) ,
\label{eq:Pchizeta-lambda}
\ee
with
\ba
\lambda(k) & = & \int \frac{d{\bf q} d{\bf k}'}{(2\pi)^3} 
\frac{({\bf k}' \cdot {\bf k})^2}{(k'^2+\mu^2)k^2} e^{i{\bf k}'\cdot{\bf q}} 
\left( e^{i{\bf k} \cdot{\bf q}} -1 \right) \nonumber \\
&& \times \, e^{-\int d{\bf k}'' \, P_{\chi\chi}(k'') 
[1-\cos({\bf k''}\cdot{\bf q})] ({\bf k}' \cdot {\bf k}'')^2/k''^4 }  . \;\;\;
\label{eq:lambda-def}
\ea
Defining the quantities $\alpha(q)$ and $\beta(q)$ by
\ba
&& \alpha(q) = \frac{4\pi}{3} \int_0^\infty dk \, P_{\chi\chi}(k) 
[ 1 - j_0(k q) - j_2(k q) ] , \;\;\;
\label{eq:alpha-bar-def} \\
&& \beta(q) = 4\pi \int_0^\infty dk \, P_{\chi\chi}(k) j_2(k q) ,
\label{eq:beta-def} 
\ea
we have
\ba
&& \int d{\bf k}'' \, P_{\chi\chi}(k'') \frac{({\bf k}\cdot{\bf k}'')^2}{k''^4} 
\left[ 1 - \cos({\bf k''}\cdot{\bf q}) \right] = \nonumber \\
&& \alpha(q) k^2 + \beta(q) k^2 \left(\frac{{\bf k}\cdot{\bf q}}{k q} \right)^2 ,
\label{eq:exponent-alpha-beta}
\ea
and $\lambda(k)$ also reads as
\ba
\lambda(k) & = & \int \frac{d{\bf q} d{\bf k}'}{(2\pi)^3} 
\frac{({\bf k}' \cdot {\bf k})^2}{(k'^2+\mu^2)k^2} e^{i{\bf k}'\cdot{\bf q}} 
\left( e^{i{\bf k} \cdot{\bf q}} -1 \right) \nonumber \\
&& \times \, e^{-\alpha(q) k'^2 - \beta(q) k'^2 ({\bf k}'\cdot{\bf q})^2/(k'q)^2} .
\label{eq:lambda-0}
\ea

The computation of the force-velocity power spectrum $P_{\theta\zeta}$ is obtained
in the same fashion by considering the correlation $C_{\theta\zeta}(Q)$, where we
replace $\chi$ in Eq.(\ref{eq:C-chizeta-def}) by $\theta$.
As in Eq.(\ref{eq:Pchizeta-lambda}), this gives
\be
P_{\theta\zeta}(k) =  P_{\theta\chi}(k) \lambda(k) ,
\label{eq:Pthetazeta-lambda}
\ee
with the same factor $\lambda(k)$.

Thus, $\lambda(k)$ plays the role of a damping factor, that will lessen the positive correlation
between the force and the displacement and velocity fields, as compared with the linear
theory where $\lambda_L=1$.

\subsubsection{Absence of infrared divergences}
\label{sec:infrared}

We note that $\lambda$ only depends on the relative displacements, as was expected from
the expression (\ref{eq:Fq}) of the gravitational force, which only depends on relative
distances. This is because we work in a Lagrangian approach and only consider equal-times
statistics (associated with the probability distribution ${\cal P}$). Then, uniform
displacements and velocities have no effect on the divergences $\chi=-\nabla_{\bf q}\cdot{\bf\Psi}$
and $\theta=-\nabla_{\bf q}\cdot{\bf v}$. This ensures that spurious infrared divergences
or large infrared contributions, which arise in Eulerian approaches and then need special care
\cite{Blas:2016sfa,Peloso:2016qdr,Senatore:2017pbn,Noda:2017tfh}, do not appear at all in our
approach. This can be seen in Eq.(\ref{eq:lambda-def}) through the fact that the argument 
of the last exponential is not the one-point displacement variance, 
given by
\ba
\alpha_\infty & = & \frac{4\pi}{3} \int_0^\infty dk \, P_{\chi\chi}(k) 
= \frac{1}{3} \langle | {\bf\Psi}({\bf 0}) |^2 \rangle \nonumber \\
& = & \lim_{q\to\infty} \frac{1}{6} \langle | {\bf\Psi}({\bf q}) - {\bf\Psi}(0) |^2 \rangle ,
\label{eq:alpha-infty}
\ea
such that
\be
\int d{\bf k}'' \, P_{\chi\chi}(k'') \frac{({\bf k}\cdot{\bf k}'')^2}{k''^4} 
= \alpha_\infty k^2 .
\ee
In contrast, Eq.(\ref{eq:lambda-def}) depends on the two-point relative 
displacement variance over Lagrangian distance $q$, associated with the factor 
$1-\cos({\bf k}''\cdot {\bf q})$. 
This factor damps the contribution of long wavelengths and regularizes the infrared 
divergences that can appear in Eulerian or different-times approaches.

\subsubsection{Behavior of the variances $\alpha(q)$ and $\beta(q)$}
\label{sec:behavior-alpha-beta}

We shall see below that we obtain a displacement power spectrum $P_{\chi\chi}(k)$ that
decays faster than $k^{-3}$ at large $k$. This implies the small-scale behaviors
\be
q\to 0 : \;\;\; \alpha = \alpha_0 q^2 + \dots , \;\;\;
\beta = \beta_0 q^2 + \dots ,
\label{eq:alpha-beta-q0}
\ee
where the dots stand for higher-order terms in $q$ and we have
\be
\alpha_0=\frac{\beta_0}{2} = \frac{2\pi}{15} \int_0^\infty dk \, k^2 P_{\chi\chi}(k) > 0 .
\label{eq:alpha0-beta0-def}
\ee
At large scales we have
\be
q\to\infty: \;\;\; \alpha \to \alpha_\infty , \;\;\; \beta\to 0 ,
\ee
provided $P_{\chi\chi}(k)$ increases more slowly than $1/k$ at low $k$.

\subsubsection{Behavior of the damping factor $\lambda(k)$}
\label{sec:behavior-lambda}

\paragraph{Linear regime --}
\label{sec:linear-lambda}

At early times, the amplitude of the power spectrum $P_{\chi\chi}$ and of the
displacement variances $\alpha$ and $\beta$ vanish. Then, the last exponential
in Eq.(\ref{eq:lambda-0}) goes to unity and the integration over ${\bf q}$ gives
$\lambda=1$,
\be
\alpha\to 0, \;\; \beta\to 0 : \;\;\; \lambda \to 1 .
\ee
Thus, we recover the linear regime with $P_{\chi\zeta}=P_{\chi\chi}$
and $P_{\theta\zeta}=P_{\theta\chi}$.

\paragraph{Large scales --}
\label{sec:large-scales-lambda}

The limit of large scales corresponds to $k\to 0$. This is not equivalent to the
limit $P_{\chi\chi}\to 0$. For instance, SPT corresponds to
expansions over powers of $P_L(k)$ \cite{Bernardeau:2001qr}, 
which corresponds to the limit $P_L \to 0$,
whereas EFT approaches \cite{Baumann:2010tm,Carrasco:2012cv}
consider the limit $k\to 0$.
This can include nonperturbative terms, such as 
$P_L(k) \frac{k^2}{k_{\rm NL}^2} e^{-1/\sigma^2}$, where $k_{\rm NL}$ is the wave number
that marks the transition to the nonlinear regime and $\sigma^2$ is a displacement variance
such as $\alpha$ and $\beta$. These terms do not scale as integer powers of $P_L$
and are beyond the reach of standard perturbative expansions because of the exponential.
They arise from shell crossing and the factor $e^{-1/\sigma^2}$ describes the
probability of shell crossing or gravitational collapse for Gaussian initial conditions.
(In EFT approaches, the nonperturbative factor is not derived
but obtained from fits to numerical simulations and inserted as a coefficient of
higher derivative operators in the effective Lagrangian or equations of motion.)
Nevertheless, as seen in appendix~\ref{sec:expression-numerical},
we recover the linear regime in the large-scale limit, 
\be
k \to 0 : \;\;\; \lambda \to 1 .
\label{eq:lambda-low-k}
\ee
This means that, as usual, the linear regime and the large-scale limit coincide
at leading order.

\paragraph{Small scales --}
\label{sec:small-scales-lambda}

As seen in appendix~\ref{sec:expression-numerical}, we have the small-scale
behavior
\be
k \to \infty: \;\;\; \lambda(k) \sim \lambda_{\infty} \ln(k) ,
\label{eq:lambda-large-k}
\ee
with
\be
\lambda_\infty = - \frac{e^{-1/(12\alpha_0)}}{6\sqrt{3\pi}\alpha_0^{3/2}} .
\label{eq:lambda-infty-def}
\ee
Thus, in the nonlinear regime the damping factor $\lambda$ decreases below unity and actually
goes to $-\infty$.
This will give rise to a strong deviation of the power spectrum $P_{\chi\chi}(k)$ from
the linear power spectrum $P_L(k)$.
The nonperturbative reach of our approach appears clearly in Eq.(\ref{eq:lambda-infty-def})
through the nonperturbative exponential factor, which vanishes at all orders of perturbation
theory in powers of $P_L$.

\subsection{Density power spectrum}
\label{sec:density-power-Gaussian}

With the Gaussian ansatz (\ref{eq:chi-theta-Fourier}), we can compute the density power
spectrum (\ref{eq:P-deltadelta-def}) exactly as for the Zeldovich approximation
\cite{Zeldovich:1969sb}.
Indeed, although the displacement field ${\bf \Psi}({\bf q})$ is no longer given by linear theory,
it is still Gaussian within this approximation and the statistical average in 
Eq.(\ref{eq:P-deltadelta-def}) is again straightforward. Using Eq.(\ref{eq:Psi-dq-variance})
we obtain for $k>0$
\be
P(k) = \int \frac{d{\bf q}}{(2\pi)^3} e^{i {\bf k}\cdot{\bf q} -\int d{\bf k}' \, P_{\chi\chi}(k') 
[ 1 - \cos( {\bf k'} \cdot {\bf q} ) ] ({\bf k} \cdot {\bf k}')^2/k'^4 } .
\label{eq:Pk-density-Pchichi}
\ee
This is the same expression as for the Zeldovich power spectrum, except that the
linear power spectrum $P_L(k)$ in the exponent is replaced by the nonlinear
power spectrum $P_{\chi\chi}(k)$.
With the notations of Eq.(\ref{eq:exponent-alpha-beta}) this also reads as
\be
P(k) = \int \frac{d{\bf q}}{(2\pi)^3} e^{i k q \mu - \alpha(q) k^2 - \beta(q) k^2 \mu^2} ,
\label{eq:Pk-alpha-beta-q}
\ee
where $\mu=({\bf k}\cdot{\bf q})/(kq)$ is the cosine of the angle between ${\bf k}$ and 
${\bf q}$. We can integrate over angles to obtain \cite{1995MNRAS.273..475S,Valageas:2007ge}
\be
P(k) = \int \frac{d q}{2\pi^2}  q^2 e^{-(\alpha+\beta)k^2} \sum_{\ell=0}^{\infty} 
\left( \frac{2\beta k}{q} \right)^{\ell} j_{\ell}(kq) .
\label{eq:Pk-density-alpha-beta}
\ee
We describe in appendix~\ref{sec:numerical-density} our numerical method
to compute the power spectrum (\ref{eq:Pk-alpha-beta-q}).

The power spectrum associated with the standard Zeldovich approximation
\cite{Zeldovich:1969sb} is also given by Eqs.(\ref{eq:Pk-density-Pchichi}) and
(\ref{eq:Pk-alpha-beta-q}), where we replace the nonlinear power $P_{\chi\chi}$ 
and variances $\{\alpha,\beta\}$ by their linear values 
\cite{1995MNRAS.273..475S,Taylor:1996ne,Valageas:2007ge},
\be
P_Z(k) = \int \frac{d{\bf q}}{(2\pi)^3} e^{i k q \mu - \alpha_L(q) k^2 - \beta_L(q) k^2 \mu^2} .
\label{eq:PkZ-alpha-beta-q}
\ee
In particular, the same numerical methods can be used for our model (\ref{eq:Pk-alpha-beta-q})
and for the Zeldovich power spectrum (\ref{eq:PkZ-alpha-beta-q}).

\section{Comparison with some other approaches}
\label{sec:comparison-theory}

In comparison with previous studies, this work is to some extent a continuation of
\cite{Valageas:2013gba}, where we developed a Lagrangian-space ansatz designed
to go beyond perturbation theory. That model matched SPT 
up to one-loop order on large scales and a halo model \cite{Cooray:2002dia} on small scales, 
by combining various elements. It included some parameters fitted to numerical
simulations (e.g., the halo mass function and the halo profiles) to recover at high $k$
a halo model defined a priori. In contrast, in the approach presented in this paper,
we do not enforce any specific behavior on either large or small scales and we have
no free parameters.

Our method is also related to \cite{McDonald:2017ths}, as it is based on a Lagrangian
approach, in order to go beyond shell crossing, and on statistical quantities instead
of individual realizations of the fields. However, \cite{McDonald:2017ths}
considered the generating functional $Z[{\bf j}]$ of correlation functions, which is then
expanded up to some finite order. This means that the equation of motion, which enters
the action, is also expanded and approximated up to some order. In addition,
\cite{McDonald:2017ths} introduce by hand an auxiliary truncation of the linear power 
spectrum, to separate the modes that are kept in the Gaussian part and in the expanded
part. In contrast, in this paper we work with the probability distribution functional
${\cal P}({\bf \Psi},{\bf v},t)$ (in practice, it is defined by the power spectra
in the Gaussian case) and we do not expand the equations of motion, which
are exactly taken into account at the full nonlinear level (but we only include
a few of them, among the infinite number of constraints obeyed by the $n$-point correlations).
Also, we do not introduce a truncation of the linear power spectrum of the displacement
field. It arises from the equations of motion themselves.

Our work is also related to \cite{Blas:2015qsi,Blas:2016sfa}, as it is based on the evolution
with time of the distribution functional ${\cal P}$ of the fields.
However, \cite{Blas:2015qsi} considered the probability distribution ${\cal P}(\delta,\theta)$
of the Eulerian density and velocity divergence fields, whereas we consider the
probability distribution of the Lagrangian displacement field. Then, they assume
a curl-free velocity field, based on the Euler equation, which breaks down beyond shell crossing.
They also perform an expansion of the probability distribution ${\cal P}(\theta)$,
written under the form $e^{-\sum_n \Gamma_n \theta^n}$, by expanding over the non-Gaussian
terms $n\geq 3$. In particular, the Gaussian part is given by $\Gamma_2=1/P_L(k)$
and the nonlinearity of the gravitational dynamics is captured by the higher orders
$\Gamma_n$, $n \geq 3$. In a fashion somewhat similar to SPT,
these higher-order vertices are obtained from recursion relations that follow from 
the evolution equation of ${\cal P}(\theta)$.
The spirit of the approach proposed in this paper is quite different in this respect.
Instead of capturing the nonlinearities of the dynamics by adding higher order terms,
such as higher powers in $P_L$ in SPT or higher-order vertices $\Gamma_n$,
the nonlinearity is already partly taken into account in the Gaussian part of the
probability distribution, as the displacement and velocity power spectra get modified
from the linear prediction.
Following the analogy with the minimization problem of a cost functional $S[\varphi(x)]$,
discussed in the introduction and in Sec.~\ref{sec:probability-distribution}, 
our strategy is not to estimate the minimum $\varphi_{\min}$
by expanding $\varphi$ around the known minimum $\varphi_0$ of a simpler
cost functional $S_0$, in powers of $S-S_0$. Instead, we look for the exact minimum
in a simpler subspace of $\{\varphi\}$. For instance, if $\varphi_0(x)=a_0 (x-b_0)^2$ is quadratic,
the strategy at lowest order is simply to let free the parabola parameters, 
$a_0$ and $b_0$, and find their new values $\{a,b\}$ that minimize the new functional $S$.
Clearly, this allows one to reach minima that are very far from the initial guess $\varphi_0$
and obtain nonpeturbative results. 
In practice, this means that we avoid explicit perturbative expansions.

Thus, we emphasize that the result (\ref{eq:Pchizeta-lambda})-(\ref{eq:lambda-def}) is
nonperturbative. Indeed, we do not expand on the displacements ${\bf\Psi}$, which are
not assumed to be small. Within the Gaussian ansatz for the probability distribution ${\cal P}$,
we perform the exact computation of the displacement-force correlation 
$\langle\chi\zeta\rangle$, using the exact expression (\ref{eq:Fq})
of the gravitational force.
Thus, our approach follows a strategy that is quite different from usual perturbative methods.
We do not expand the equations of motion either, which are kept at a fully nonlinear level
as in (\ref{eq:dPchichi-deta})-(\ref{eq:dPthetatheta-deta}), but we only include the lowest-order
ones. Then, the approximation scheme, or truncation, occurs instead at the level of the trial 
distribution ${\cal P}$. 

For the simplest Gaussian ansatz considered in this paper, this program is easy
to complete, as exact computations are easily performed for Gaussian fields.
However, for higher orders, that is, for more complex ansatze that go beyond the Gaussian,
this may represent a much more difficult task. Indeed, to fulfil the nonperturbative
promise of this approach, we would need again to compute exactly quantities such as
$\langle {\bf\Psi} {\bf F} \rangle$. This may prove much more difficult for non-Gaussian
probability distributions and represent a drawback of this approach.

In terms of the density field, Eq.(\ref {eq:Pk-density-Pchichi}) coincides with a truncated 
Zeldovich approximation \cite{Coles:1992vr}. 
However, in our case the truncation is not set a priori with 
a cutoff that follows from an educated guess or a fit to numerical simulation. 
Instead, the cutoff $\lambda(k)$ is generated by the equations of motion themselves 
and there are no free parameters to be fitted to simulations.
This represents a significant improvement over most previous analytical approaches,
which either fail to regularize small-scale divergences (such as SPT)
or introduce counterterms with an amplitude that must be measured in simulations
(such as EFT methods).
  
As in EFT methods \cite{Baumann:2010tm,Carrasco:2012cv}, the ultraviolet divergences,
or artificially large contributions, associated with the continuous rise of the linear 
density fluctuations on small scales, are tamed. In EFT this is done by introducing 
counterterms to the SPT diagrammatic computations, which arise from new operators 
in the Lagrangian or the equations of motion. The latter are expected to describe the 
effects of multistreaming that are not included in the hydrodynamical equations of motion.
They are obtained from systematic large-scale expansions, but with free coefficients that
must be fitted to numerical simulations. In our approach, as we shall see in the next
sections, the displacement linear power spectrum is damped at high $k$ by the
factor $\lambda(k)$ in Eq.(\ref{eq:Pchizeta-lambda}). As in the truncated 
Zeldovich approximation \cite{Coles:1992vr}, this removes ultraviolet divergences
and provides an implicit regularization. For instance, we shall see that in 
Sec.~\ref{sec:numerical-n=0} that even when the standard Zeldovich power spectrum
does not exist, because of such ultraviolet divergences, our approach remains
well defined. In contrast with EFT methods, this does not involve free parameters and
new operators, and this self-regularization directly follows from the equations of motion.

\section{Self-similar dynamics}
\label{sec:self-similar}

\subsection{Differential equations for power spectra}
\label{sec:differential-self}

To illustrate our approach, we consider in this section the simpler case of the Einstein-de Sitter
cosmology, $\Omega_{\rm m}=1$, where $D(t) = a(t)$, and the initial linear power
spectrum is a power law,
\be
P_L(k) \propto k^n .
\label{eq:n-Pk-def}
\ee
Then, because Newtonian gravity is scale free, it is well known that the dynamics are
self-similar \cite{Peebles1980} and statistics no longer depend on time once they are 
expressed in units of the nonlinear wave number $k_{\rm NL}(t)$ that marks the transition 
to the nonlinear regime.
Defining for instance $k_{\rm NL}(\eta)$ by
\be
4\pi k_{\rm NL}^3 P_L(k_{\rm NL},\eta) = 1 ,
\ee
the linear power spectrum can be written as
\be
P_L(k,\eta) = \frac{e^{2\eta}}{4\pi k_0^3} \left( \frac{k}{k_0} \right)^n 
= \frac{1}{4\pi k_{\rm NL}^3} \left( \frac{k}{k_{\rm NL}} \right)^n ,
\ee
where $k_0$ defines the normalization of the linear power spectrum and $k_{\rm NL}(\eta)$
is given by
\be
k_{\rm NL}(\eta) = k_0 e^{-2\eta/(n+3)} .
\label{eq:k-NL-def}
\ee
Then, all power spectra have the self-similar form
\be
P(k,\eta) = \frac{1}{4\pi k^3} {\cal D}\left( \frac{k}{k_{\rm NL}(\eta)} \right) 
= \frac{1}{4\pi k^3} {\cal D}\left( \frac{k}{k_0} e^{2\eta/(n+3)} \right) ,
\label{eq:Pk-self-similar}
\ee
with the scaling function ${\cal D}$ that only depends on the ratio $k/k_{\rm NL}$.
In the linear regime we have ${\cal D}_L(x) = x^{n+3}$.
The self-similar evolution (\ref{eq:Pk-self-similar}) implies the relation
\be
\frac{\partial P}{\partial\eta} = \frac{2}{n+3} \left( k \frac{\partial P}{\partial k} + 3 P \right) .
\label{eq:deta-dk-self}
\ee
This exact relation allows us to replace time derivatives of statistical quantities by
spatial derivatives. Using also $\Omega_{\rm}/f^2=1$ in the Einstein-de Sitter
cosmology, Eqs.(\ref{eq:dPchichi-deta})-(\ref{eq:dPthetatheta-deta}) simplify as
\ba
&& \frac{2}{n+3} \left( \frac{\partial P_{\chi\chi}}{\partial\ln k} 
+ 3 P_{\chi\chi} \right) = 2 P_{\chi\theta} , 
\label{eq:dPchichi-dlnk} \\
&& \frac{2}{n+3} \left( \frac{\partial P_{\chi\theta}}{\partial\ln k} 
+ 3 P_{\chi\theta} \right) = P_{\theta\theta} - \frac{1}{2} P_{\chi\theta} 
+ \frac{3}{2} P_{\chi\zeta} , \hspace{0.5cm}
\label{eq:dPchitheta-dlnk} \\
&& \frac{2}{n+3} \left( \frac{\partial P_{\theta\theta}}{\partial\ln k} 
+ 3 P_{\theta\theta} \right) = - P_{\theta\theta} + 3 P_{\theta\zeta} .
\label{eq:dPthetatheta-dlnk}
\ea
Introducing the 3D power $\Delta^2(k,\eta)$ per logarithmic interval of wave number by
\be
\Delta^2(k,\eta) = 4\pi k^3 P(k,\eta) = {\cal D}\left(\frac{k}{k_{\rm NL}}\right) ,
\label{eq:Delta2-scaling-n}
\ee
and the wave number scaling coordinate $u$,
\be
u = (n+3) \ln\left(\frac{k}{k_{\rm NL}}\right) ,
\label{eq:u-def}
\ee
the system (\ref{eq:dPchichi-dlnk})-(\ref{eq:dPthetatheta-dlnk}) writes
\ba
&& {\cal D}_{\chi\chi}' = {\cal D}_{\chi\theta} , 
\label{eq:dDchichi-du} \\
&& {\cal D}_{\chi\theta}' = - \frac{1}{4} {\cal D}_{\chi\theta} 
+ \frac{1}{2} {\cal D}_{\theta\theta} + \frac{3}{4} {\cal D}_{\chi\zeta} , 
\label{eq:dDchitheta-du} \\
&& {\cal D}_{\theta\theta}' = - \frac{1}{2} {\cal D}_{\theta\theta} 
+ \frac{3}{2} {\cal D}_{\theta\zeta} ,
\label{eq:dDthetatheta-du}
\ea
where the prime denotes the derivative with respect to $u$.
The linear regime corresponds to all ${\cal D}_{\star\star}$ identical with
\be
{\cal D}_L(u) = e^u .
\label{eq:DL-def}
\ee
Thanks to the self-similarity (\ref{eq:Pk-self-similar}), the system of partial 
differential equations (\ref{eq:dPchichi-deta})-(\ref{eq:dPthetatheta-deta}) 
has been transformed into a system of ordinary differential equations.
These equations are exact but require the force cross-power spectra
$P_{\chi\zeta}$ and $P_{\theta\zeta}$ to form a closed system.

\subsection{Curl-free Gaussian ansatz}
\label{sec:curl-free-Gaussian-self}

Within the curl-free Gaussian ansatz presented in Sec.~\ref{sec:curl-free-Gaussian},
we can close the system (\ref{eq:dDchichi-du})-(\ref{eq:dDthetatheta-du}) 
thanks to Eqs.(\ref{eq:Pchizeta-lambda}) and (\ref{eq:Pthetazeta-lambda}).
This gives
\ba
&& {\cal D}_{\chi\chi}' = {\cal D}_{\chi\theta} , 
\label{eq:dDchichi-du-lambda} \\
&& {\cal D}_{\chi\theta}' = \frac{3}{4} \lambda {\cal D}_{\chi\chi}
- \frac{1}{4} {\cal D}_{\chi\theta} + \frac{1}{2} {\cal D}_{\theta\theta} , 
\label{eq:dDchitheta-du-lambda} \\
&& {\cal D}_{\theta\theta}' = \frac{3}{2} \lambda {\cal D}_{\chi\theta} 
- \frac{1}{2} {\cal D}_{\theta\theta} .
\label{eq:dDthetatheta-du-lambda}
\ea
By combining these three equations we can eliminate ${\cal D}_{\chi\theta}$
and ${\cal D}_{\theta\theta}$ to obtain a third-order linear equation over
${\cal D}_{\chi\chi}$,
\be
{\cal D}_{\chi\chi}''' + \frac{3}{4} {\cal D}_{\chi\chi}'' + \left( \frac{1}{8} - \frac{3\lambda}{2} \right)
{\cal D}_{\chi\chi}' - \left( \frac{3\lambda}{8} + \frac{3\lambda'}{4} \right)
{\cal D}_{\chi\chi}  =  0 .
\ee
The general solution of this equation is \cite{Polyanin2017}
\be
{\cal D}_{\chi\chi}(u) = e^{-u/4} \left[ c_1 y_1(u)^2 + c_2 y_1(u) y_2(u) + c_3 y_2(u)^2 \right] ,
\label{eq:D-chichi-y1y2}
\ee
where $c_i$ are constants and $y_i(u)$ are two independent solutions of the
second-order linear differential equation
\be
y'' - \frac{24\lambda+1}{64} y = 0 .
\label{eq:y-ODE-self}
\ee
On large scales, that is, for large negative $u$, we have $\lambda=1$ and the two 
independent solutions are $y_1=e^{5u/8}$ and $y_2=e^{-5u/8}$.
The matching to the linear regime (\ref{eq:DL-def}) implies $c_2=c_3=0$
in Eq.(\ref{eq:D-chichi-y1y2}). Therefore, we obtain
\be
{\cal D}_{\chi\chi}(u) = e^{-u/4} y(u)^2 \geq 0 ,
\label{eq:D-chichi-y2}
\ee
where $y(u)$ is the solution of Eq.(\ref{eq:y-ODE-self}) with the boundary condition
at large negative $u$
\be
u \to -\infty: \;\;\; y(u) = e^{5u/8} .
\label{eq:yu-boundary-self}
\ee
On small scales, that is, for $u\to \infty$, we have from Eq.(\ref{eq:lambda-large-k})
$\lambda \sim -|\lambda_\infty| u/(n+3)$. This gives the asymptotic behavior
\be
u \to \infty: \;\;\; y(u) = c_1 {\rm Ai}(- \gamma_\infty u) + c_2 {\rm Bi}(- \gamma_\infty u)  ,
\ee
where ${\rm Ai}(x)$ and ${\rm Bi}(x)$ are the Airy functions of the first and second kind,
$c_i$ are constants and $\gamma_\infty$ is given by
\be
\gamma_\infty = \left( \frac{3 |\lambda_\infty|}{8(n+3)} \right)^{1/3} .
\ee
From the asymptotic behaviors of the Airy functions we obtain
\ba
k \gg k_{\rm NL} : && \Delta^2_{\chi\chi}(k,\eta) \sim
\left( \frac{k}{k_{\rm NL}} \right)^{-(n+3)/4} \nonumber \\
&& \times [ \ln(k/k_{\rm NL}) ]^{-1/2} [ c_1 \sin\psi + c_2 \cos\psi ]^2 , \hspace{0.5cm}
\label{eq:D-chichi-nonlinear-self}
\ea
where $c_i$ are constants and $\psi(k,\eta)$ is given at leading order by      
\be
\psi(k,\eta) \sim \sqrt{\frac{|\lambda_\infty|}{6}} (n+3) [ \ln(k/k_{\rm NL}) ]^{3/2} .
\label{eq:psi-oscill-def}
\ee
From Eqs.(\ref{eq:dDchichi-du-lambda})-(\ref{eq:dDchitheta-du-lambda}) and
Eq.(\ref{eq:D-chichi-y2}) we obtain for the other power spectra
\ba
&& {\cal D}_{\chi\theta}(u) = e^{-u/4} \left( - \frac{y^2}{4} + 2 y y' \right)  , 
\label{eq:D-chitheta-f2}\\
&& {\cal D}_{\theta\theta}(u) = e^{-u/4} \left( \frac{y}{4} - 2 y' \right)^2 \geq 0 .
\label{eq:D-thetatheta-f2}
\ea
Omitting the sine and cosine factors, this gives the small-scale behaviors
\ba
&& k \gg k_{\rm NL} : \;\;\;  \Delta^2_{\chi\theta}(k,\eta) \sim 
\left( \frac{k}{k_{\rm NL}} \right)^{-(n+3)/4} ,
\label{eq:D-chitheta-y2}
\\
&& \Delta^2_{\theta\theta}(k,\eta) \sim \left( \frac{k}{k_{\rm NL}} \right)^{-(n+3)/4}
[ \ln(k/k_{\rm NL}) ]^{1/2} .
\label{eq:D-thetatheta-y2}
\ea
Thus, at leading order the three logarithmic power spectra decay as
$\Delta^2_{**}(k) \propto k^{-(n+3)/4}$, and the power spectra decays faster
than $k^{-3}$, as $P_{**}(k) \propto k^{-3-(n+3)/4}$.
This leads to the universal behavior (\ref{eq:alpha-beta-q0}), independently of the
exponent $n$ of the linear power spectrum.

A remarkable feature of the solutions (\ref{eq:D-chichi-y2}) and 
(\ref{eq:D-chitheta-y2})-(\ref{eq:D-thetatheta-y2}) is that the auto power spectra
$\Delta^2_{\chi\chi}$ and $\Delta^2_{\theta\theta}$ are always positive, whereas
the cross power spectrum $\Delta^2_{\chi\theta}$ can change sign.
By definition, auto power spectra must be positive, but this property is often violated
in approximation schemes, such as perturbative expansions. 
Indeed, terms of successive orders can become increasingly large with alternating signs 
on nonlinear scales, and the sign of the prediction depends on the truncation order
if the series has not converged yet.

In our approach, even though we performed the simplest Gaussian approximation
in Sec.~\ref{sec:curl-free-Gaussian}, the auto power spectra
$\Delta^2_{\chi\chi}$ and $\Delta^2_{\theta\theta}$ are always positive.
This was not obvious from the differential system 
(\ref{eq:dPchichi-dlnk})-(\ref{eq:dPthetatheta-dlnk}) and was not explicitly enforced
by additional constraints.
This could signal the robustness of our approach.
It may follow from the fact that we keep track of the exact equations of motion 
(\ref{eq:dPchichi-dlnk})-(\ref{eq:dPthetatheta-dlnk}), and that the cross-power spectra
$P_{\chi\zeta}$ and $P_{\theta\zeta}$ are exactly computed from an ansatz
that is always physical, albeit different from the true particle distribution 
(we did not obtain the exact solution of the gravitational dynamics).
Indeed, the Gaussian ansatz of Sec.~\ref{sec:curl-free-Gaussian} corresponds to
a physical distribution of particles and velocities so that force cross power spectra
derived in this manner do not hide any inconsistencies.
(This is not the case for approaches that start directly at the level of the correlation
functions, where it is not always known whether there exists a distribution of particles
that provides a physical realization of the ansatz used for the density correlations.
Then, this ansatz may contain some inconsistencies,
that may be harmful or not, depending on the quantities and regimes of interest.) 
However, even though we have a physical ansatz at each time, we do not follow 
the exact dynamics. Therefore, there is no guarantee that our integration of 
(\ref{eq:dPchichi-dlnk})-(\ref{eq:dPthetatheta-dlnk}) avoids all inconsistencies.
Nevertheless, this approach is clearly a step in the direction towards self-consistency
and it appears to be sufficient to ensure positivity of displacement and velocity
auto power spectra.

\subsection{Numerical computation}
\label{sec:numerical-self}

To obtain the power spectra ${\cal D}_{**}(u)$ we need to solve the differential equation
(\ref{eq:y-ODE-self}), where $\lambda(u)$ depends on ${\cal D}_{\chi\chi}(u)$
through Eq.(\ref{eq:lambda-0}). This gives a nonlinear system of equations, which we solve
by an iterative procedure.
Starting with an initial guess for ${\cal D}_{\chi\chi}(u)$, which converges to the 
linear regime (\ref{eq:DL-def}) for $u \ll -1$ and decays as $e^{-u/4}$ for 
$u \gg 1$, we compute from Eq.(\ref{eq:lambda-0}) the damping factor $\lambda(u)$.
The expressions that we use in practice for the numerical computations are given
in appendix~\ref{sec:expression-numerical}.
Then, we obtain $y(u)$ from Eq.(\ref{eq:y-ODE-self}), and the power spectra
from (\ref{eq:D-chichi-y2}) and (\ref{eq:D-chitheta-f2})-(\ref{eq:D-thetatheta-f2}).
Next, we repeat the procedure, computing $\lambda$, $y$ and ${\cal D}_{**}$
from this new power spectrum ${\cal D}_{\chi\chi}$.
We iterate until the damping factor $\lambda$ and the power spectra have converged.
Finally, from the displacement power spectrum $P_{\chi\chi}$ we obtain the
density power spectrum $P(k)$ from Eq.(\ref{eq:Pk-density-Pchichi}).
Our numerical procedure is described in appendix~\ref{sec:numerical-density}.

We normalize the linear power spectra by $k_{\rm NL}=1$,
\be
k_{\rm NL} = 1 : \;\;\; P_L(k) = \frac{k^n}{4\pi} .
\label{eq:PL-n-norm}
\ee
This also means that the Lagrangian scale $q_{\rm NL}$ that marks the transition to the
nonlinear regime is of order unity.
We have noticed in Sec.~\ref{sec:behavior-alpha-beta} that the variances $\{\alpha,\beta\}$ 
show the universal behavior (\ref{eq:alpha-beta-q0}) at small $q$, because the nonlinear
power spectrum $P_{\chi\chi}(k)$ decays faster than $k^{-3}$ from 
Eq.(\ref{eq:D-chichi-nonlinear-self}).
At large separation $q$, the displacement variances 
(\ref{eq:alpha-bar-def})-(\ref{eq:beta-def}) are governed by the low-$k$ part of the
power spectrum $P_{\chi\chi}(k)$, which converges to the linear power spectrum $P_L(k)$.
This gives
\be
q \gg q_{\rm NL} : \;\;\; \beta(q) \propto q^{-(n+1)} \;\;\; \mbox{for} \;\;\; -3<n<1 ,
\label{eq:beta-large-q}
\ee
and
\ba
&& \alpha(q) \propto q^{-(n+1)} \to \infty \;\;\; \mbox{for} \;\;\; -3<n<-1 , \nonumber \\
&& \alpha(q) \to \alpha_\infty \;\;\; \mbox{for} \;\;\; -1<n<1 . 
\label{eq:alpha-large-q}
\ea

\subsubsection{Self-similar case with $n=0$}
\label{sec:numerical-n=0}

\begin{figure}
\begin{center}
\epsfxsize=8.8 cm \epsfysize=6 cm {\epsfbox{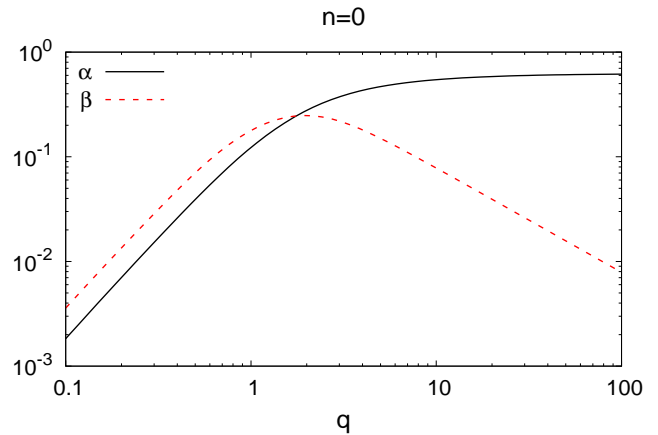}}
\end{center}
\caption{Variances $\alpha(q)$ and $\beta(q)$ defined by the nonlinear power spectrum
$P_{\chi\chi}(k)$ in the power-law case $n=0$.}
\label{fig_alphabeta_n0}
\end{figure}

We first consider for illustration the case $n=0$.
We show in Fig.~\ref{fig_alphabeta_n0} the variances $\alpha(q)$ and $\beta(q)$ 
defined by the final nonlinear power spectrum $P_{\chi\chi}(k)$, once the iterative
procedure explained above has converged.
We recover the small-scale quadratic behavior (\ref{eq:alpha-beta-q0}) and the
large-scale behavior (\ref{eq:beta-large-q})-(\ref{eq:alpha-large-q}), with a transition
at $q_{\rm NL} \sim 1$.

\begin{figure}
\begin{center}
\epsfxsize=8.8 cm \epsfysize=6 cm {\epsfbox{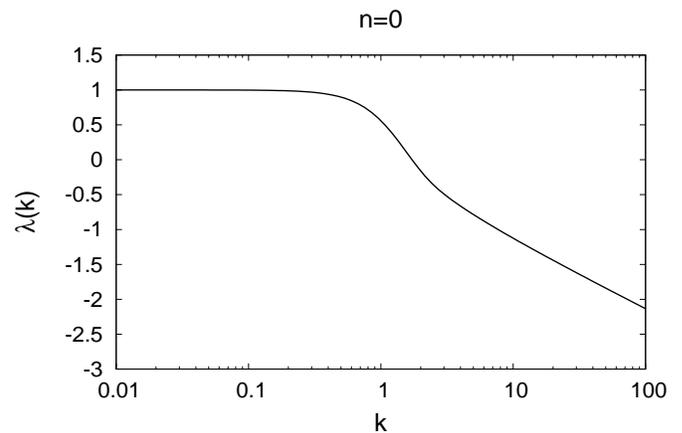}}
\end{center}
\caption{Damping factor $\lambda(k)$ for the power-law case $n=0$.}
\label{fig_lambdak_n0}
\end{figure}

We display in Fig.~\ref{fig_lambdak_n0} the damping factor $\lambda(k)$ of 
Eq.(\ref{eq:lambda-0}). For the numerical computation we use the expressions given in
the appendix~\ref{sec:expression-numerical}.
In agreement with Eqs.(\ref{eq:lambda-low-k}) and (\ref{eq:lambda-large-k}),
at low $k$ it goes to unity while at high $k$ it goes to $-\infty$ as $-\ln(k)$.
The transition occurs around $k_{\rm NL}=1$.

\begin{figure}
\begin{center}
\epsfxsize=8.8 cm \epsfysize=6 cm {\epsfbox{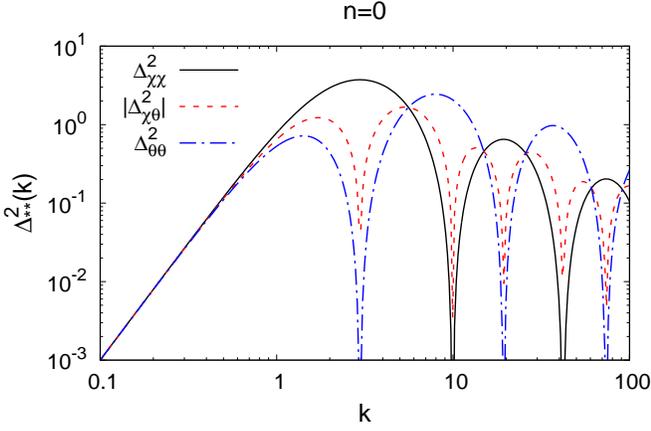}}
\end{center}
\caption{Displacement and velocity logarithmic power spectra $\Delta^2_{\chi\chi}$,
$|\Delta^2_{\chi\theta}|$ and $\Delta^2_{\theta\theta}$, for the initial power-law case $n=0$.}
\label{fig_Deltak_n0}
\end{figure}

We show in Fig.~\ref{fig_Deltak_n0} the displacement and velocity logarithmic power spectra 
$\Delta^2_{\chi\chi}$, $\Delta^2_{\chi\theta}$ and $\Delta^2_{\theta\theta}$, from
Eqs.(\ref{eq:D-chichi-y2}), (\ref{eq:D-chitheta-f2}) and (\ref{eq:D-thetatheta-f2}).
At low $k$, all power spectra converge to the linear power spectrum
$\Delta^2_L(k) = k^3$, for the normalization (\ref{eq:PL-n-norm}).
The cross power spectrum $\Delta^2_{\chi\theta}$ changes sign and we show its
absolute value.  
The power spectra $\Delta^2_{\chi\chi}$ and $\Delta^2_{\theta\theta}$
are always positive, despite their spikes to small but nonzero values.
This means that in the oscillating factor, as in Eq.(\ref{eq:D-chichi-nonlinear-self}), 
one of the coefficients $c_i$ is significantly greater than the other, 
so that $\Delta^2_{\chi\chi}$ almost reaches zero as $\sin^2\psi$ or $\cos^2\psi$.
Thus, within decaying envelopes, $y(u)$ oscillates as $\cos(\psi-\psi_0)$, 
where $\psi_0$ is a constant and $\psi$ was defined in Eq.(\ref{eq:D-thetatheta-y2}),
while $y'(u)$ oscillates in quadrature as $\sin(\psi-\psi_0)$.
This implies that, within decaying envelopes, 
$\Delta^2_{\chi\chi} \propto y^2 \propto \cos^2(\psi-\psi_0)$ and 
$\Delta^2_{\theta\theta} \propto y'^2 \propto \sin^2(\psi-\psi_0)$ oscillate in quadrature.
Since $y' \gg y$, we have $\Delta^2_{\chi\theta} \propto y y'$ and it oscillates twice 
faster, as $\sin[2(\psi-\psi_0)]$.
We can check these phase differences in Fig.~\ref{fig_Deltak_n0}.
The envelope of these power spectra decays as $k^{-3/4}$ at high $k$,
in agreement with Eqs.(\ref{eq:D-chichi-nonlinear-self}), (\ref{eq:D-chitheta-y2}) and
(\ref{eq:D-thetatheta-y2}).

\begin{figure}
\begin{center}
\epsfxsize=8.8 cm \epsfysize=6 cm {\epsfbox{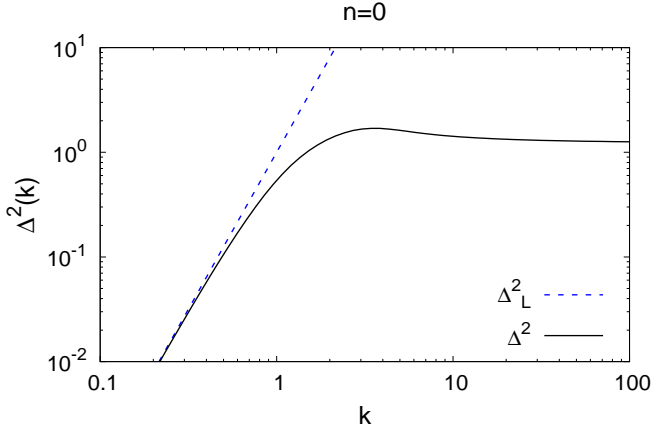}}
\end{center}
\caption{Density logarithmic power spectrum $\Delta^2(k)$.
We show the linear power $\Delta^2_L$ (dashed line) and the nonlinear power
$\Delta^2$ (solid line).}
\label{fig_DeltakZ_n0}
\end{figure}

We show in Fig.~\ref{fig_DeltakZ_n0} the nonlinear density power spectrum $\Delta^2$
from Eq.(\ref{eq:Pk-alpha-beta-q}), as well as the linear power spectrum $\Delta^2_L=k^3$.
Our numerical computation is described in appendix~\ref{sec:alpha-infty-finite-PZ}.
Note that this case $n=0$ corresponds to a linear power spectrum with a lot of power 
on small scales.
Then, the linear variance $\alpha_L(q)$ defined by Eq.(\ref{eq:alpha-bar-def}) where
we replace $P_{\chi\chi}(k)$ by $P_L(k)$ is infinite.
This implies that the Zeldovich power spectrum (\ref{eq:PkZ-alpha-beta-q}) does not
exist. Indeed, the standard Zeldovich approximation does not modifiy the linear displacement
field and does not cure small-scale divergences that are already present in the
linear theory. More generally, the Zeldovich power spectrum (\ref{eq:PkZ-alpha-beta-q}) only
exists for $-3<n<-1$ \cite{Taylor:1996ne,Valageas:2007ge}, where there is no small-scale
divergence.
This is often cured by using a truncated Zeldovich approximation \cite{Coles:1992vr}, 
where the initial linear power spectrum is truncated beyond $k_{\rm NL}$ so that 
$\alpha_L(q)$ is finite and one can compute a Zeldovich power spectrum 
(\ref{eq:PkZ-alpha-beta-q}). This requires introducing an ad-hoc cutoff parameter,
which may be fitted to numerical simulations.

Our approach leads to a density power spectrum that coincides with such a truncated 
Zeldovich approximation, but the cutoff is not introduced by hand. 
It is obtained from the equations of motion, as explained in the previous sections, 
through the computation of the damping factor $\lambda(k)$ and its impact on the 
nonlinear displacement power spectrum $P_{\chi\chi}$.
Another difference from the truncated Zeldovich approximation is that 
we obtain different results for the velocity power spectra $P_{\chi\theta}$ and 
$P_{\theta\theta}$.

As for the truncated Zeldovich approximations with a strong enough cutoff,
the logarithmic density power spectrum $\Delta^2(k)$ shows a constant
asymptote at high $k$, of order unity.
This is because $\Delta^2_{\chi\chi}$ decreases at high $k$, as found in 
Eq.(\ref{eq:D-chichi-nonlinear-self}).
This avoids that spurious power on nonlinear scales for the displacement field 
completely erases small-scale structures and the density power spectrum,
as found in the standard (nontruncated) Zeldovich approximation, where
$\Delta^2_Z(k)$ typically decreases at high $k$.
For this case $n=0$ we note however that the nonlinear density power spectrum
is below the linear power spectrum on mildly nonlinear scales. 
This is due to the saturation at $\Delta^2 \sim 1$ in the nonlinear regime,
whereas the linear power spectrum $\Delta^2_L \propto k^3$ shows a very steep
rise with $k$ for these initial conditions that show a lot of power on small scales.
This agrees with the fact that the damping factor $\lambda(k)$ shown in 
Fig.~\ref{fig_lambdak_n0} is everywhere below unity.

Thus, our approach provides a significant improvement over the standard
Zeldovich approximation. This suggests that the general spirit of our method
goes in the right direction: letting the power spectra of the displacement and velocity fields 
free, instead of setting them equal to the linear power spectrum, and obtaining their values 
from constraints derived from the equations of motion gives a better description of the
system.
However, our Gaussian ansatz cannot give the continuing increase of $\Delta^2(k)$
on nonlinear scales, typically associated with the ``1-halo'' term in halo models 
\cite{Cooray:2002dia,Valageas:2013gba} and the formation of high-density virialized halos.
It is likely that this would require going beyond the Gaussian and taking into
account higher-order correlations (at least three-point correlations).

\subsubsection{Self-similar case with $n=-2$}
\label{sec:numerical-n=-2}

\begin{figure}
\begin{center}
\epsfxsize=8.8 cm \epsfysize=6 cm {\epsfbox{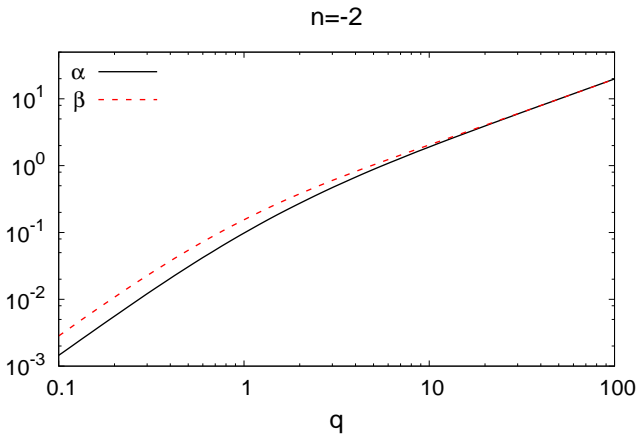}}
\end{center}
\caption{Variances $\alpha(q)$ and $\beta(q)$ defined by the nonlinear power spectrum
$P_{\chi\chi}(k)$ in the power-law case $n=-2$.}
\label{fig_alphabeta_n-2}
\end{figure}

In the case $n=-2$, the initial density and displacement fields show a lot of power on large
scales. Then, the variance of the linear displacement difference over separation $q$ grows 
linearly with $q$, in agreement with Eqs.(\ref{eq:beta-large-q})-(\ref{eq:alpha-large-q}). 
With the normalization (\ref{eq:PL-n-norm}), we obtain
\be
\alpha_L(q) = \beta_L(q) = \frac{\pi}{16} q ,
\label{eq:alphaL-betaL-n-2}
\ee
while $\alpha_{L\infty}$ is infinite.
This infrared divergence also applies to the nonlinear displacement field, whose power spectrum
converges to the linear power spectrum at low $k$. Thus, we have
\be
q \gg q_{\rm NL} : \;\;\; \alpha(q) = \frac{\pi}{16} q + \dots, 
\;\;\; \beta(q) = \frac{\pi}{16} q + \dots ,
\label{eq:alpha-beta-n-2}
\ee
where the dots stand for subleading terms, and $\alpha_\infty = + \infty$.
We can check the asymptotic behaviors (\ref{eq:alpha-beta-q0}) and (\ref{eq:alpha-beta-n-2})
in Fig.~\ref{fig_alphabeta_n-2}.

\begin{figure}
\begin{center}
\epsfxsize=8.8 cm \epsfysize=6 cm {\epsfbox{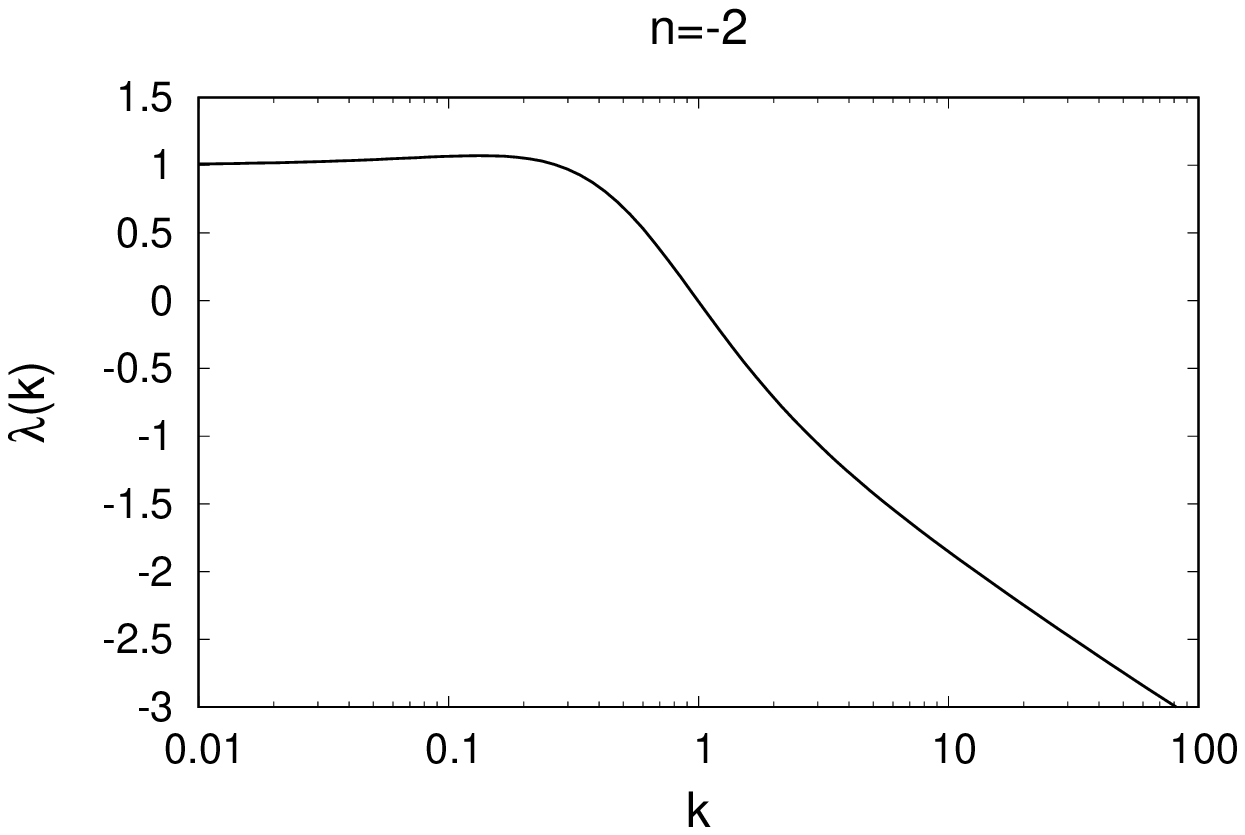}}
\end{center}
\caption{Damping factor $\lambda(k)$ for the power-law case $n=-2$.}
\label{fig_lambdak_n-2}
\end{figure}

\begin{figure}
\begin{center}
\epsfxsize=8.8 cm \epsfysize=6 cm {\epsfbox{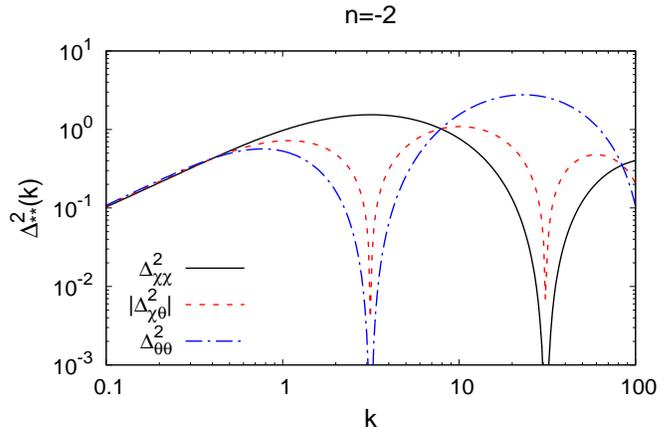}}
\end{center}
\caption{Displacement and velocity logarithmic power spectra $\Delta^2_{\chi\chi}$,
$|\Delta^2_{\chi\theta}|$ and $\Delta^2_{\theta\theta}$, for the initial power-law case $n=-2$.}
\label{fig_Deltak_n-2}
\end{figure}

The damping factor $\lambda(k)$ again goes to unity at low $k$ and to $-\infty$
as $-\ln(k)$ at high $k$, as seen in Fig.~\ref{fig_lambdak_n-2}.
The comparison with Fig.~\ref{fig_lambdak_n0} shows that the amplitude of $\lambda(k)$
at fixed wave number $k$ is greater for $n=-2$ than for $n=0$, in the nonlinear regime.
Nevertheless, the decay and the oscillation rate of the displacement and velocity power
spectra $\Delta^2_{\chi\chi}$, $\Delta^2_{\chi\theta}$ and $\Delta^2_{\theta\theta}$ are slower
than for the case $n=0$ in terms of wave number, as seen in Fig.~\ref{fig_Deltak_n-2}.
This agrees with the fact that these logarithmic power spectra now decrease
as $k^{-1/4}$ instead of $k^{-3/4}$, as shown by Eqs.(\ref{eq:D-chichi-nonlinear-self}), 
(\ref{eq:D-chitheta-y2}) and (\ref{eq:D-thetatheta-y2}).
We can also see that $\lambda(k)$ now grows slightly above unity at $k \sim 0.2$ before
its decreases in the highly nonlinear regime. This means that, in contrast with
the case $n=0$, there is now a small amplification of structure formation as compared
with the linear theory on the weakly nonlinear scale $k \sim 0.2$.

\begin{figure}
\begin{center}
\epsfxsize=8.8 cm \epsfysize=6 cm {\epsfbox{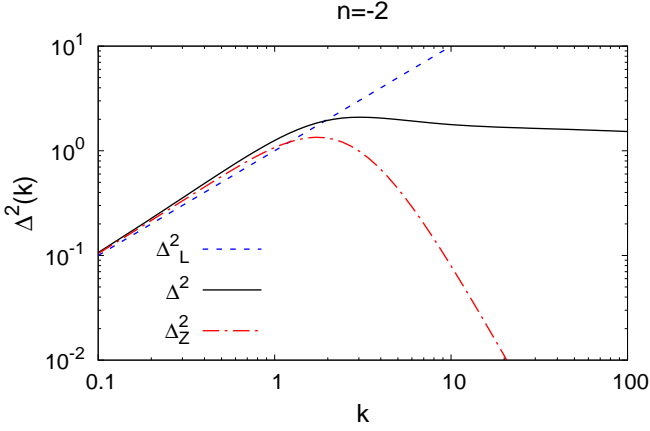}}
\end{center}
\caption{Density logarithmic power spectrum $\Delta^2(k)$.
We show the linear power spectrum $\Delta^2_L$ (dashed line), our nonlinear power spectrum
$\Delta^2$ (solid line) and the Zeldovich power spectrum $\Delta^2_Z$ (dot-dashed line).}
\label{fig_DeltakZ_n-2}
\end{figure}

We show in Fig.~\ref{fig_DeltakZ_n-2} the nonlinear density power spectrum $\Delta^2$
from Eq.(\ref{eq:Pk-alpha-beta-q}), as well as the linear prediction $\Delta^2_L$
and the Zeldovich power spectrum $\Delta^2_Z$.
Indeed, because the linear variances $\alpha_L(q)$ and $\beta_L(q)$ are now finite, 
the standard Zeldovich power spectrum (without truncation) exists. 
In fact, for this power-law case $n=-2$ it is possible to compute analytically 
the Zeldovich power spectrum (\ref{eq:PkZ-alpha-beta-q}). 
For the normalization (\ref{eq:PL-n-norm}), this gives \cite{Valageas:2007ge}
\ba
&& P_Z(k) = \frac{256}{\pi k^2} \Biggl \lbrace \frac{4}{(64+\pi^2 k^2)^2} 
+ \frac{3 \pi k (\sqrt{64+\pi^2 k^2}-8)}{(64+\pi^2 k^2)^{5/2}} \nonumber \\
&& \times \frac{ {\rm ArcTan}(\pi k/\sqrt{128+\pi^2 k^2-16\sqrt{64+\pi^2 k^2}})}
{\sqrt{128+\pi^2 k^2-16\sqrt{64+\pi^2 k^2}}} \nonumber \\
&& + \frac{3 \pi k (\sqrt{64+\pi^2 k^2}+8)}{(64+\pi^2 k^2)^{5/2}} \nonumber \\
&& \times \frac{ {\rm ArcTan}(\pi k/\sqrt{128+\pi^2 k^2+16\sqrt{64+\pi^2 k^2}})}
{\sqrt{128+\pi^2 k^2+16\sqrt{64+\pi^2 k^2}}} \Biggl \rbrace ,
\label{eq:PZ-n-2}
\ea
with the asymptotic behaviors
\be
k \to 0 : \;\;\; P_Z(k) = \frac{1}{4\pi k^2} + \frac{3\pi}{256 k} + \dots ,
\label{eq:PZ-n-2-low-k}
\ee
\be
k \to \infty: \;\;\; P_Z(k) = \frac{128 (8+3\pi)}{\pi^5 k^6} - \frac{4096 (32+15\pi)}{\pi^7 k^8} + \dots
\ee
We describe in appendix~\ref{sec:alpha-infty-infinite-PZ} our numerical computation of the
nonlinear density power spectrum (\ref{eq:Pk-alpha-beta-q}). 
Again, we recover the universal plateau at high $k$ of the nonlinear logarithmic power 
spectrum $\Delta^2$, due to the decay of the displacement logarithmic power 
spectrum $\Delta^2_{\chi\chi}$ within our Gaussian ansatz.
In contrast, the nontruncated Zeldovich power spectrum decays as
$\Delta^2_Z(k) \propto k^{-3}$, because of the artificially large power on small scales 
in the linear displacement field.
Thus, our approach improves over the nontruncated Zeldovich approximation.
It also improves over the truncated Zeldovich approximation, as there is no need
to introduce an ad-hoc truncation with free parameters.
In agreement with the slight increase above unity of the damping factor $\lambda(k)$
at $k \sim 0.2$, and in contrast with the case $n=0$, we now find that the nonlinear
power spectrum $\Delta^2$ rises above the linear prediction on weakly nonlinear scales,
$k \lesssim 1$. This feature is also seen in the standard Zeldovich approximation
(\ref{eq:PZ-n-2-low-k}). This is slightly more apparent in the case of our nonlinear
power spectrum $\Delta^2$ because it asymptotes to a constant value at high $k$ instead
of decreasing as $k^{-3}$.
The comparison with the case $n=0$ shows that such detailed features depend on the
shape of the initial linear power spectrum.
This is consistent with the fact that, within SPT, the one-loop
correction to the density power spectrum is positive for $n \lesssim -1.4$
and negative for $n\gtrsim -1.4$
\cite{Makino:1991rp,Scoccimarro:1996se,Scoccimarro:1996jy}.

\section{$\Lambda$-CDM cosmology}
\label{sec:LCDM}

\subsection{Integration of the curl-free Gaussian ansatz}
\label{sec:integration-LCDM}

We now consider the realistic case of the $\Lambda$-CDM cosmology with 
a linear CDM power spectrum that is not a power law.
Then, we must go back to the system of partial differential equations 
(\ref{eq:dPchichi-deta})-(\ref{eq:dPthetatheta-deta}).
With the curl-free Gaussian ansatz presented in Sec.~\ref{sec:curl-free-Gaussian},
this reads

\ba
&& \frac{\partial\Delta^2_{\chi\chi}}{\partial\eta} = 2 \Delta^2_{\chi\theta} , 
\label{eq:dDeltachichi-deta} \\
&& \frac{\partial\Delta^2_{\chi\theta}}{\partial\eta}  = \frac{3\Omega_{\rm m}}{2 f^2} 
\lambda \Delta^2_{\chi\chi} + \left(1- \frac{3\Omega_{\rm m}}{2 f^2}\right) 
\Delta^2_{\chi\theta} + \Delta^2_{\theta\theta} , \hspace{0.7cm}
\label{eq:dDeltachitheta-deta} \\
&& \frac{\partial\Delta^2_{\theta\theta}}{\partial\eta} = \frac{3\Omega_{\rm m}}{f^2} 
\lambda \Delta^2_{\chi\theta} + \left(2- \frac{3\Omega_{\rm m}}{f^2}\right) 
\Delta^2_{\theta\theta}  ,
\label{eq:dDeltathetatheta-deta}
\ea
where we introduced the logarithmic power $\Delta^2(k,\eta)$ as in
(\ref{eq:Delta2-scaling-n}). 
For the self-similar cases studied in Sec.~\ref{sec:self-similar}, we reduced the
problem to the set of one-dimensional scaling functions ${\cal D}_{**}(u)$ and we could
solve the associated system of ordinary differential equations.
In the general case (\ref{eq:dDeltachichi-deta})-(\ref{eq:dDeltathetatheta-deta}),
thanks to the factorizations (\ref{eq:Pchizeta-lambda}) and 
(\ref{eq:Pthetazeta-lambda}), we again obtain a system of ordinary differential equations.
Indeed, different wave numbers $k$ decouple, once we consider $\lambda(k,\eta)$ as
an external control function, and we can now solve over the time $\eta$ at fixed $k$.
We can again eliminate $\Delta^2_{\chi\theta}$ and $\Delta^2_{\theta\theta}$
to obtain the third-order linear equation
\ba
&& \frac{\partial^3\Delta^2_{\chi\chi}}{\partial\eta^3} 
+ \left( \frac{9\Omega_{\rm m}}{2 f^2}-3\right) \frac{\partial^2\Delta^2_{\chi\chi}}{\partial\eta^2}
+ \Biggl [ 2+ \frac{\partial}{\partial\eta} \left( \frac{3\Omega_{\rm m}}{2 f^2} \right) \nonumber \\
&& - \frac{6\Omega_{\rm m}}{f^2} (1+\lambda) + \frac{9\Omega_{\rm m}^2}{2 f^4} \Biggl ]
\frac{\partial\Delta^2_{\chi\chi}}{\partial\eta} + \Biggl [ - \frac{\partial}{\partial\eta} 
\left( \frac{3\Omega_{\rm m}}{f^2} \lambda \right) \nonumber \\
&& + \frac{6\Omega_{\rm m}}{f^2} \lambda
-  \frac{9\Omega_{\rm m}^2}{f^4} \lambda \Biggl ] \Delta^2_{\chi\chi} = 0 .
\ea
Again, the general solution of this ordinary differential equation over $\eta$, at fixed $k$,
is \cite{Polyanin2017}
\be
\Delta^2_{\chi\chi}(k,\eta) = c_1 y_1(\eta)^2 + c_2 y_1(\eta) y_2(\eta) + c_3 y_2(\eta)^2 ,
\ee
where $c_i$ are integration constants that depend on $k$ and $y_i(\eta)$ are two 
independent solutions of the second-order linear differential equation
\be
y'' + \left( \frac{3\Omega_{\rm m}}{2 f^2} - 1 \right) y' - \frac{3\Omega_{\rm m}}{2 f^2} \lambda y 
= 0 ,
\label{eq:y-ODE}
\ee
where the prime denotes the derivative with respect to $\eta$.
Because $\lambda(k,\eta)$ depends on both $k$ and $\eta$, the functions $y_i(\eta)$
also depend on $k$, understood here as a parameter.
At early times, in the matter dominated era, we must recover the linear regime,
\be
\eta\to -\infty: \;\;\; \Delta^2_L(k,\eta) = e^{2\eta} \Delta^2_{L0}(k) .
\ee
In this regime, we also have $\Omega_{\rm m}/f^2 \to 1$ and $\lambda\to 1$,
which gives the two solutions $y_1 \propto e^{\eta}$ and $y_2\propto e^{-3\eta/2}$.
Therefore, the matching to the linear regime at early times gives $c_2=c_3=0$ and
we obtain
\be
\Delta^2_{\chi\chi}(k,\eta) =  y(\eta)^2 \, \Delta^2_{L0}(k) \geq 0 ,
\label{eq:Delta2-chichi-y2}
\ee
where $y(\eta)$ is the solution of Eq.(\ref{eq:y-ODE}) with the boundary condition at large
negative $\eta$
\be
\eta\to -\infty: \;\;\; y(\eta) = e^{\eta} .
\ee
This gives for the other power spectra
\be
\Delta^2_{\chi\theta}(k,\eta) =  y y' \, \Delta^2_{L0}(k) , \;\;\;
\Delta^2_{\theta\theta}(k,\eta) =  y'^2 \, \Delta^2_{L0}(k) \geq 0 .
\label{eq:Delta2-thetatheta-yp2}
\ee
Over a limited range of wave numbers and times, the dynamics can be approximated
by a self-similar evolution with an effective index $n$. Then, from Eq.(\ref{eq:k-NL-def})
and $\lambda \sim - |\lambda_{\infty}| \ln(k/k_{\rm NL})$ we obtain $\lambda \sim - |\lambda_{\infty}| 2\eta/(n+3)$.
This gives $y(\eta) \sim e^{-\eta/4} [ {\rm Ai}(-\eta) + {\rm Bi}(-\eta) ]$,
where we omit numerical factors in the bracket and in the argument of the Airy functions.
This gives
\be
\eta \gg \eta_{\rm NL} : \;\;\; \Delta^2_{\chi\chi}(k,\eta) \sim e^{-\eta/2} 
[ {\rm Ai}(-\eta) + {\rm Bi}(-\eta) ]^2 ,
\ee
where $\eta_{\rm NL}(k)$ is the time that marks the entry of the wave number $k$ into the
nonlinear regime.
At leading order, this gives for all logarithmic power spectra
\be
\eta \gg \eta_{\rm NL} : \;\;\; \Delta^2_{**}(k,\eta) \sim e^{-\eta/2} ,
\ee
which agrees with Eqs.(\ref{eq:D-chichi-nonlinear-self}), (\ref{eq:D-chitheta-y2}) and
(\ref{eq:D-thetatheta-y2}), using Eq.(\ref{eq:k-NL-def}). Since this nonlinear decay with time 
does not depend on the index $n$, it should be quite robust and a good approximation
for the $\Lambda$-CDM cosmology, with a smooth initial power spectrum.
In a similar fashion, the power spectra should decay with wave number as
\be
k \gg k_{\rm NL} : \;\;\; \Delta^2_{**}(k,\eta) \sim k^{-(n+3)/4} ,
\ee
where $n$ is the local effective exponent of the linear power spectrum.

Remarkably, Eqs.(\ref{eq:Delta2-chichi-y2}) and (\ref{eq:Delta2-thetatheta-yp2})
show that the positivity of the auto-power spectra $\Delta^2_{\chi\chi}$
and $\Delta^2_{\theta\theta}$ is still ensured in the general case, for any cosmology
and initial power spectrum. As noticed in Sec.~\ref{sec:curl-free-Gaussian}, 
such positivity constraints are not respected in most perturbative schemes or 
approximation methods. This is related to the nonperturbative character of our approach,
which does not truncate the equations of motion. Morever, the approximation needed to
close our system, entering at the level of the force cross power spectra, is computed
in an exact manner from a physical Gaussian ansatz. That is, although the Gaussian
distribution of particles is only an approximate ansatz, the force cross power spectra
associated with this distribution are exactly computed and as such satisfy all physical
requirements associated with the constraint that they can be derived from a physical state
(e.g., with positive matter density, conservation of matter, ...).

In fact, the solutions (\ref{eq:Delta2-chichi-y2}) and (\ref{eq:Delta2-thetatheta-yp2})
do not directly rely on the Gaussian ansatz, but on the equality of the damping
factors associated with the cross power spectra of both the displacement and the velocity
with the force. Thus, defining $\lambda_{\chi\zeta}(k,\eta)$ and 
$\lambda_{\theta\zeta}(k,\eta)$ by the ratios
\be
\lambda_{\chi\zeta}(k,\eta) \equiv \frac{P_{\chi\zeta}}{P_{\chi\chi}} , \;\;\;
\lambda_{\theta\zeta}(k,\eta) \equiv \frac{P_{\theta\zeta}}{P_{\theta\chi}} ,
\ee
the solutions (\ref{eq:Delta2-chichi-y2}) and (\ref{eq:Delta2-thetatheta-yp2})
hold as long as $\lambda_{\chi\zeta}=\lambda_{\theta\zeta}$, and we denote $\lambda$
their common value.
This equality may remain a good approximation beyond the Gaussian ansatz and 
we have seen that it ensures the positivity of the auto power spectra
$\Delta^2_{\chi\chi}$ and $\Delta^2_{\theta\theta}$.
However, for the exact non-Gaussian dynamics we generically expect $\lambda_{\chi\zeta}$ 
and $\lambda_{\theta\zeta}$ to differ in the nonlinear regime. 
Unfortunately, we could not find an explicit solution of the system 
(\ref{eq:dDeltachichi-deta})-(\ref{eq:dDeltathetatheta-deta}) when 
$\lambda_{\chi\zeta} \neq \lambda_{\theta\zeta}$. 
In that case, the requirements $\Delta^2_{\chi\chi}\geq 0$ and $\Delta^2_{\theta\theta}\geq 0$
may provide some constraints on the pair $\{\lambda_{\chi\zeta},\lambda_{\theta\zeta}\}$.
However, it is not obvious whether this can be written in a simple explicit form.

\subsection{Numerical computation}
\label{sec:numerical-LCDM}

As for the self-similar case studied in Sec.~\ref{sec:numerical-self},
we compute the solution of 
Eqs.(\ref{eq:dDeltachichi-deta})-(\ref{eq:dDeltathetatheta-deta}) by an iterative scheme.
We start with an initial guess for the power spectrum $\Delta^2_{\chi\chi}(k,\eta)$,
which is equal to the linear power spectrum in the linear regime where
$\Delta^2_L \leq 1$, and decays for instance as $1/k$ at higher wavenumbers.
This is stored as an initial guess on a 2D grid in $\{k,\eta\}$.
Then, we compute the variances $\alpha(q,\eta)$ and 
$\beta(q,\eta)$ from Eqs.(\ref{eq:alpha-bar-def})-(\ref{eq:beta-def}).
This gives the damping factor $\lambda(k,\eta)$ from Eq.(\ref{eq:lambda-0}),
using again the numerical method described in the appendix~\ref{sec:expression-numerical}.
Next, we compute the functions $y(\eta)$, for all grid-points $k$, from
Eq.(\ref{eq:y-ODE}). 
This provides the updated nonlinear displacement and velocity power spectra 
$\Delta^2_{\chi\chi}$, $\Delta^2_{\chi\theta}$ and $\Delta^2_{\theta\theta}$
through Eqs.(\ref{eq:Delta2-chichi-y2}) and (\ref{eq:Delta2-thetatheta-yp2}).
Next, we repeat the procedure, computing $\{\alpha,\beta,\lambda\}$ from the new
$\Delta^2_{\chi\chi}$ and next the new spectra $\Delta^2_{**}$.
We iterate until convergence.
Finally, from the displacement power spectrum $P_{\chi\chi}$ we obtain the
density power spectrum $P(k)$ from Eq.(\ref{eq:Pk-density-Pchichi}), using again
the numerical method described in the appendix~\ref{sec:alpha-infty-finite-PZ}.
We do not make the approximation $\Omega_{\rm}/f^2 \simeq 1$ that is often used
in analytical studies and we exactly integrate Eq.(\ref{eq:y-ODE}) over time.

\begin{figure}
\begin{center}
\epsfxsize=8.8 cm \epsfysize=6 cm {\epsfbox{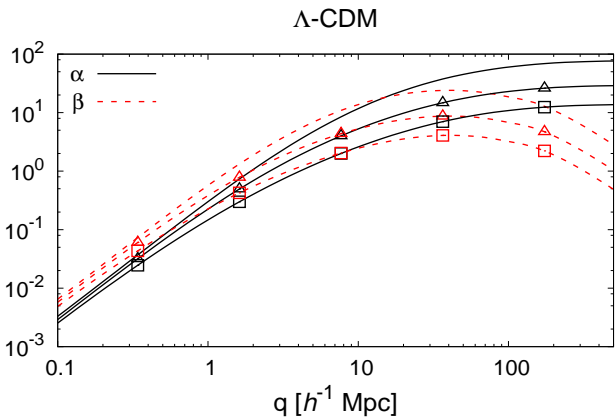}}
\end{center}
\caption{Variances $\alpha(q)$ and $\beta(q)$ defined by the nonlinear power spectrum
$P_{\chi\chi}(k)$ for the $\Lambda$-CDM cosmology. We show the results at redshifts $z=0$
(lines without symbols), $z=1$ (triangles) and $z=2$ (squares).}
\label{fig_alphabeta_LCDM}
\end{figure}

We show the variances $\alpha(q)$ and $\beta(q)$ at redshifts $z=0$, $1$ and $2$
in Fig.~\ref{fig_alphabeta_LCDM}. Again, we have the quadratic behavior 
(\ref{eq:alpha-beta-q0}) on small scales. At large distances, the displacement
variances are governed by the low-$k$ part of the displacement power spectrum,
which converges to the linear power spectrum with $P_L(k) \propto k^n$
and $n\simeq 0.96$ for the $\Lambda$-CDM cosmology.
In this respect, we are in the same class of initial conditions as for the
self-similar case with $n=0$; $\alpha(q)$ goes to a finite value $\alpha_\infty$ 
whereas $\beta(q)$ decreases as $q^{-(n+1)}$, as in 
Eqs.(\ref{eq:beta-large-q})-(\ref{eq:alpha-large-q}).
The small change in the shape of the functions $\alpha(q)$ and $\beta(q)$ with redshift
is due to the fact that the $\Lambda$-CDM linear power spectrum is curved,
with the local slope ranging from $n\simeq 0.96$ at low $k$ to
$n \simeq -3$ at high $k$.
In particular, the infinite-separation variance $\alpha_\infty$ of 
Eq.(\ref{eq:alpha-infty}) is governed by the scale $k_{-1}$ where the local exponent
is $n=-1$. This is significantly larger than the scale $k_{\rm NL}$
associated with the nonlinear transition of the matter density power spectrum.
Thus, in contrast with the self-similar case $n=0$, $\alpha_\infty$ can be significantly
greater than $1/k_{\rm NL}^2$, especially at high $z$.
In contrast with some Eulerian-space perturbative schemes, this is not a problem
for our approach as it only depends on relative displacements, 
as seen in Eq.(\ref{eq:Pk-density-Pchichi}), and it is independent of the value
of $\alpha_\infty$. This is clear from the fact that our approach can also be applied
to the self-similar case $n=-2$ where $\alpha_\infty$ is infinite, 
see Sec.~\ref{sec:numerical-n=-2} and appendix~\ref{sec:alpha-infty-infinite-PZ}.

\begin{figure}
\begin{center}
\epsfxsize=8.8 cm \epsfysize=6 cm {\epsfbox{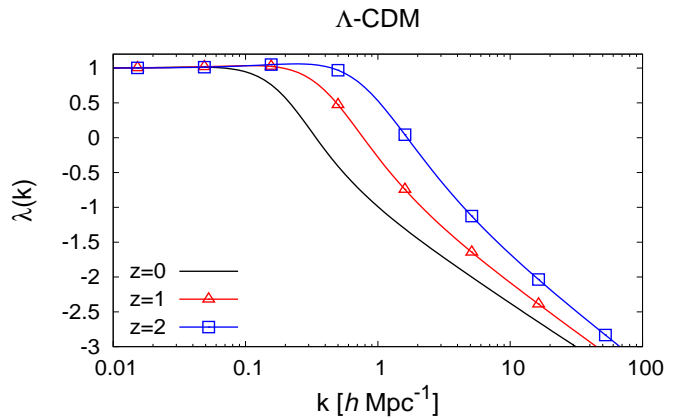}}
\end{center}
\caption{Damping factor $\lambda(k)$ for the $\Lambda$-CDM cosmology,
at redshifts $z=0$ (lines without symbols), $z=1$ (triangles) and $z=2$ (squares).}
\label{fig_lambdak_LCDM}
\end{figure}

We display in Fig.~\ref{fig_lambdak_LCDM} the damping factor $\lambda(k)$ at redshifts 
$z=0$, $1$ and $2$. Again, it goes to unity at low $k$ and to $-\infty$ as $-\ln(k)$
at high $k$. At high redshift $z=2$, where the effective exponent on weakly nonlinear
scales is $n\simeq -2$, we distinguish a small excursion above unity for
$\lambda(k)$ around $k \sim 0.3 h {\rm Mpc}^{-1}$, in agreement with the self-similar
case $n=-2$ shown in Fig.~\ref{fig_lambdak_n-2}.
At low redshift $z=0$, where $n \simeq -1.5$, $\lambda(k)$ remains below unity.
This agrees with the behavior found for the self-similar
case $n=0$ shown in Fig.~\ref{fig_lambdak_n0}.

\begin{figure}
\begin{center}
\epsfxsize=8.8 cm \epsfysize=6 cm {\epsfbox{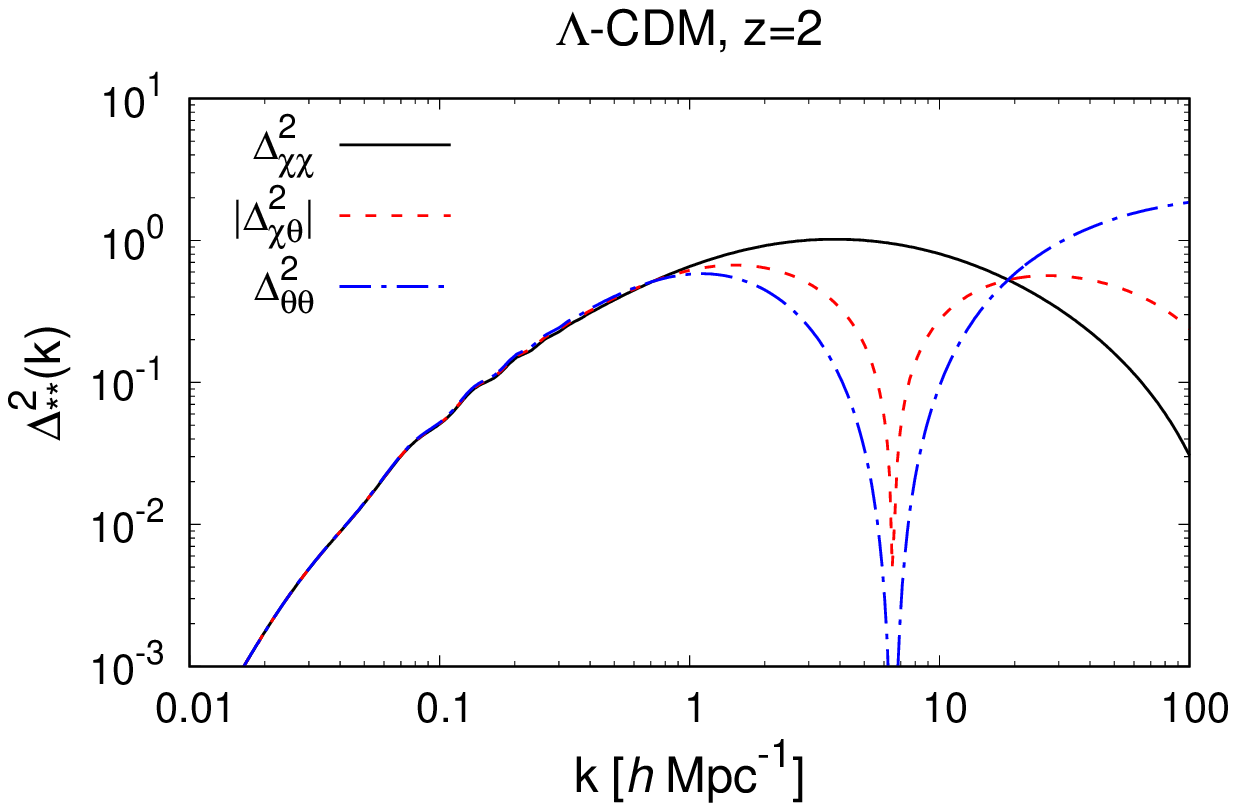}}\\
\epsfxsize=8.8 cm \epsfysize=6 cm {\epsfbox{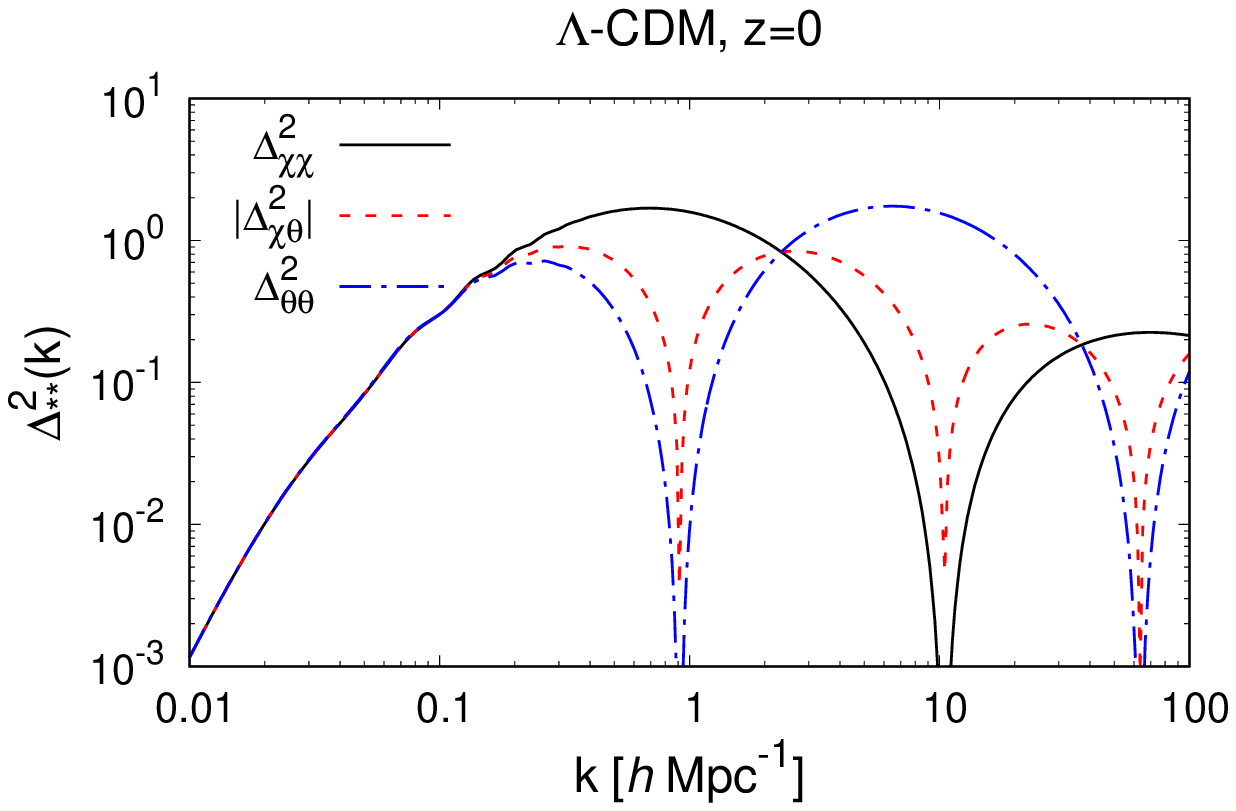}}
\end{center}
\caption{Displacement and velocity logarithmic power spectra $\Delta^2_{\chi\chi}$,
$|\Delta^2_{\chi\theta}|$ and $\Delta^2_{\theta\theta}$, for the $\Lambda$-CDM cosmology.}
\label{fig_Deltak_LCDM}
\end{figure}

We show the displacement and velocity power spectra in Fig.~\ref{fig_Deltak_LCDM}.
In agreement with the analysis in Sec.~\ref{sec:integration-LCDM}, on nonlinear
scales the power spectra $\Delta^2_{\chi\chi}$ and $\Delta^2_{\theta\theta}$ 
oscillate in quadrature, whereas $\Delta^2_{\chi\theta}$ oscillates twice faster.
Their envelope decays as $\sim k^{-(n+3)/4}$.
Again, the evolution with redshift can be understood from the change of the effective
exponent $n$ at the scales that are turning nonlinear.
At high redshift $z=2$, where $n \simeq -2$, we recover a slow decay with a large 
oscillation period over wave number, while at low redshift $z=0$, where $n \simeq -1.5$,
we obtain a stronger decay and faster oscillations.
This agrees with Eqs.(\ref{eq:D-chichi-nonlinear-self})-(\ref{eq:psi-oscill-def})
and with the comparison of Figs.~\ref{fig_Deltak_n0} and \ref{fig_Deltak_n-2}.

\begin{figure}
\begin{center}
\epsfxsize=8.8 cm \epsfysize=6 cm {\epsfbox{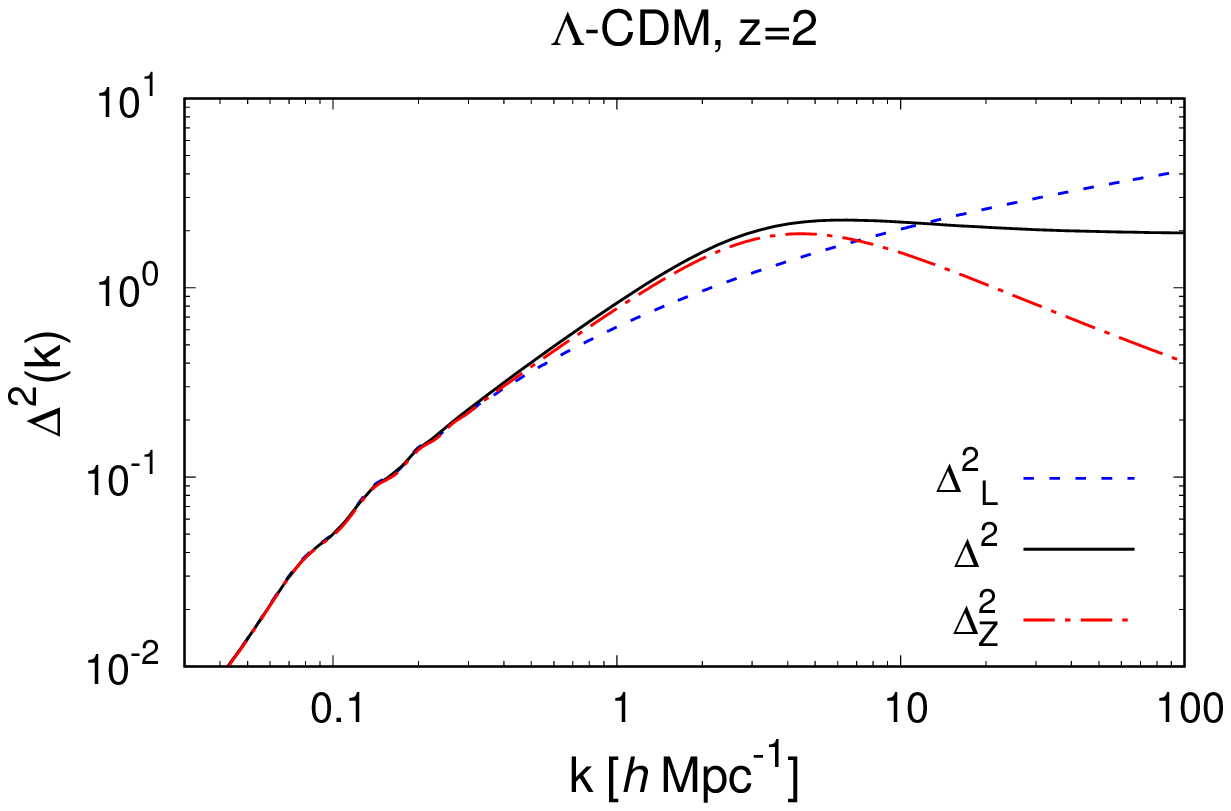}}\\
\epsfxsize=8.8 cm \epsfysize=6 cm {\epsfbox{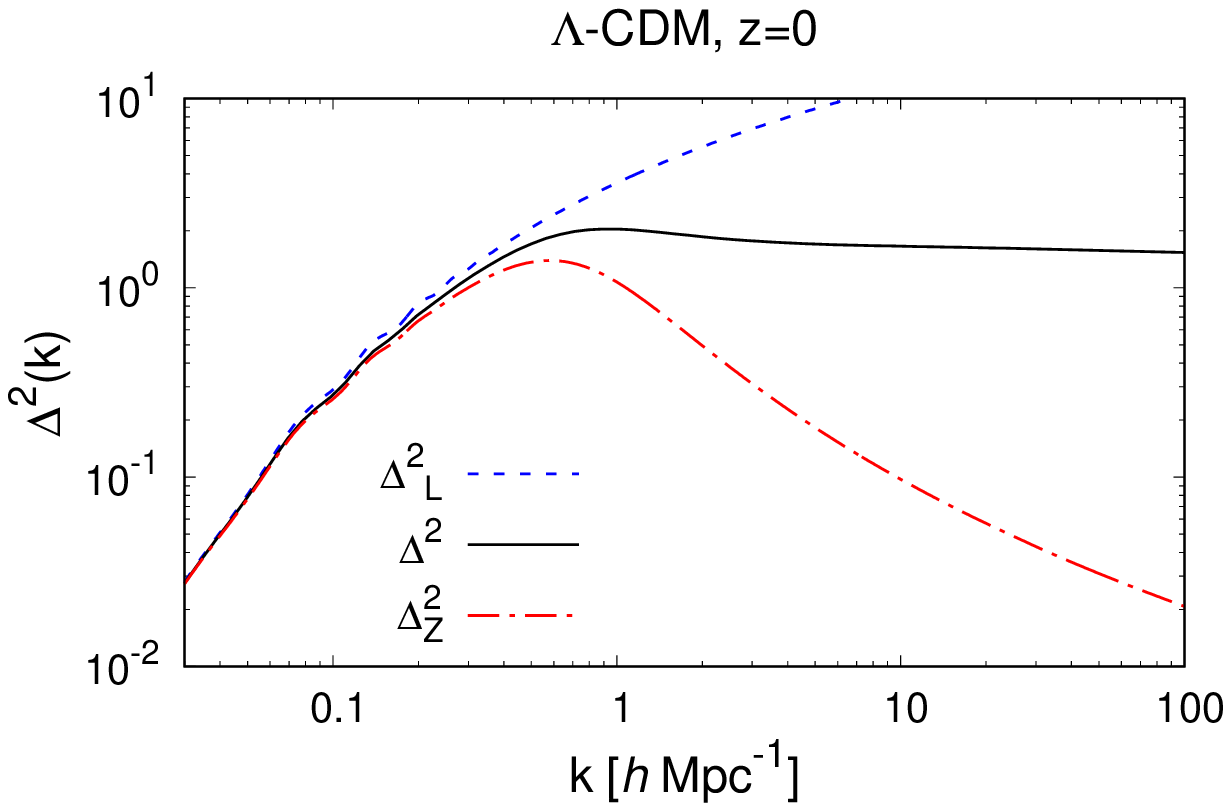}}
\end{center}
\caption{Density logarithmic power spectrum $\Delta^2(k)$ for the $\Lambda$-CDM cosmology.
We show the linear power spectrum $\Delta^2_L$ (dashed line), our nonlinear power spectrum
$\Delta^2$ (solid line) and the Zeldovich power spectrum $\Delta^2_Z$ (dot-dashed line).}
\label{fig_DeltakZ_LCDM}
\end{figure}

We compare in Fig.~\ref{fig_DeltakZ_LCDM} the nonlinear matter density power spectrum
$\Delta^2$ with the linear prediction $\Delta^2_L$ and the Zeldovich power spectrum
$\Delta^2_Z$.
Again, we find that $\Delta^2$ roughly follows $\Delta^2_Z$ on weakly nonlinear scales
and next asymptotes to a constant $\Delta^2 \sim 1$ at high $k$, whereas $\Delta^2_Z$
decays as $k^{-3}$.
As for the other statistics, the detailed behavior with redshift reflects the change
of the effective index $n$. At $z=2$ we find that both $\Delta^2$ and $\Delta^2_Z$ 
rise above the linear power spectrum $\Delta^2_L$ on weakly nonlinear scales,
$k \sim 2 h {\rm Mpc}^{-1}$, whereas at $z=0$ they remain below $\Delta^2_L$.
This agrees with the comparison of Figs.~\ref{fig_DeltakZ_n0} and \ref{fig_DeltakZ_n-2}.

\subsection{Comparison with numerical simulations}
\label{sec:simulations}

\subsubsection{Matter density power spectrum}
\label{sec:simulations-P(k)}

\begin{figure*}
\begin{center}
\epsfxsize=8.8 cm \epsfysize=6 cm {\epsfbox{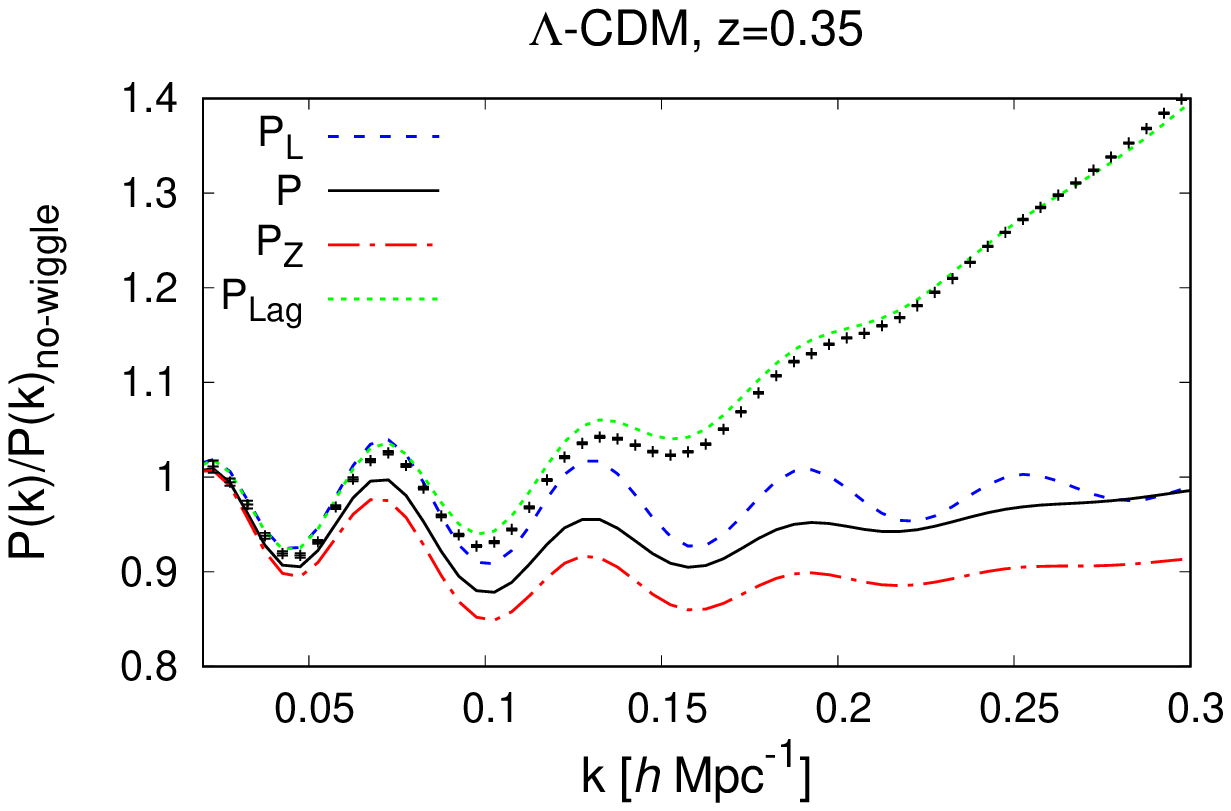}}
\epsfxsize=8.8 cm \epsfysize=6 cm {\epsfbox{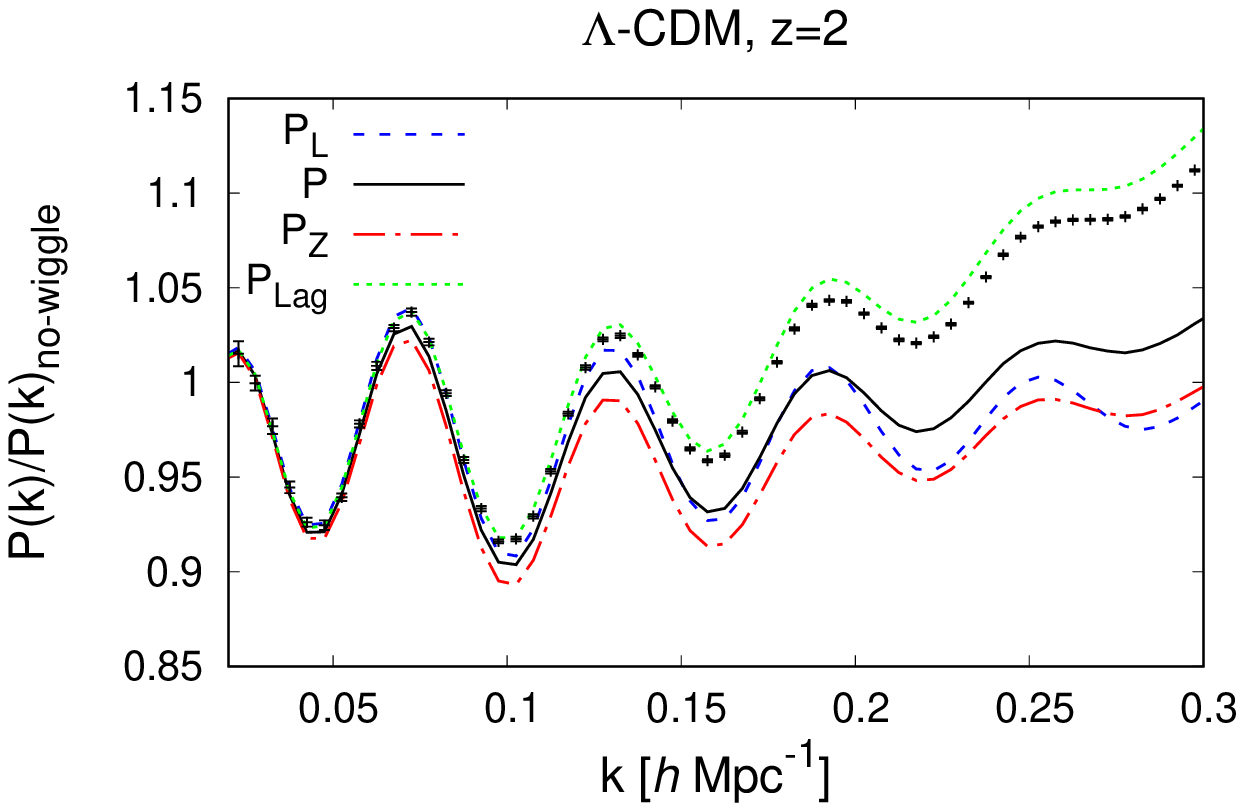}}
\\
\epsfxsize=8.8 cm \epsfysize=6 cm {\epsfbox{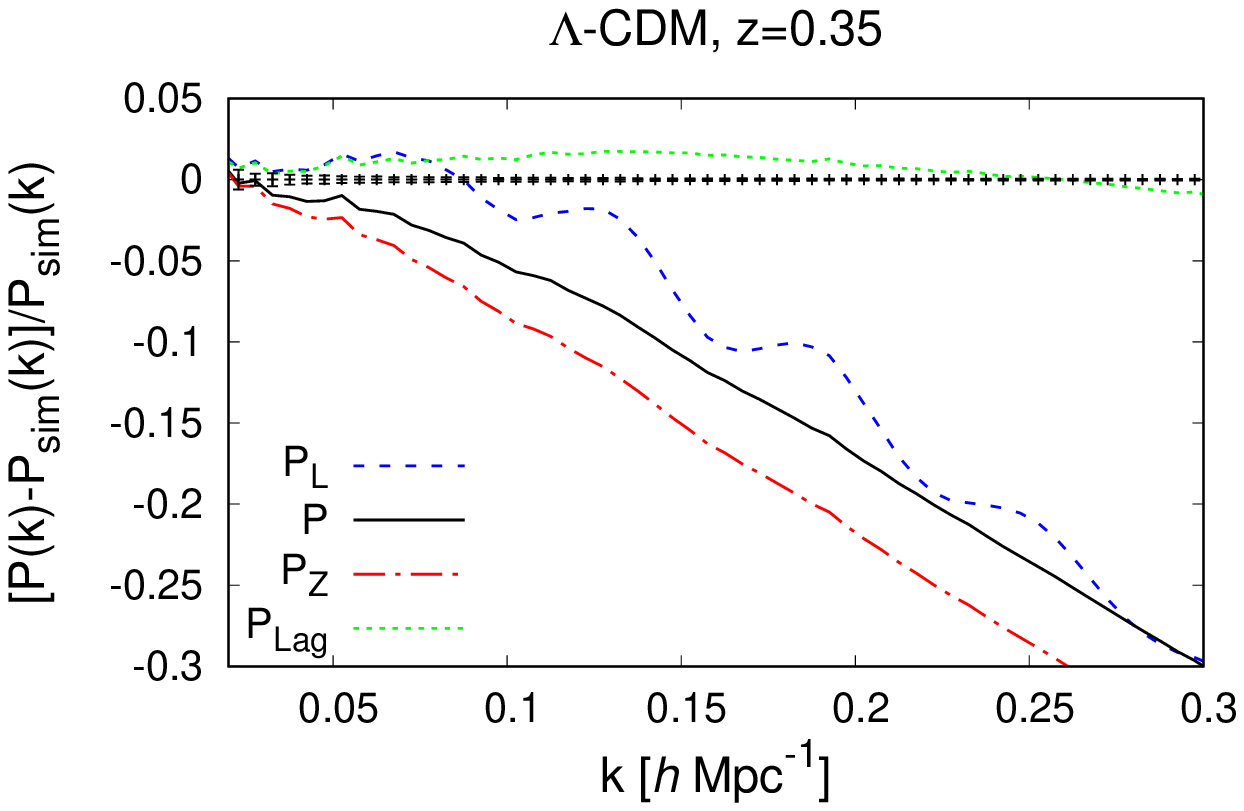}}
\epsfxsize=8.8 cm \epsfysize=6 cm {\epsfbox{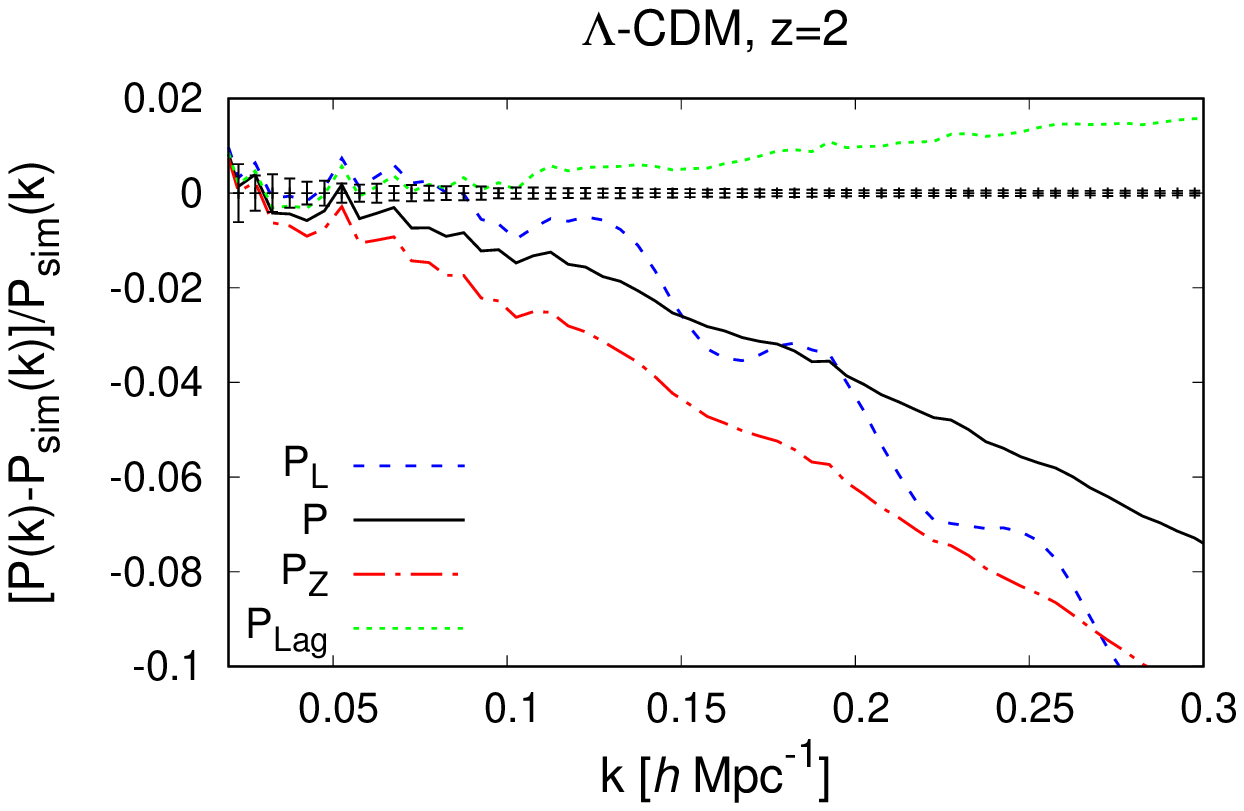}}
\end{center}
\caption{{\it Upper panels:} matter density power spectrum divided by a reference 
no-wiggle linear power spectrum, at redshifts $z=0.35$ and $z=2$. 
We show the predictions of linear theory ``$P_L$''(blue dashed line),
our nonlinear model ``$P$'' (black solid line), the standard Zeldovich approximation 
``$P_Z$'' (red dot-dashed line), a Lagrangian model ``$P_{\rm Lag}$'' 
(green dotted line) and the numerical simulations (black crosses).
{\it Lower panels:} relative deviation of these density power spectra from the
numerical simulations. The error bars centered on zero are the numerical simulations
statistical error bars.}
\label{fig_dPk_LCDM}
\end{figure*}

Finally, we compare the predictions of our curl-free Gaussian ansatz with numerical
simulations of the large-scale matter density field in the $\Lambda$-CDM cosmology,
which were presented in \cite{Valageas:2010yw} and \cite{Taruya:2012ut}.
Since our Gaussian model cannot describe highly nonlinear scales, as explained
in the previous sections, we focus on large quasilinear scales associated with the
baryon acoustic peak.
We show the matter density power spectrum in Fig.~\ref{fig_dPk_LCDM}.
To distinguish more clearly the baryon acoustic oscillations and the different
models, we plot in the upper panels the ratio of the density power spectra 
by a reference no-wiggle linear power spectrum that does not contain baryon 
acoustic oscillations. In the lower panels, we directly plot the relative deviation
from the numerical simulations.

As we can see in the upper panels, our result $P(k)$ is similar to the standard
Zeldovich approximation on these large scales. This agrees with the results of
Fig.~\ref{fig_DeltakZ_LCDM} and the fact that on such large scales the effective
truncation of the displacement power spectrum on nonlinear scales does not have 
a great impact. Thus, the damping of the oscillations at higher $k$, as compared
with the linear power spectrum, is similar in both models. However, in agreement
with the results of previous sections, the amplitude of the power spectrum
given by our model is somewhat larger than for the standard Zeldovich approximation.
Thus, the modification of the displacement field on nonlinear scales only leads
to a broad-band change to the density power spectrum on BAO scales.
In agreement with Fig.~\ref{fig_DeltakZ_LCDM}, our power spectrum remains
below the linear theory up to $k\leq 0.3 h/{\rm Mpc}$ at $z=0.35$ while it raises
above the linear theory at $k\simeq 0.2 h/{\rm Mpc}$ at $z=2$.
In terms of the absolute value of the density power spectrum, our model is not
competitive with other approaches that can reach percent-level accuracy on these
scales, as shown for instance by the comparison with the Lagrangian model
$P_{\rm Lag}(k)$ developed in \cite{Valageas:2013gba}.
Indeed, this older model is correct up to one-loop order, while matching the halo
model on highly nonlinear scales, which ensures a reasonably good accuracy.
In contrast, as for the standard Zeldovich approximation, the Gaussian model 
presented in this paper does not match with SPT at one-loop order.
This is due to the use of our Gaussian ansatz. To ensure a correct one-loop order,
we should extend this Gaussian ansatz and include three-point correlations. 
This would in turn involve additional constraint equations to the system
(\ref{eq:dDeltachichi-deta})-(\ref{eq:dDeltathetatheta-deta}), associated with
the evolution of the bispectrum.
We leave such an extension to future works.

The lower panels show more clearly that while the relative deviations from the
numerical simulations are of the same order of magnitude for the linear theory,
our model and the Zeldovich approximation, the oscillations found for the
linear prediction disappear for both our model and the Zeldovich approximation.
This is because nonlinear mode couplings damp the initial baryon acoustic oscillations.
Therefore, relative to the flatter nonlinear result (given by the numerical simulations),
the linear power spectrum shows oscillations at high $k$. In contrast, the nonlinear
damping of the oscillations is well recovered by our model and the Zeldovich approximation,
so that the relative deviation is flat.
This suggests that both our model and the Zeldovich approximation could be efficiently
used to study the BAO features of the density power spectrum. One simply needs to
extract the oscillations from the data (e.g. through a high-pass filter), as in
\cite{Noda:2017tfh,Noda:2019pzd}, or to add
to the analytical predictions a smooth low-order polynomial, with one or two free 
parameters, that describes the smooth drift of the amplitude.

\subsubsection{Matter density correlation function}
\label{sec:simulations-xi(x)}

\begin{figure}
\begin{center}
\epsfxsize=8.8 cm \epsfysize=6 cm {\epsfbox{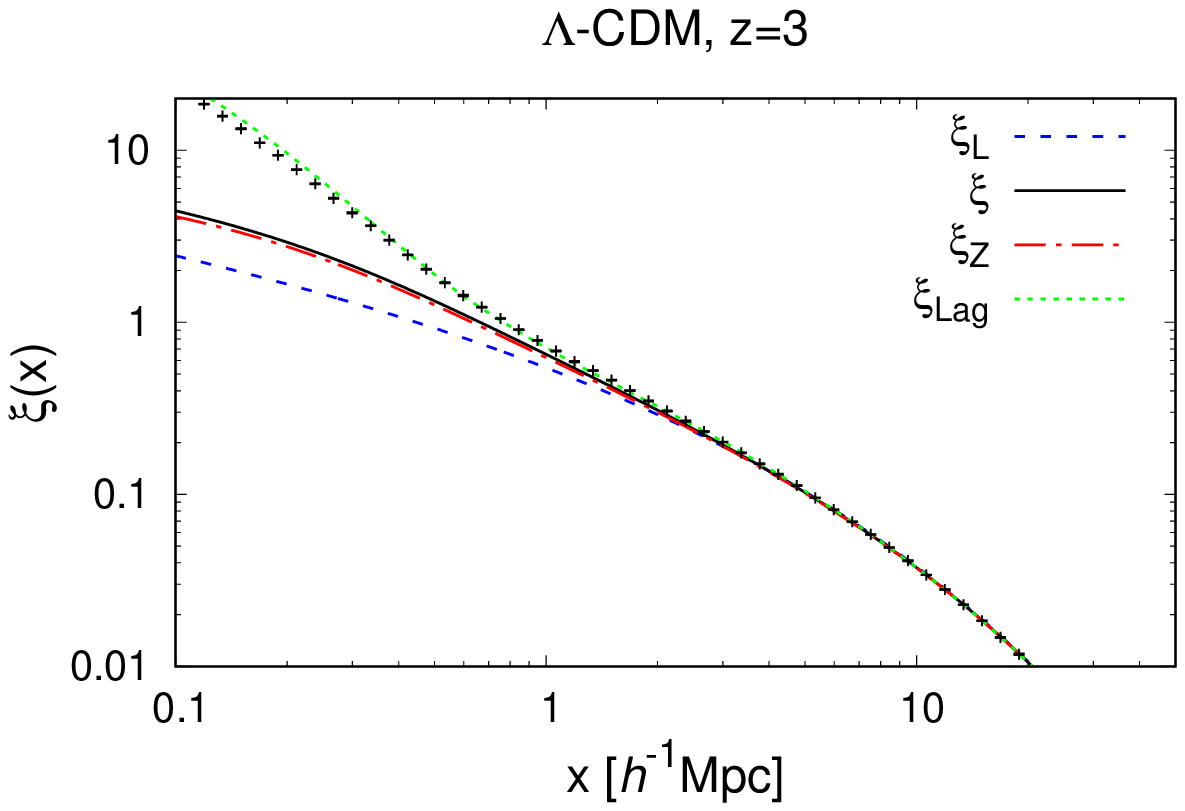}}
\epsfxsize=8.8 cm \epsfysize=6 cm {\epsfbox{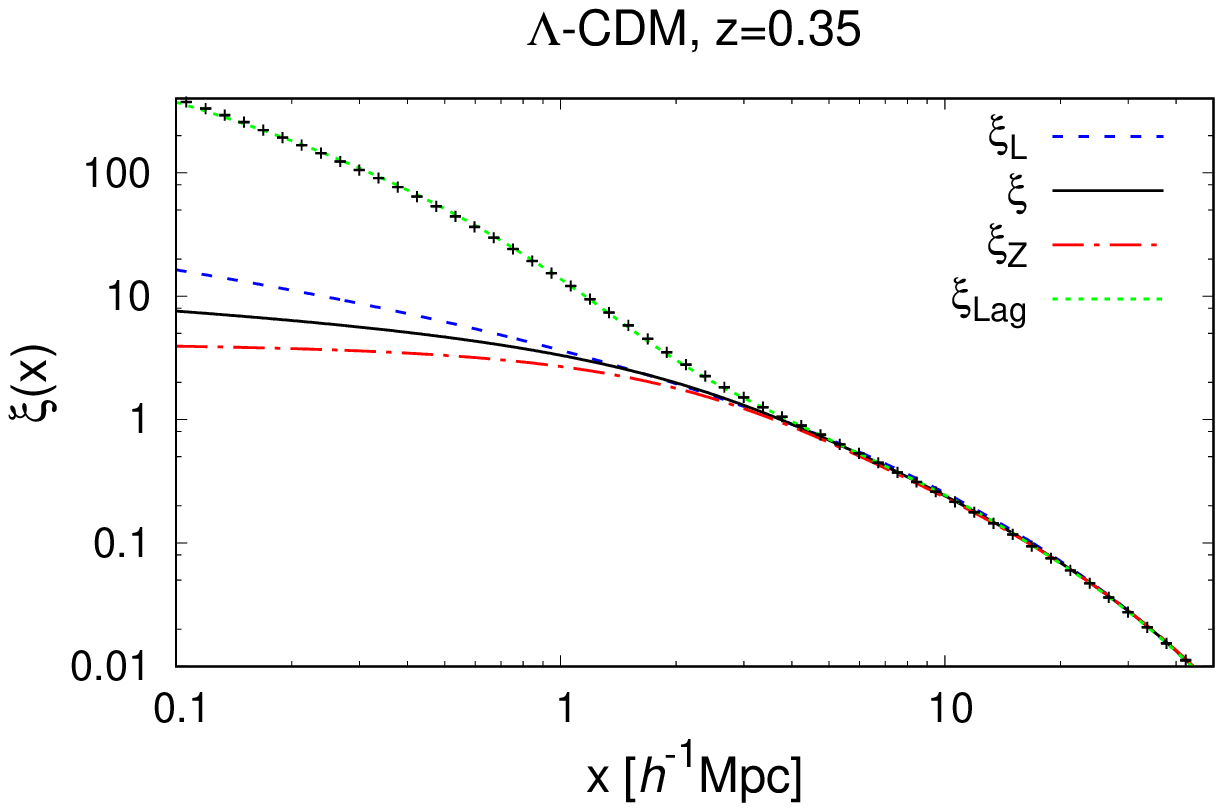}}
\end{center}
\caption{Matter density correlation function at redshifts $z=3$ and $z=0.35$. 
As in Fig.~\ref{fig_dPk_LCDM}, we show the predictions of linear theory 
``$\xi_L$''(blue dashed line), our nonlinear model ``$\xi$'' (black solid line), 
the standard Zeldovich approximation ``$\xi_Z$'' (red dot-dashed line), a Lagrangian 
model ``$\xi_{\rm Lag}$'' (green dotted line) and the numerical simulations (black crosses).}
\label{fig_lxi_LCDM}
\end{figure}

\begin{figure*}
\begin{center}
\epsfxsize=8.8 cm \epsfysize=6 cm {\epsfbox{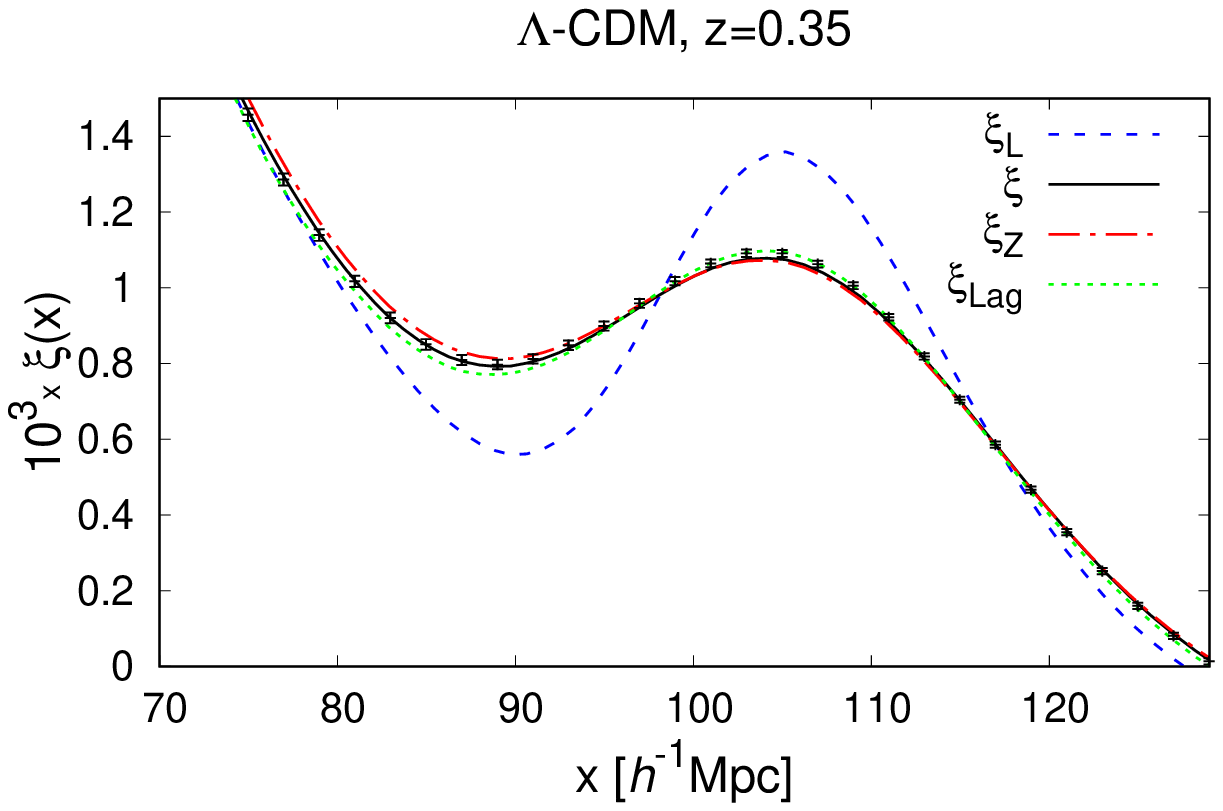}}
\epsfxsize=8.8 cm \epsfysize=6 cm {\epsfbox{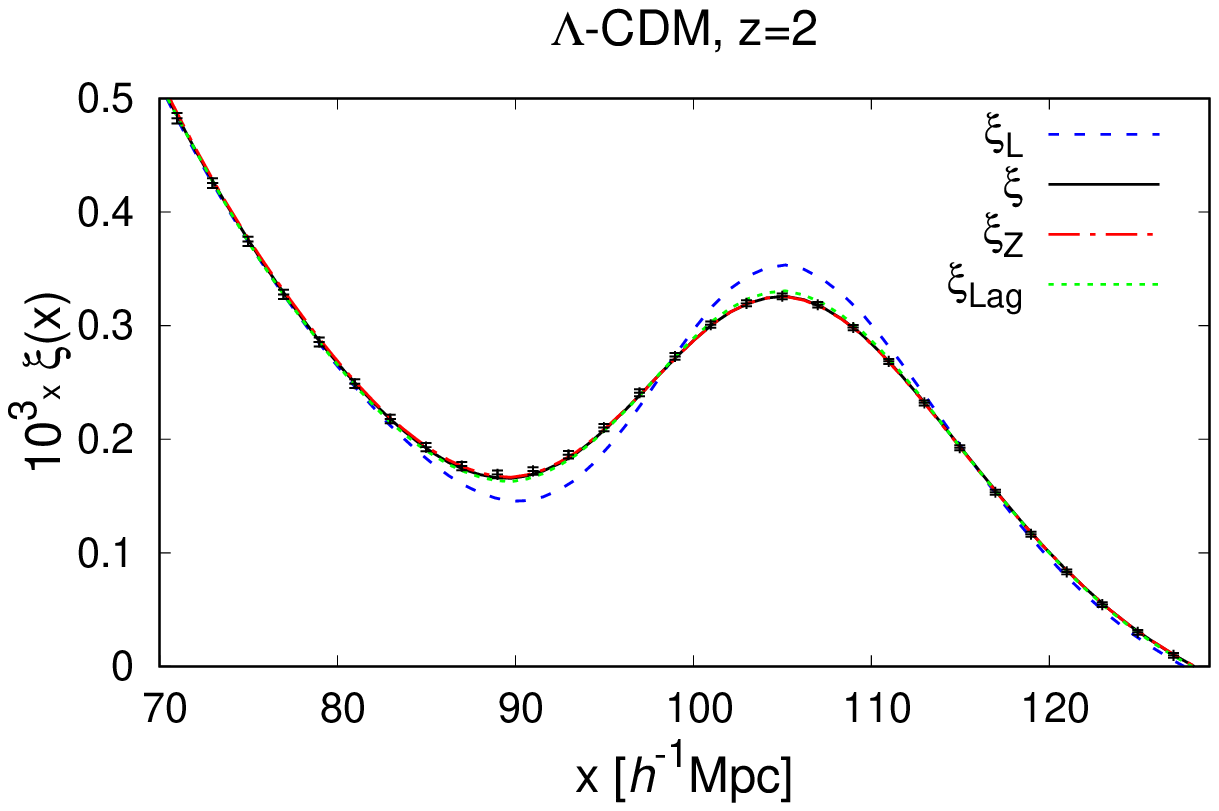}}
\\
\epsfxsize=8.8 cm \epsfysize=6 cm {\epsfbox{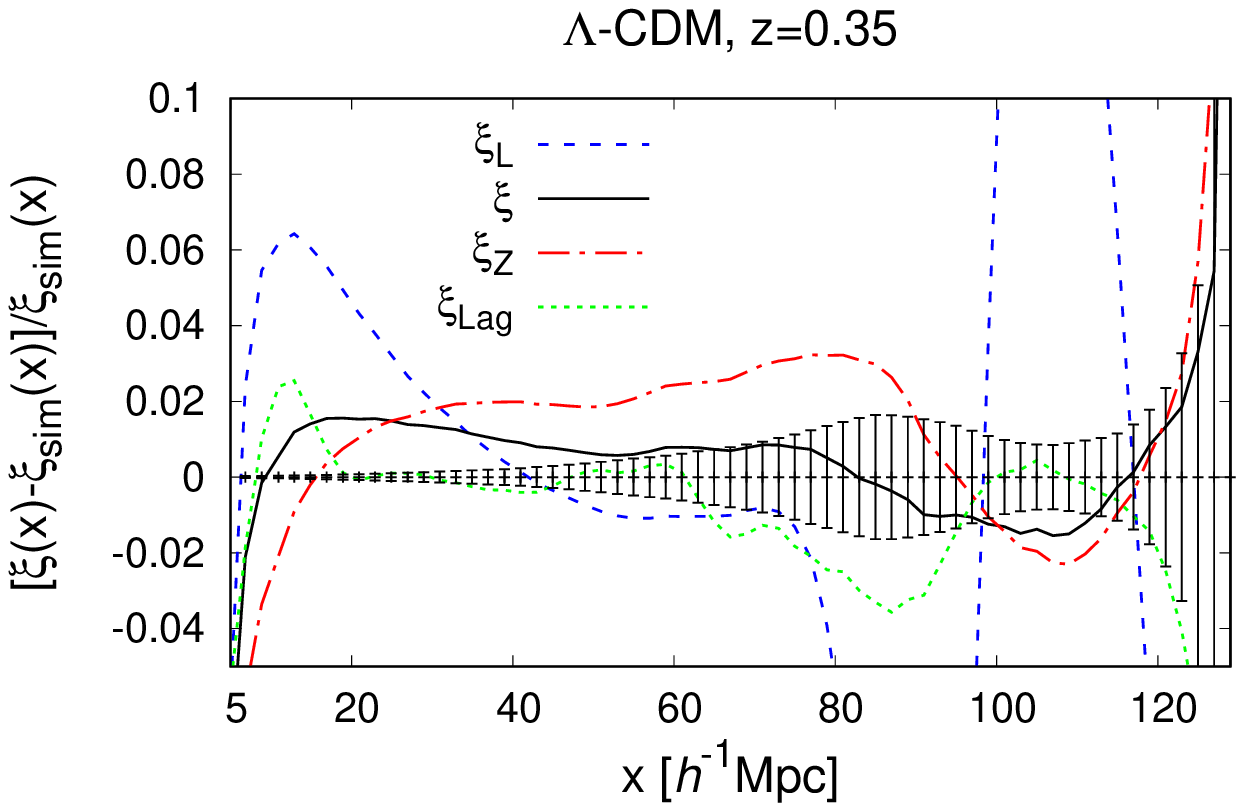}}
\epsfxsize=8.8 cm \epsfysize=6 cm {\epsfbox{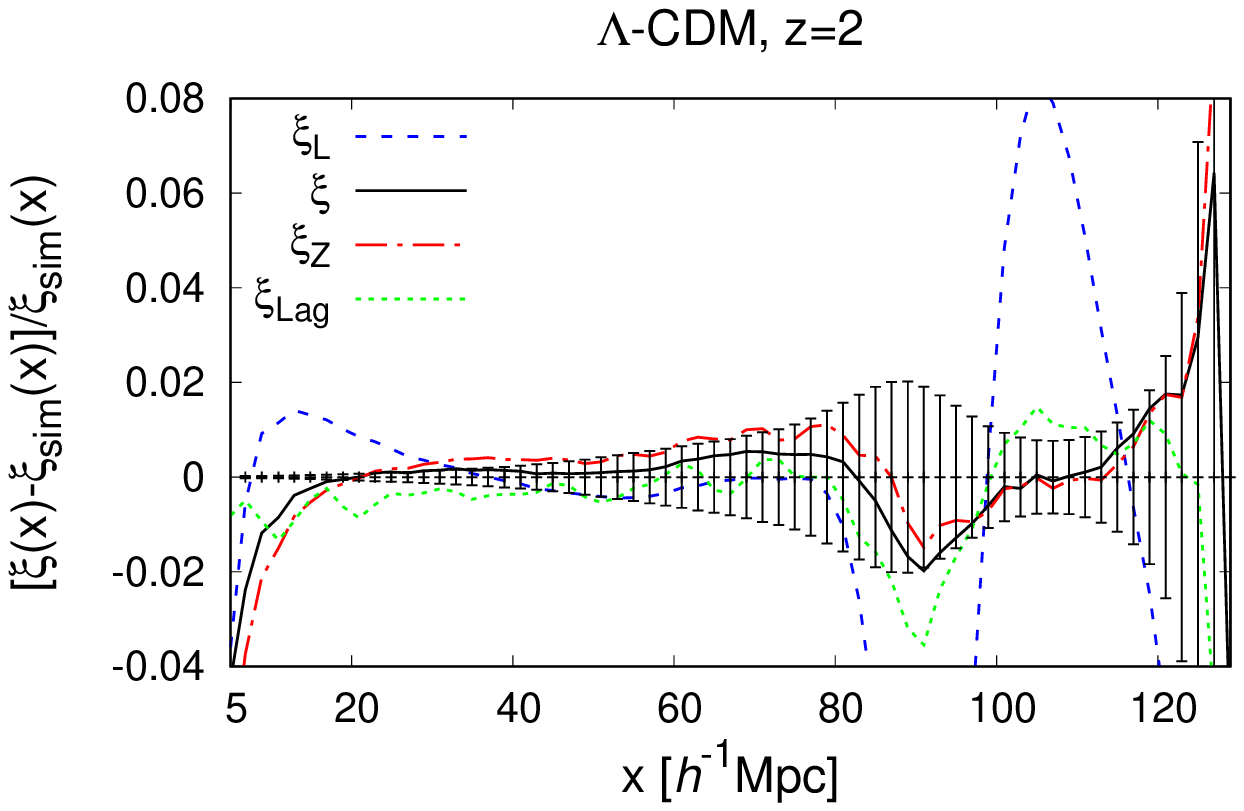}}
\end{center}
\caption{{\it Upper panels:} matter density correlation function at redshifts $z=0.35$ 
and $z=2$. As in Fig.~\ref{fig_dPk_LCDM}, we show the predictions of linear theory 
``$\xi_L$''(blue dashed line), our nonlinear model ``$\xi$'' (black solid line), 
the standard Zeldovich approximation ``$\xi_Z$'' (red dot-dashed line), a Lagrangian 
model ``$\xi_{\rm Lag}$'' (green dotted line) and the numerical simulations (black crosses).
{\it Lower panels:} relative deviation of these density correlation functions from the
numerical simulations. The error bars centered on zero are the numerical simulations
statistical error bars.}
\label{fig_dxi_LCDM}
\end{figure*}

We next consider the matter density correlation function in Figs.~\ref{fig_lxi_LCDM}
and \ref{fig_dxi_LCDM}, for the same models and redshifts.
It is computed from the power spectra by integrating
\be
\xi(x) = 4\pi \int_0^{\infty} dk \, k^2 P(k) j_0(kx) .
\label{eq:xi-Pk-def}
\ee
As seen in Fig.~\ref{fig_lxi_LCDM}, on large scales all curves converge to the linear
theory. 
Whereas the Zeldovich approximation gives a constant correlation at low $x$,
because its power spectrum decays faster than $k^{-3}$ at high $k$,
our model gives a logarithmic growth at low $x$, because its power spectrum decays as 
$k^{-3}$ at high $k$.
However, neither model nor the Zeldovich approximation can describe highly nonlinear
scales associated with virialized halos. In particular, these methods are not competitive
as compared with the Lagrangian model of \cite{Valageas:2013gba}.

We focus on the BAO peak in the upper panels of Fig.~\ref{fig_dxi_LCDM},
while in the lower panels we show the relative deviations with respect 
to numerical simulations, from weakly nonlinear scales up to the BAO scales. 
The growth  of all relative deviations and of the simulation error bars 
at $x \gtrsim 120 h^{-1} {\rm Mpc}$ is due to the fact
that the correlation function vanishes at $x \sim 130 h^{-1} {\rm Mpc}$.
Because analytical predictions and numerical simulations do not recover the exact 
position of this zero-crossing the relative deviation diverges at this point 
but this is not a good measure of the validity of approximation schemes.

We recover the fact that the Zeldovich approximation provides a great improvement 
over the linear prediction for the BAO peak
\cite{Okamura:2011nu,Valageas:2013hxa,Vlah:2014nta}.
Its accuracy is better than $3\%$ accuracy on these scales and redshifts.
Our Gaussian model gives similar results, with an improved accuracy below $2\%$
on BAO scales, $70 < x < 120 h^{-1} {\rm Mpc}$.
It actually fares slightly better than the Lagrangian model of \cite{Valageas:2013gba},
which however gave a much better prediction for the power spectrum in 
Fig.~\ref{fig_dPk_LCDM}. 
In agreement with the discussion above, this means that the information associated
with the position and shape of the baryon acoustic peak in the correlation function
is related to the frequency and damping of the baryon acoustic oscillations in
the power spectrum and is mostly independent of any additional smooth drift.
This agrees with the results of \cite{Noda:2017tfh}, who also find that the BAO oscillatory
features of the power spectrum are mostly governed by the long-range displacements
(infrared effects), which are automatically taken into account by Lagrangian approaches,
while the broadband shape are affected by small-scale processes, 
see also \cite{Valageas:2013hxa,Baldauf:2015xfa}.

Thus, our model provides the BAO peak of the density correlation function to better
than $2\%$, without any free parameter. Its accuracy is actually the same as that
of the numerical simulations.
On smaller scales, the Lagrangian model of \cite{Valageas:2013gba} is usually more
accurate, but we find that our model agrees with the numerical simulations to better
than $2\%$ down to $7 h^{-1} {\rm Mpc}$, at $z \geq 0.35$.
This level of accuracy is better than most other approaches.
Eulerian perturbation schemes, like SPT or EFT, do not give a well-defined correlation 
function because they predict a power spectrum that grows artificially fast at high $k$.
Lagrangian approaches that go beyond the Zeldovich approximation by including
higher-order cumulants, such as convolution Lagrangian perturbation theory, 
do not significantly improve over the Zeldovich approximation and can become worse
below $30 h^{-1}{\rm Mpc}$ \cite{Vlah:2014nta}.

The better agreement with the configuration-space correlation function $\xi(x)$
than with the matter power spectrum $P(k)$ shows that the former is a more robust
statistics in the nonlinear regime \cite{Valageas:2013hxa,Tassev:2013rta}. 
Indeed, in contrast with linear scales,
where different Fourier modes are uncorrelated, we can expect nonlinear processes
that are local in space to generate weaker correlations between different scales in configuration
space than in Fourier space. Then, the power spectrum being the Fourier transform of
the correlation function, it receives contributions from
the correlation function at all scales.
The model presented in this paper is also more naturally suited to 
configuration-space statistics at it is based on the displacement field, hence on
a Lagrangian approach where we follow particle trajectories in spacetime.
Indeed, as for the Zeldovich approximation and its extensions to low-order
cumulants, it is known that Lagrangian-space formulations are rather efficient
for the two-point correlation functions 
\cite{Matsubara:2007wj,Okamura:2011nu,Valageas:2013hxa,Vlah:2014nta}.

In Lagrangian approaches that go beyond the Zeldovich approximation by taking 
into account higher-order cumulants and are exact up to one or two-loop order,
it has been found that higher orders can actually worsen the agreement with simulations
on intermediate scales and BAO scales at low redshifts \cite{Okamura:2011nu,Vlah:2014nta}.
This is partly due to the fact that within such Lagrangian approaches, as in the
standard Zeldovich approximation, particles do not remain trapped inside nonlinear
density fluctuations. This free streaming erases small-scale structures and 
leads to an underestimate of the matter density fluctuations.
Adding higher orders in a perturbative manner does not solve this issue and can actually
worsen the problem, as the amplitude of the displacement field is further increased on small
scales. The systematic better agreement obtained by our approach is due to the
effective truncation of the displacement power spectrum at high $k$.
As for the truncated Zeldovich approximation, this partly cures the erasing of nonlinear structures
and provides a better model for the large-scale density field 
\cite{Coles:1992vr}.

\section{Conclusion}
\label{sec:Conclusion}

In this paper, we have presented a new approach to model the gravitational dynamics
of large-scale structures. The aim is to avoid introducing free parameters, which
need to be fitted to numerical simulations, and to go beyond perturbation theory.
To do so, we work within a Lagrangian framework. This allows us to use the 
exact equations of motion, which remain valid beyond shell crossing, if we neglect
baryonic effects.
Then, we propose to use these equations of motion as constraints on the evolution
of the probability distribution functional ${\cal P}$ of the displacement and velocity fields.
Thus, the approximation only enters at the level of the description of this
distribution ${\cal P}$. In this paper we focus on a Gaussian ansatz, but
in principle we could consider more complex distributions ${\cal P}$,
which would involve additional parameters in addition to the power spectra
(e.g., low-order cumulants). As the ansatz used for ${\cal P}$ becomes more
complex, one increases the number of constraints
derived from the equations of motion of the particles, so as to fully determine ${\cal P}$
(e.g., the evolution equations of low-order correlation functions).
In this fashion, one can hope to systematically increase the accuracy of the predictions.
However, the complexity beyond the Gaussian case may prove difficult for practical
computations. We leave this investigation for future works.

Already at the Gaussian level for ${\cal P}$, we have found that this approach
leads to interesting results. Because we use exact equations of motion
(in fact, a subset of the infinite sequence that determines the exact probability
distribution) we go beyond perturbation theory. Then, the displacement-field
power spectrum becomes damped on nonlinear scales, with a truncation that is not
put by hand but arises from the dynamics. In particular, the damping factor 
$\lambda_\infty \sim e^{-1/(12\alpha_0)}$ is nonperturbative and shows the
characteristic exponential factor associated with the probability to form nonlinear
structures (for Gaussian initial conditions).
Moreover, in contrast with the Zeldovich approximation, the displacement and
velocity power spectra are different.

An interesting feature is that the auto power spectra are automatically positive,
as they should be, while cross power spectra can change sign.
This positivity property is not put by hand and appears naturally in our framework,
while it is often broken in perturbative schemes.
This nice behavior is likely related to the fact that, at each time, the probability distribution
${\cal P}$ is well defined and all quantities are exactly computed from this distribution.
Thus, they follow from physical particle distributions. This ensures that they do not
lead to theoretical inconsistencies (such as negative matter densities, or inconsistent
higher-order correlations).

We find that, both for self-similar dynamics and the realistic $\Lambda$-CDM cosmology,
the equations of motion for the displacement and velocity power spectra can be integrated
in terms of basic functions $y(k,\eta)$ that describe the amount of damping. This reduction
provides explicit expressions that ensure the positivity discussed above and also simplifies
the computations.

Already at this Gaussian level, this approach improves over the standard
Zeldovich approximation. It generates a self-truncation at high $k$, so that we obtain
a finite prediction even when the standard Zeldovich approximation does not exist,
as for linear power spectra with a lot of power at high $k$, $P_L(k) \propto k^n$
with $n \geq -1$.
It also improves over the truncated Zeldovich approximation as the truncation is not
put by hand and does not need to be fitted to simulations.
In a sense, this method obtains the ``best'' Gaussian approximation to the
gravitational dynamics, as selected by the equations of motion.

We have first discussed the predictions obtained for self-similar dynamics, to understand
how the exponent $n$ of the linear power spectrum affects the results.
Then, we have considered the realistic $\Lambda$-CDM cosmology. There, the qualitative
features can be understood from the change with redshift of the effective exponent $n$.
The comparison with numerical simulations shows that our results are probably not
competitive with other methods for the density power spectrum on BAO scales,
because of the failure to faithfully recover the smooth amplification of the nonlinear power 
spectrum. However, the damping of the BAO is well recovered and the method could
be useful if one is able to extract the oscillatory pattern from the data, in the spirit
of \cite{Noda:2017tfh,Noda:2019pzd}. Alternatively,
one can add a couple of free nuisance parameters to the model to take care of this
smooth component.
The agreement with simulations is much better for the configuration-space correlation function.
This is expected as we use a Lagrangian approach and the Zeldovich approximation is already
known to significantly improve over linear theory for these statistics. Our prediction improves
somewhat further over the Zeldovich approximation and we obtain an accuracy to better than
$2\%$ from BAO scales down to $7 h^{-1} {\rm Mpc}$ at $z \geq 0.35$.
Although some other methods may prove more accurate, the fact that there is no free
parameter to be marginalized over could make this approach competitive in terms of
constraining power on cosmological scenarios. We leave such an investigation for future works.

Among the natural extensions of this work, one could consider redshift-space statistics. 
However, the main question is whether one can efficiently go beyond the Gaussian
ansatz used in this paper. One possibility would be to include a few low-order
cumulants, such as the bispectrum, or to add higher-order terms in the probability
distribution itself. An alternative would be to consider nonlinear functionals of Gaussian
fields. We plan to investigate such issues in future studies.

\acknowledgments
The author would like to thank Takahiro Nishimichi for the use of his numerical simulations.

\appendix

\section{Expressions of $\lambda(k)$ for numerical computations}
\label{sec:expression-numerical}

For numerical computations, it is convenient to decompose the last exponential 
in Eq.(\ref{eq:lambda-0}) over its large-$q$ part and the remainder, as
\be
\lambda(k) = \lambda_0(k) + \lambda_1(k) , 
\label{eq:lambda-split-0-1}
\ee
with
\be
\lambda_0(k) = \int \frac{d{\bf q} d{\bf k}'}{(2\pi)^3} 
\frac{({\bf k}'\cdot{\bf k})^2}{(k'^2+\mu^2)k^2} 
\left( e^{i{\bf k} \cdot{\bf q}} -1 \right) e^{i{\bf k}'\cdot{\bf q} -\alpha_\infty k'^2} 
\label{eq:lambda-0-def}
\ee
and
\ba
&& \hspace{-0.5cm} \lambda_1(k) = \int \frac{d{\bf q} d{\bf k}'}{(2\pi)^3} 
\frac{({\bf k}'\cdot{\bf k})^2}{k'^2 k^2} \left( e^{i{\bf k} \cdot{\bf q}} -1 \right)
e^{i{\bf k}'\cdot{\bf q}}  \nonumber \\
&& \times \left[ e^{-\alpha(q) k'^2 - \beta(q) k'^2 ({\bf k}'\cdot{\bf q})^2/(k'q)^2} 
- e^{-\alpha_\infty k'^2} \right] .
\label{eq:lambda-1-def}
\ea
In Eq.(\ref{eq:lambda-1-def}) we have already taken the limit $\mu\to 0$,
as it is regular thanks to the vanishing of the last bracket for $k'\to 0$.
The integration over ${\bf q}$ gives at once for $\lambda_0$
\be
\lambda_0(k) = e^{-\alpha_\infty k^2} ,
\label{eq:lambda0-result}
\ee
where we took the limit $\mu\to 0$ at the end.
For $\lambda_1$, it is convenient to take the angular average 
of the expression (\ref{eq:lambda-1-def}) over the direction ${\bf \Omega}$ of
${\bf k}$, taking advantage of the fact that $\lambda_1(k)$ does not depend on 
${\bf \Omega}$. Using the property
\ba
\int \frac{d{\bf\Omega}}{4\pi} ({\bf n}\cdot{\bf k}_2)^2 \; e^{i k_1 {\bf n}\cdot{\bf q}} 
& = & \frac{k_2^2}{3} \biggl [ j_0(k_1 q) + j_2(k_1 q) \nonumber \\
&& \times \left( 1 - 3 \left(\frac{{\bf k}_2\cdot{\bf q}}{k_2 q}\right)^2 \right) \biggl ] 
, \hspace{0.5cm}
\ea
and integrating next over the angles of ${\bf q}$ and ${\bf k}'$, we obtain
\ba
\lambda_1(k) & = & \frac{2}{3\pi} \int_0^{\infty} dq dk'  q^2 k'^2 
\int_0^1 du \cos(k' q u)  \nonumber \\
&& \times \left[ j_0(k q) + j_2(k q) (1-3u^2) - 1 \right] \nonumber \\
&& \times \left[ e^{-\alpha k'^2 - \beta k'^2 u^2} - e^{-\alpha_\infty k'^2} \right] .
\label{eq:lambda1-1} 
\ea
Using the property \cite{Gradshteyn1965}
\be
\int_0^{\infty} dx \, \cos(ax) x^2 e^{-p^2 x^2} = \sqrt{\pi} \frac{2 p^2-a^2}{8 p^5} 
e^{-a^2/(4 p^2)} ,
\ee
the integration over $k'$ yields
\ba
&& \hspace{-0.3cm} \lambda_1(k) = \frac{1}{12\sqrt{\pi}} 
\int_0^{\infty} \hspace{-0.2cm} dq \, q^2
\int_0^1 \hspace{-0.1cm} du \, \Big[ j_0(k q) -1 + j_2(k q) \nonumber \\
&& \times (1-3u^2) \Big] \biggl\lbrace \frac{2\alpha+(2\beta-q^2) u^2}{(\alpha+\beta u^2)^{5/2}} 
e^{-q^2u^2/[4(\alpha+\beta u^2)]} \nonumber \\
&& - \frac{2\alpha_\infty-q^2u^2}{\alpha_\infty^{5/2}} e^{-q^2u^2/(4\alpha_\infty)} \biggl\rbrace .
\label{eq:lambda1-2} 
\ea
Changing variable from $u$ to $t=q^2u^2/[4(\alpha+\beta u^2)]$, we can integrate over $t$ 
most factors and we obtain
\ba
&& \hspace{-0.3cm} \lambda_1(k) = \int_0^{\infty} \!\!\! dq \biggl \lbrace  
\frac{j_0(k q) -1}{6 \sqrt{\pi}} q^2 \Big[ 
\frac{e^{-q^2/[4(\alpha+\beta)]}}{\alpha\sqrt{\alpha+\beta}} 
- \frac{e^{-q^2/(4\alpha_\infty)}}{\alpha_\infty^{3/2}} \Big] \nonumber \\
&& + j_2(k q) \Big[ \frac{2}{q} {\rm Erf}\Big(\frac{q}{2\sqrt{\alpha+\beta}}\Big) 
- \frac{2}{q} {\rm Erf}\Big(\frac{q}{2\sqrt{\alpha_\infty}}\Big) \nonumber \\
&& - \frac{2 e^{-q^2/[4(\alpha+\beta)]}}{\sqrt{\pi}\sqrt{\alpha+\beta}} 
\Big(1+\frac{(2\alpha-\beta)q^2}{12\alpha(\alpha+\beta)}\Big) 
+ \frac{2 e^{-q^2/(4\alpha_\infty)}}{\sqrt{\pi}\sqrt{\alpha_\infty}} \nonumber \\
&& \times \Big(1+\frac{q^2}{6\alpha_\infty}\Big) - \frac{8\beta}{\sqrt{\pi}q}
\int_0^{q^2/[4(\alpha+\beta)]} \!\!\! dt 
\frac{(1-2t)t^{3/2} e^{-t}}{q^2-4\beta t} \Big] \biggl \rbrace , \nonumber \\
&&
\label{eq:lambda1-3} 
\ea
where ${\rm Erf}(x)=\frac{2}{\sqrt{\pi}} \int_0^x dx \, e^{-x^2}$ is the error function.
This can also be written in terms of the complementary error function,
${\rm Erfc}(x) = 1-{\rm Erf}(x) = \frac{2}{\sqrt{\pi}} \int_x^{\infty} dx \, e^{-x^2}$, as
\ba
&& \hspace{-0.3cm} \lambda_1(k) = \int_0^{\infty} \!\!\! dq \biggl \lbrace  
\frac{j_0(k q) -1}{6 \sqrt{\pi}} q^2 \Big[ 
\frac{e^{-q^2/[4(\alpha+\beta)]}}{\alpha\sqrt{\alpha+\beta}} 
- \frac{e^{-q^2/(4\alpha_\infty)}}{\alpha_\infty^{3/2}} \Big] \nonumber \\
&& + j_2(k q) \Big[ - \frac{2}{q} {\rm Erfc}\Big(\frac{q}{2\sqrt{\alpha+\beta}}\Big) 
+ \frac{2}{q} {\rm Erfc}\Big(\frac{q}{2\sqrt{\alpha_\infty}}\Big) \nonumber \\
&& - \frac{2 e^{-q^2/[4(\alpha+\beta)]}}{\sqrt{\pi}\sqrt{\alpha+\beta}} 
\Big(1+\frac{(2\alpha-\beta)q^2}{12\alpha(\alpha+\beta)}\Big) 
+ \frac{2 e^{-q^2/(4\alpha_\infty)}}{\sqrt{\pi}\sqrt{\alpha_\infty}} \nonumber \\
&& \times \Big(1+\frac{q^2}{6\alpha_\infty}\Big) - \frac{8\beta}{\sqrt{\pi}q}
\int_0^{q^2/[4(\alpha+\beta)]} \!\!\! dt 
\frac{(1-2t)t^{3/2} e^{-t}}{q^2-4\beta t} \Big] \biggl \rbrace . \nonumber \\
&&
\label{eq:lambda1-4} 
\ea

From Eq.(\ref{eq:lambda1-4}) we can see that $\lambda_1(k) \propto k^2$
at low $k$, while Eq.(\ref{eq:lambda0-result}) gives $\lambda_0 \to 1$.
This gives the large-scale limit (\ref{eq:lambda-low-k}).

At large $k$, $\lambda_0(k)$ vanishes while the factors in Eq.(\ref{eq:lambda1-3})
associated with $[j_0(kq)-1]$, paired with the second term in the following bracket, 
or with $j_2(kq)$ go to a constant.
Then, Eq.(\ref{eq:lambda1-3}) is dominated by the factor $[j_0(kq)-1]$ associated with the  
first term in the following bracket.
Using the small-scale behaviors (\ref{eq:alpha-beta-q0}) we obtain Eq.(\ref{eq:lambda-large-k}).

For linear density fields with a lot of power on large scales, where $P_L(k)$ grows at least
as fast as $1/k$ for $k\to 0$, the variance $\alpha_\infty$ is infinite. This occurs for
the self-similar case $n=-2$ studied in Sec.~\ref{sec:numerical-n=-2}.
Then, we still perform the decomposition (\ref{eq:lambda-split-0-1}) but $\alpha_\infty$
is now an arbitrary parameter that we take of order $1/k_{\rm NL}^2$.
Indeed, the decomposition (\ref{eq:lambda-split-0-1}) is only used for numerical
convenience and the result $\lambda=\lambda_0 + \lambda_1$ does not depend
on the choice of $\alpha_\infty$. We checked that our numerical result does not change
as we vary $\alpha_\infty$ over two orders of magnitude.
Then, all expressions above still apply.

\section{Numerical computation of the density power spectrum}
\label{sec:numerical-density}

\subsection{Case where $\alpha_{\infty}$ is finite}
\label{sec:alpha-infty-finite-PZ}

Following \cite{Valageas:2010rx}, for the numerical computation of the density power 
spectrum (\ref{eq:Pk-density-Pchichi}), it is convenient to also use the expression obtained 
by expanding the oscillating part of the exponent in Eq.(\ref{eq:Pk-density-Pchichi})
\cite{Crocce:2005xz}. Using the one-point variance $\alpha_{\infty}$ introduced in
Eq.(\ref{eq:alpha-infty}), this gives
\ba
P(k) & = & e^{-\alpha_{\infty}k^2} \int \frac{d{\bf q}}{(2\pi)^3} e^{i {\bf k}\cdot{\bf q}}
\sum_{n=0}^{\infty} \frac{1}{n!} \nonumber \\
&& \times \left[ \int d{\bf k}' \, P_{\chi\chi}(k') 
\frac{({\bf k}\cdot{\bf k}')^2}{k'^4} e^{i {\bf k}'\cdot{\bf q}} \right]^n .
\label{eq:Pk-density-expanded-cosine}
\ea
In particular, the zeroth-order term vanishes for $k \neq 0$ and the linear and quadratic terms
give
\be
P(k) = e^{-\alpha_{\infty}k^2} [ P_{\chi\chi}(k) + P_{22}(k) + \dots ] ,
\label{eq:PZel-PL-P22}
\ee
where the dots stand for terms that are cubic or higher powers in $P_{\chi\chi}(k)$ and
\ba
P_{22}(k) & = & \int d{\bf k}_1 d{\bf k}_2 \; \delta_D({\bf k}_1+{\bf k}_2-{\bf k})
P_{\chi\chi}(k_1) P_{\chi\chi}(k_2) \nonumber \\
&& \times \frac{({\bf k}_1\cdot{\bf k})^2 ({\bf k}_2\cdot{\bf k})^2}{2 k_1^4 k_2^4} .
\ea
Then, we subtract the two terms of Eq.(\ref{eq:PZel-PL-P22}) from the expansion
(\ref{eq:Pk-density-alpha-beta}) to obtain
\ba
&& \hspace{-0.2cm} P(k) = e^{-\alpha_{\infty}k^2} [ P_{\chi\chi} + P_{22} ] 
+ e^{-\alpha_{\infty}k^2} \int \frac{d q}{2\pi^2}  q^2 \nonumber \\
&& \hspace{-0.2cm} \times \Biggl \lbrace e^{(\alpha_{\infty}-\alpha-\beta) k^2} 
\sum_{\ell=0}^{\infty} \left( \frac{2\beta k}{q} \right)^{\ell} j_{\ell}(kq) \nonumber \\
&& \hspace{-0.2cm} - \left[ 1 + (\alpha_{\infty}-\alpha-\beta) k^2 
+ \frac{1}{2} (\alpha_{\infty}-\alpha-\beta)^2 k^4 \right] j_0(kq) \nonumber \\
&& \hspace{-0.2cm} - \left[ 1 + (\alpha_{\infty}-\alpha-\beta) k^2 \right] \frac{2\beta k}{q} j_1(kq) 
- \left( \frac{2\beta k}{q} \right)^2 j_2(kq) \Biggl \rbrace .  \nonumber \\
&& \label{eq:Pk-series-jl}
\ea
This ensures that the integral over $q$ is regular and shows a fast convergence at large $q$.

Here we assumed that $\alpha_\infty$ is finite. In particular, $\beta \to 0$ for $q\to\infty$,
so that higher orders in $\ell$ in the series in Eq.(\ref{eq:Pk-series-jl}) are strongly
suppressed. This corresponds to the power spectrum $n=0$, in the self-similar
case studied in Sec.~\ref{sec:numerical-n=0}, and to the $\Lambda$-CDM cosmology
studied in Sec.~\ref{sec:LCDM}.

\subsection{Case $n=-2$ where $\alpha_{\infty}$ is infinite}
\label{sec:alpha-infty-infinite-PZ}

When $\alpha_\infty$ is infinite, as for the case $n=-2$ studied in Sec.~\ref{sec:numerical-n=-2},
we modify the approach leading to Eq.(\ref{eq:Pk-series-jl}).
Instead of subtracting a term $e^{-\alpha_{\infty}k^2} [ P_{\chi\chi} + P_{22} ]$ from
the expression (\ref{eq:Pk-alpha-beta-q}), we simply subtract the Zeldovich power spectrum
(\ref{eq:PkZ-alpha-beta-q}), which has the same form except that the nonlinear variances
$\alpha$ and $\beta$ are replaced by the linear-theory variances $\alpha_L$ and
$\beta_L$. For $n=-2$ they are given by Eq.(\ref{eq:alphaL-betaL-n-2}).
Thus, we write
\ba
P(k) & = & P_Z(k) + \int \frac{d q}{2\pi^2}  q^2 \int_0^1 d\mu \; \cos(kq\mu) \Biggl [ 
e^{-(\alpha+\beta\mu^2)k^2} \nonumber \\
&& - e^{-\pi (1+\mu^2) k^2 q/16} \Biggl ] .
\ea
Integrating over the angle cosine $\mu$ as for Eq.(\ref{eq:Pk-density-alpha-beta}),
we obtain
\ba
P(k) & = & P_Z(k) + \int \frac{d q}{2\pi^2}  q^2 \sum_{\ell=0}^{\infty} j_{\ell}(kq) 
\Biggl [ e^{-(\alpha+\beta) k^2} \left( \frac{2\beta k}{q} \right)^{\ell} \nonumber \\
&& - e^{-\pi k^2 q/8} \left( \frac{\pi k}{8} \right)^{\ell} \Biggl ] .
\ea
This again ensures fast numerical computations.
The Zeldovich power spectrum part $P_Z(k)$ is easily computed from the analytical
expression (\ref{eq:PZ-n-2}).

\bibliography{ref1}

\begin{thebibliography}{75}
\expandafter\ifx\csname natexlab\endcsname\relax\def\natexlab#1{#1}\fi
\expandafter\ifx\csname bibnamefont\endcsname\relax
  \def\bibnamefont#1{#1}\fi
\expandafter\ifx\csname bibfnamefont\endcsname\relax
  \def\bibfnamefont#1{#1}\fi
\expandafter\ifx\csname citenamefont\endcsname\relax
  \def\citenamefont#1{#1}\fi
\expandafter\ifx\csname url\endcsname\relax
  \def\url#1{\texttt{#1}}\fi
\expandafter\ifx\csname urlprefix\endcsname\relax\def\urlprefix{URL }\fi
\providecommand{\bibinfo}[2]{#2}
\providecommand{\eprint}[2][]{\url{#2}}

\bibitem[{\citenamefont{Eisenstein et~al.}(2007)\citenamefont{Eisenstein, Seo,
  and White}}]{Eisenstein:2006nj}
\bibinfo{author}{\bibfnamefont{D.~J.} \bibnamefont{Eisenstein}},
  \bibinfo{author}{\bibfnamefont{H.-j.} \bibnamefont{Seo}}, \bibnamefont{and}
  \bibinfo{author}{\bibfnamefont{M.~J.} \bibnamefont{White}},
  \bibinfo{journal}{Astrophys. J.} \textbf{\bibinfo{volume}{664}},
  \bibinfo{pages}{660} (\bibinfo{year}{2007}), \eprint{astro-ph/0604361}.

\bibitem[{\citenamefont{Eisenstein et~al.}(1998)\citenamefont{Eisenstein, Hu,
  and Tegmark}}]{Eisenstein:1998tu}
\bibinfo{author}{\bibfnamefont{D.~J.} \bibnamefont{Eisenstein}},
  \bibinfo{author}{\bibfnamefont{W.}~\bibnamefont{Hu}}, \bibnamefont{and}
  \bibinfo{author}{\bibfnamefont{M.}~\bibnamefont{Tegmark}},
  \bibinfo{journal}{Astrophys. J.} \textbf{\bibinfo{volume}{504}},
  \bibinfo{pages}{L57} (\bibinfo{year}{1998}), \eprint{astro-ph/9805239}.

\bibitem[{\citenamefont{Munshi et~al.}(2008)\citenamefont{Munshi, Valageas,
  Van~Waerbeke, and Heavens}}]{Munshi:2006fn}
\bibinfo{author}{\bibfnamefont{D.}~\bibnamefont{Munshi}},
  \bibinfo{author}{\bibfnamefont{P.}~\bibnamefont{Valageas}},
  \bibinfo{author}{\bibfnamefont{L.}~\bibnamefont{Van~Waerbeke}},
  \bibnamefont{and} \bibinfo{author}{\bibfnamefont{A.}~\bibnamefont{Heavens}},
  \bibinfo{journal}{Phys. Rept.} \textbf{\bibinfo{volume}{462}},
  \bibinfo{pages}{67} (\bibinfo{year}{2008}), \eprint{astro-ph/0612667}.

\bibitem[{\citenamefont{Ross et~al.}(2017)}]{Ross:2016gvb}
\bibinfo{author}{\bibfnamefont{A.~J.} \bibnamefont{Ross}} \bibnamefont{et~al.}
  (\bibinfo{collaboration}{BOSS}), \bibinfo{journal}{Mon. Not. Roy. Astron.
  Soc.} \textbf{\bibinfo{volume}{464}}, \bibinfo{pages}{1168}
  (\bibinfo{year}{2017}), \eprint{1607.03145}.

\bibitem[{\citenamefont{Blake et~al.}(2011)}]{Blake:2011en}
\bibinfo{author}{\bibfnamefont{C.}~\bibnamefont{Blake}} \bibnamefont{et~al.},
  \bibinfo{journal}{Mon. Not. Roy. Astron. Soc.}
  \textbf{\bibinfo{volume}{418}}, \bibinfo{pages}{1707} (\bibinfo{year}{2011}),
  \eprint{1108.2635}.

\bibitem[{\citenamefont{Martini et~al.}(2018)}]{Martini:2018kdj}
\bibinfo{author}{\bibfnamefont{P.}~\bibnamefont{Martini}} \bibnamefont{et~al.}
  (\bibinfo{collaboration}{DESI}), \bibinfo{journal}{Proc. SPIE Int. Soc. Opt.
  Eng.} \textbf{\bibinfo{volume}{10702}}, \bibinfo{pages}{107021F}
  (\bibinfo{year}{2018}), \eprint{1807.09287}.

\bibitem[{\citenamefont{Laureijs et~al.}(2011)}]{Laureijs:2011gra}
\bibinfo{author}{\bibfnamefont{R.}~\bibnamefont{Laureijs}} \bibnamefont{et~al.}
  (\bibinfo{collaboration}{EUCLID}) (\bibinfo{year}{2011}), \eprint{1110.3193}.

\bibitem[{\citenamefont{Abell et~al.}(2009)}]{Abell:2009aa}
\bibinfo{author}{\bibfnamefont{P.~A.} \bibnamefont{Abell}} \bibnamefont{et~al.}
  (\bibinfo{collaboration}{LSST Science, LSST Project}) (\bibinfo{year}{2009}),
  \eprint{0912.0201}.

\bibitem[{\citenamefont{Goroff et~al.}(1986)\citenamefont{Goroff, Grinstein,
  Rey, and Wise}}]{Goroff:1986ep}
\bibinfo{author}{\bibfnamefont{M.~H.} \bibnamefont{Goroff}},
  \bibinfo{author}{\bibfnamefont{B.}~\bibnamefont{Grinstein}},
  \bibinfo{author}{\bibfnamefont{S.~J.} \bibnamefont{Rey}}, \bibnamefont{and}
  \bibinfo{author}{\bibfnamefont{M.~B.} \bibnamefont{Wise}},
  \bibinfo{journal}{Astrophys. J.} \textbf{\bibinfo{volume}{311}},
  \bibinfo{pages}{6} (\bibinfo{year}{1986}).

\bibitem[{\citenamefont{Bernardeau et~al.}(2002)\citenamefont{Bernardeau,
  Colombi, Gaztanaga, and Scoccimarro}}]{Bernardeau:2001qr}
\bibinfo{author}{\bibfnamefont{F.}~\bibnamefont{Bernardeau}},
  \bibinfo{author}{\bibfnamefont{S.}~\bibnamefont{Colombi}},
  \bibinfo{author}{\bibfnamefont{E.}~\bibnamefont{Gaztanaga}},
  \bibnamefont{and}
  \bibinfo{author}{\bibfnamefont{R.}~\bibnamefont{Scoccimarro}},
  \bibinfo{journal}{Phys. Rept.} \textbf{\bibinfo{volume}{367}},
  \bibinfo{pages}{1} (\bibinfo{year}{2002}), \eprint{astro-ph/0112551}.

\bibitem[{\citenamefont{Crocce and
  Scoccimarro}(2006{\natexlab{a}})}]{Crocce:2005xy}
\bibinfo{author}{\bibfnamefont{M.}~\bibnamefont{Crocce}} \bibnamefont{and}
  \bibinfo{author}{\bibfnamefont{R.}~\bibnamefont{Scoccimarro}},
  \bibinfo{journal}{Phys. Rev.} \textbf{\bibinfo{volume}{D73}},
  \bibinfo{pages}{063519} (\bibinfo{year}{2006}{\natexlab{a}}),
  \eprint{astro-ph/0509418}.

\bibitem[{\citenamefont{Valageas}(2007{\natexlab{a}})}]{Valageas:2006bi}
\bibinfo{author}{\bibfnamefont{P.}~\bibnamefont{Valageas}},
  \bibinfo{journal}{Astron. Astrophys.} \textbf{\bibinfo{volume}{465}},
  \bibinfo{pages}{725} (\bibinfo{year}{2007}{\natexlab{a}}),
  \eprint{astro-ph/0611849}.

\bibitem[{\citenamefont{Bernardeau et~al.}(2008)\citenamefont{Bernardeau,
  Crocce, and Scoccimarro}}]{Bernardeau:2008fa}
\bibinfo{author}{\bibfnamefont{F.}~\bibnamefont{Bernardeau}},
  \bibinfo{author}{\bibfnamefont{M.}~\bibnamefont{Crocce}}, \bibnamefont{and}
  \bibinfo{author}{\bibfnamefont{R.}~\bibnamefont{Scoccimarro}},
  \bibinfo{journal}{Phys. Rev.} \textbf{\bibinfo{volume}{D78}},
  \bibinfo{pages}{103521} (\bibinfo{year}{2008}), \eprint{0806.2334}.

\bibitem[{\citenamefont{Taruya et~al.}(2012)\citenamefont{Taruya, Bernardeau,
  Nishimichi, and Codis}}]{Taruya:2012ut}
\bibinfo{author}{\bibfnamefont{A.}~\bibnamefont{Taruya}},
  \bibinfo{author}{\bibfnamefont{F.}~\bibnamefont{Bernardeau}},
  \bibinfo{author}{\bibfnamefont{T.}~\bibnamefont{Nishimichi}},
  \bibnamefont{and} \bibinfo{author}{\bibfnamefont{S.}~\bibnamefont{Codis}},
  \bibinfo{journal}{Phys. Rev.} \textbf{\bibinfo{volume}{D86}},
  \bibinfo{pages}{103528} (\bibinfo{year}{2012}), \eprint{1208.1191}.

\bibitem[{\citenamefont{Carlson et~al.}(2009)\citenamefont{Carlson, White, and
  Padmanabhan}}]{Carlson:2009it}
\bibinfo{author}{\bibfnamefont{J.}~\bibnamefont{Carlson}},
  \bibinfo{author}{\bibfnamefont{M.}~\bibnamefont{White}}, \bibnamefont{and}
  \bibinfo{author}{\bibfnamefont{N.}~\bibnamefont{Padmanabhan}},
  \bibinfo{journal}{Phys. Rev.} \textbf{\bibinfo{volume}{D80}},
  \bibinfo{pages}{043531} (\bibinfo{year}{2009}), \eprint{0905.0479}.

\bibitem[{\citenamefont{Valageas}(2011)}]{Valageas:2010rx}
\bibinfo{author}{\bibfnamefont{P.}~\bibnamefont{Valageas}},
  \bibinfo{journal}{Astron. Astrophys.} \textbf{\bibinfo{volume}{526}},
  \bibinfo{pages}{A67} (\bibinfo{year}{2011}), \eprint{1009.0106}.

\bibitem[{\citenamefont{Blas et~al.}(2014)\citenamefont{Blas, Garny, and
  Konstandin}}]{Blas:2013aba}
\bibinfo{author}{\bibfnamefont{D.}~\bibnamefont{Blas}},
  \bibinfo{author}{\bibfnamefont{M.}~\bibnamefont{Garny}}, \bibnamefont{and}
  \bibinfo{author}{\bibfnamefont{T.}~\bibnamefont{Konstandin}},
  \bibinfo{journal}{JCAP} \textbf{\bibinfo{volume}{1401}}, \bibinfo{pages}{010}
  (\bibinfo{year}{2014}), \eprint{1309.3308}.

\bibitem[{\citenamefont{Valageas}(2013)}]{Valageas:2013hxa}
\bibinfo{author}{\bibfnamefont{P.}~\bibnamefont{Valageas}},
  \bibinfo{journal}{Phys. Rev.} \textbf{\bibinfo{volume}{D88}},
  \bibinfo{pages}{083524} (\bibinfo{year}{2013}), \eprint{1308.6755}.

\bibitem[{\citenamefont{Pichon and Bernardeau}(1999)}]{Pichon:1999tk}
\bibinfo{author}{\bibfnamefont{C.}~\bibnamefont{Pichon}} \bibnamefont{and}
  \bibinfo{author}{\bibfnamefont{F.}~\bibnamefont{Bernardeau}},
  \bibinfo{journal}{Astron. Astrophys.} \textbf{\bibinfo{volume}{343}},
  \bibinfo{pages}{663} (\bibinfo{year}{1999}), \eprint{astro-ph/9902142}.

\bibitem[{\citenamefont{Pueblas and Scoccimarro}(2009)}]{Pueblas:2008uv}
\bibinfo{author}{\bibfnamefont{S.}~\bibnamefont{Pueblas}} \bibnamefont{and}
  \bibinfo{author}{\bibfnamefont{R.}~\bibnamefont{Scoccimarro}},
  \bibinfo{journal}{Phys. Rev.} \textbf{\bibinfo{volume}{D80}},
  \bibinfo{pages}{043504} (\bibinfo{year}{2009}), \eprint{0809.4606}.

\bibitem[{\citenamefont{Pietroni et~al.}(2012)\citenamefont{Pietroni, Mangano,
  Saviano, and Viel}}]{Pietroni:2011iz}
\bibinfo{author}{\bibfnamefont{M.}~\bibnamefont{Pietroni}},
  \bibinfo{author}{\bibfnamefont{G.}~\bibnamefont{Mangano}},
  \bibinfo{author}{\bibfnamefont{N.}~\bibnamefont{Saviano}}, \bibnamefont{and}
  \bibinfo{author}{\bibfnamefont{M.}~\bibnamefont{Viel}},
  \bibinfo{journal}{JCAP} \textbf{\bibinfo{volume}{1201}}, \bibinfo{pages}{019}
  (\bibinfo{year}{2012}), \eprint{1108.5203}.

\bibitem[{\citenamefont{Baumann et~al.}(2012)\citenamefont{Baumann, Nicolis,
  Senatore, and Zaldarriaga}}]{Baumann:2010tm}
\bibinfo{author}{\bibfnamefont{D.}~\bibnamefont{Baumann}},
  \bibinfo{author}{\bibfnamefont{A.}~\bibnamefont{Nicolis}},
  \bibinfo{author}{\bibfnamefont{L.}~\bibnamefont{Senatore}}, \bibnamefont{and}
  \bibinfo{author}{\bibfnamefont{M.}~\bibnamefont{Zaldarriaga}},
  \bibinfo{journal}{JCAP} \textbf{\bibinfo{volume}{1207}}, \bibinfo{pages}{051}
  (\bibinfo{year}{2012}), \eprint{1004.2488}.

\bibitem[{\citenamefont{Carrasco et~al.}(2012)\citenamefont{Carrasco,
  Hertzberg, and Senatore}}]{Carrasco:2012cv}
\bibinfo{author}{\bibfnamefont{J.~J.~M.} \bibnamefont{Carrasco}},
  \bibinfo{author}{\bibfnamefont{M.~P.} \bibnamefont{Hertzberg}},
  \bibnamefont{and} \bibinfo{author}{\bibfnamefont{L.}~\bibnamefont{Senatore}},
  \bibinfo{journal}{JHEP} \textbf{\bibinfo{volume}{09}}, \bibinfo{pages}{082}
  (\bibinfo{year}{2012}), \eprint{1206.2926}.

\bibitem[{\citenamefont{D'Amico et~al.}(2019)\citenamefont{D'Amico, Gleyzes,
  Kokron, Markovic, Senatore, Zhang, Beutler, and Gil-Marn}}]{DAmico:2019fhj}
\bibinfo{author}{\bibfnamefont{G.}~\bibnamefont{D'Amico}},
  \bibinfo{author}{\bibfnamefont{J.}~\bibnamefont{Gleyzes}},
  \bibinfo{author}{\bibfnamefont{N.}~\bibnamefont{Kokron}},
  \bibinfo{author}{\bibfnamefont{D.}~\bibnamefont{Markovic}},
  \bibinfo{author}{\bibfnamefont{L.}~\bibnamefont{Senatore}},
  \bibinfo{author}{\bibfnamefont{P.}~\bibnamefont{Zhang}},
  \bibinfo{author}{\bibfnamefont{F.}~\bibnamefont{Beutler}}, \bibnamefont{and}
  \bibinfo{author}{\bibfnamefont{H.}~\bibnamefont{Gil-Marn}}
  (\bibinfo{year}{2019}), \eprint{1909.05271}.

\bibitem[{\citenamefont{Senatore}(2015)}]{Senatore:2014eva}
\bibinfo{author}{\bibfnamefont{L.}~\bibnamefont{Senatore}},
  \bibinfo{journal}{JCAP} \textbf{\bibinfo{volume}{1511}}, \bibinfo{pages}{007}
  (\bibinfo{year}{2015}), \eprint{1406.7843}.

\bibitem[{\citenamefont{Perko et~al.}(2016)\citenamefont{Perko, Senatore,
  Jennings, and Wechsler}}]{Perko:2016puo}
\bibinfo{author}{\bibfnamefont{A.}~\bibnamefont{Perko}},
  \bibinfo{author}{\bibfnamefont{L.}~\bibnamefont{Senatore}},
  \bibinfo{author}{\bibfnamefont{E.}~\bibnamefont{Jennings}}, \bibnamefont{and}
  \bibinfo{author}{\bibfnamefont{R.~H.} \bibnamefont{Wechsler}}
  (\bibinfo{year}{2016}), \eprint{1610.09321}.

\bibitem[{\citenamefont{Lewandowski et~al.}(2015)\citenamefont{Lewandowski,
  Perko, and Senatore}}]{Lewandowski:2014rca}
\bibinfo{author}{\bibfnamefont{M.}~\bibnamefont{Lewandowski}},
  \bibinfo{author}{\bibfnamefont{A.}~\bibnamefont{Perko}}, \bibnamefont{and}
  \bibinfo{author}{\bibfnamefont{L.}~\bibnamefont{Senatore}},
  \bibinfo{journal}{JCAP} \textbf{\bibinfo{volume}{1505}}, \bibinfo{pages}{019}
  (\bibinfo{year}{2015}), \eprint{1412.5049}.

\bibitem[{\citenamefont{Zeldovich}(1970)}]{Zeldovich:1969sb}
\bibinfo{author}{\bibfnamefont{{\relax Ya}.~B.} \bibnamefont{Zeldovich}},
  \bibinfo{journal}{Astron. Astrophys.} \textbf{\bibinfo{volume}{5}},
  \bibinfo{pages}{84} (\bibinfo{year}{1970}).

\bibitem[{\citenamefont{Buchert}(1992)}]{Buchert:1992ya}
\bibinfo{author}{\bibfnamefont{T.}~\bibnamefont{Buchert}},
  \bibinfo{journal}{Mon. Not. Roy. Astron. Soc.}
  \textbf{\bibinfo{volume}{254}}, \bibinfo{pages}{729} (\bibinfo{year}{1992}).

\bibitem[{\citenamefont{Bouchet et~al.}(1992)\citenamefont{Bouchet,
  Juszkiewicz, Colombi, and Pellat}}]{Bouchet:1992uh}
\bibinfo{author}{\bibfnamefont{F.~R.} \bibnamefont{Bouchet}},
  \bibinfo{author}{\bibfnamefont{R.}~\bibnamefont{Juszkiewicz}},
  \bibinfo{author}{\bibfnamefont{S.}~\bibnamefont{Colombi}}, \bibnamefont{and}
  \bibinfo{author}{\bibfnamefont{R.}~\bibnamefont{Pellat}},
  \bibinfo{journal}{Astrophys. J.} \textbf{\bibinfo{volume}{394}},
  \bibinfo{pages}{L5} (\bibinfo{year}{1992}).

\bibitem[{\citenamefont{Buchert and Ehlers}(1993)}]{Buchert:1993xz}
\bibinfo{author}{\bibfnamefont{T.}~\bibnamefont{Buchert}} \bibnamefont{and}
  \bibinfo{author}{\bibfnamefont{J.}~\bibnamefont{Ehlers}},
  \bibinfo{journal}{Mon. Not. Roy. Astron. Soc.}
  \textbf{\bibinfo{volume}{264}}, \bibinfo{pages}{375} (\bibinfo{year}{1993}).

\bibitem[{\citenamefont{Bouchet et~al.}(1995)\citenamefont{Bouchet, Colombi,
  Hivon, and Juszkiewicz}}]{Bouchet:1994xp}
\bibinfo{author}{\bibfnamefont{F.~R.} \bibnamefont{Bouchet}},
  \bibinfo{author}{\bibfnamefont{S.}~\bibnamefont{Colombi}},
  \bibinfo{author}{\bibfnamefont{E.}~\bibnamefont{Hivon}}, \bibnamefont{and}
  \bibinfo{author}{\bibfnamefont{R.}~\bibnamefont{Juszkiewicz}},
  \bibinfo{journal}{Astron. Astrophys.} \textbf{\bibinfo{volume}{296}},
  \bibinfo{pages}{575} (\bibinfo{year}{1995}), \eprint{astro-ph/9406013}.

\bibitem[{\citenamefont{Matsubara}(2008)}]{Matsubara:2007wj}
\bibinfo{author}{\bibfnamefont{T.}~\bibnamefont{Matsubara}},
  \bibinfo{journal}{Phys. Rev.} \textbf{\bibinfo{volume}{D77}},
  \bibinfo{pages}{063530} (\bibinfo{year}{2008}), \eprint{0711.2521}.

\bibitem[{\citenamefont{Vlah et~al.}(2015)\citenamefont{Vlah, Seljak, and
  Baldauf}}]{Vlah:2014nta}
\bibinfo{author}{\bibfnamefont{Z.}~\bibnamefont{Vlah}},
  \bibinfo{author}{\bibfnamefont{U.}~\bibnamefont{Seljak}}, \bibnamefont{and}
  \bibinfo{author}{\bibfnamefont{T.}~\bibnamefont{Baldauf}},
  \bibinfo{journal}{Phys. Rev.} \textbf{\bibinfo{volume}{D91}},
  \bibinfo{pages}{023508} (\bibinfo{year}{2015}), \eprint{1410.1617}.

\bibitem[{\citenamefont{Matsubara}(2015)}]{Matsubara:2015ipa}
\bibinfo{author}{\bibfnamefont{T.}~\bibnamefont{Matsubara}},
  \bibinfo{journal}{Phys. Rev.} \textbf{\bibinfo{volume}{D92}},
  \bibinfo{pages}{023534} (\bibinfo{year}{2015}), \eprint{1505.01481}.

\bibitem[{\citenamefont{Taruya and Colombi}(2017)}]{Taruya:2017ohk}
\bibinfo{author}{\bibfnamefont{A.}~\bibnamefont{Taruya}} \bibnamefont{and}
  \bibinfo{author}{\bibfnamefont{S.}~\bibnamefont{Colombi}},
  \bibinfo{journal}{Mon. Not. Roy. Astron. Soc.}
  \textbf{\bibinfo{volume}{470}}, \bibinfo{pages}{4858} (\bibinfo{year}{2017}),
  \eprint{1701.09088}.

\bibitem[{\citenamefont{McDonald and Vlah}(2018)}]{McDonald:2017ths}
\bibinfo{author}{\bibfnamefont{P.}~\bibnamefont{McDonald}} \bibnamefont{and}
  \bibinfo{author}{\bibfnamefont{Z.}~\bibnamefont{Vlah}},
  \bibinfo{journal}{Phys. Rev.} \textbf{\bibinfo{volume}{D97}},
  \bibinfo{pages}{023508} (\bibinfo{year}{2018}), \eprint{1709.02834}.

\bibitem[{\citenamefont{Valageas}(2004)}]{Valageas:2003gm}
\bibinfo{author}{\bibfnamefont{P.}~\bibnamefont{Valageas}},
  \bibinfo{journal}{Astron. Astrophys.} \textbf{\bibinfo{volume}{421}},
  \bibinfo{pages}{23} (\bibinfo{year}{2004}), \eprint{astro-ph/0307008}.

\bibitem[{\citenamefont{Tassev}(2011)}]{Tassev:2010us}
\bibinfo{author}{\bibfnamefont{S.}~\bibnamefont{Tassev}},
  \bibinfo{journal}{JCAP} \textbf{\bibinfo{volume}{1110}}, \bibinfo{pages}{022}
  (\bibinfo{year}{2011}), \eprint{1012.0282}.

\bibitem[{\citenamefont{Widrow and Kaiser}(1993)}]{Widrow:1993qq}
\bibinfo{author}{\bibfnamefont{L.~M.} \bibnamefont{Widrow}} \bibnamefont{and}
  \bibinfo{author}{\bibfnamefont{N.}~\bibnamefont{Kaiser}},
  \bibinfo{journal}{Astrophys. J.} \textbf{\bibinfo{volume}{416}},
  \bibinfo{pages}{L71} (\bibinfo{year}{1993}).

\bibitem[{\citenamefont{Uhlemann et~al.}(2014)\citenamefont{Uhlemann, Kopp, and
  Haugg}}]{Uhlemann:2014npa}
\bibinfo{author}{\bibfnamefont{C.}~\bibnamefont{Uhlemann}},
  \bibinfo{author}{\bibfnamefont{M.}~\bibnamefont{Kopp}}, \bibnamefont{and}
  \bibinfo{author}{\bibfnamefont{T.}~\bibnamefont{Haugg}},
  \bibinfo{journal}{Phys. Rev.} \textbf{\bibinfo{volume}{D90}},
  \bibinfo{pages}{023517} (\bibinfo{year}{2014}), \eprint{1403.5567}.

\bibitem[{\citenamefont{Taruya and Hiramatsu}(2008)}]{Taruya:2007xy}
\bibinfo{author}{\bibfnamefont{A.}~\bibnamefont{Taruya}} \bibnamefont{and}
  \bibinfo{author}{\bibfnamefont{T.}~\bibnamefont{Hiramatsu}},
  \bibinfo{journal}{Astrophys. J.} \textbf{\bibinfo{volume}{674}},
  \bibinfo{pages}{617} (\bibinfo{year}{2008}), \eprint{0708.1367}.

\bibitem[{\citenamefont{Pietroni}(2008)}]{Pietroni:2008jx}
\bibinfo{author}{\bibfnamefont{M.}~\bibnamefont{Pietroni}},
  \bibinfo{journal}{JCAP} \textbf{\bibinfo{volume}{0810}}, \bibinfo{pages}{036}
  (\bibinfo{year}{2008}), \eprint{0806.0971}.

\bibitem[{\citenamefont{Anselmi and Pietroni}(2012)}]{Anselmi:2012cn}
\bibinfo{author}{\bibfnamefont{S.}~\bibnamefont{Anselmi}} \bibnamefont{and}
  \bibinfo{author}{\bibfnamefont{M.}~\bibnamefont{Pietroni}},
  \bibinfo{journal}{JCAP} \textbf{\bibinfo{volume}{1212}}, \bibinfo{pages}{013}
  (\bibinfo{year}{2012}), \eprint{1205.2235}.

\bibitem[{\citenamefont{Davis and Peebles}(1977)}]{Davis_1977}
\bibinfo{author}{\bibfnamefont{M.}~\bibnamefont{Davis}} \bibnamefont{and}
  \bibinfo{author}{\bibfnamefont{P.~J.~E.} \bibnamefont{Peebles}},
  \bibinfo{journal}{The Astrophysical Journal Supplement Series}
  \textbf{\bibinfo{volume}{34}}, \bibinfo{pages}{425} (\bibinfo{year}{1977}),
  ISSN \bibinfo{issn}{1538-4365},
  \urlprefix\url{http://dx.doi.org/10.1086/190456}.

\bibitem[{\citenamefont{Blas et~al.}(2016{\natexlab{a}})\citenamefont{Blas,
  Garny, Ivanov, and Sibiryakov}}]{Blas:2015qsi}
\bibinfo{author}{\bibfnamefont{D.}~\bibnamefont{Blas}},
  \bibinfo{author}{\bibfnamefont{M.}~\bibnamefont{Garny}},
  \bibinfo{author}{\bibfnamefont{M.~M.} \bibnamefont{Ivanov}},
  \bibnamefont{and}
  \bibinfo{author}{\bibfnamefont{S.}~\bibnamefont{Sibiryakov}},
  \bibinfo{journal}{JCAP} \textbf{\bibinfo{volume}{1607}}, \bibinfo{pages}{052}
  (\bibinfo{year}{2016}{\natexlab{a}}), \eprint{1512.05807}.

\bibitem[{\citenamefont{Valageas}(2008)}]{Valageas:2007su}
\bibinfo{author}{\bibfnamefont{P.}~\bibnamefont{Valageas}},
  \bibinfo{journal}{Astron. Astrophys.} \textbf{\bibinfo{volume}{484}},
  \bibinfo{pages}{79} (\bibinfo{year}{2008}), \eprint{0711.3407}.

\bibitem[{\citenamefont{Kiessling}(2003)}]{Kiessling:1999eq}
\bibinfo{author}{\bibfnamefont{M.~K.~H.} \bibnamefont{Kiessling}},
  \bibinfo{journal}{Adv. Appl. Math.} \textbf{\bibinfo{volume}{31}},
  \bibinfo{pages}{132} (\bibinfo{year}{2003}), \eprint{astro-ph/9910247}.

\bibitem[{\citenamefont{{Gabrielli} et~al.}(2009)\citenamefont{{Gabrielli},
  {Joyce}, and {Sicard}}}]{2009PhRvE..80d1108G}
\bibinfo{author}{\bibfnamefont{A.}~\bibnamefont{{Gabrielli}}},
  \bibinfo{author}{\bibfnamefont{M.}~\bibnamefont{{Joyce}}}, \bibnamefont{and}
  \bibinfo{author}{\bibfnamefont{F.}~\bibnamefont{{Sicard}}},
  \bibinfo{journal}{\pre} \textbf{\bibinfo{volume}{80}}, \bibinfo{eid}{041108}
  (\bibinfo{year}{2009}), \eprint{0812.4249}.

\bibitem[{\citenamefont{Gabrielli et~al.}(2010)\citenamefont{Gabrielli, Joyce,
  Marcos, and Sicard}}]{Gabrielli:2010uk}
\bibinfo{author}{\bibfnamefont{A.}~\bibnamefont{Gabrielli}},
  \bibinfo{author}{\bibfnamefont{M.}~\bibnamefont{Joyce}},
  \bibinfo{author}{\bibfnamefont{B.}~\bibnamefont{Marcos}}, \bibnamefont{and}
  \bibinfo{author}{\bibfnamefont{F.}~\bibnamefont{Sicard}},
  \bibinfo{journal}{J. Statist. Phys.} \textbf{\bibinfo{volume}{141}},
  \bibinfo{pages}{970} (\bibinfo{year}{2010}), \eprint{1003.5680}.

\bibitem[{\citenamefont{Bernardeau and Valageas}(2008)}]{Bernardeau:2008ss}
\bibinfo{author}{\bibfnamefont{F.}~\bibnamefont{Bernardeau}} \bibnamefont{and}
  \bibinfo{author}{\bibfnamefont{P.}~\bibnamefont{Valageas}},
  \bibinfo{journal}{Phys. Rev.} \textbf{\bibinfo{volume}{D78}},
  \bibinfo{pages}{083503} (\bibinfo{year}{2008}), \eprint{0805.0805}.

\bibitem[{\citenamefont{Valageas}(2014)}]{Valageas:2013zda}
\bibinfo{author}{\bibfnamefont{P.}~\bibnamefont{Valageas}},
  \bibinfo{journal}{Phys. Rev.} \textbf{\bibinfo{volume}{D89}},
  \bibinfo{pages}{123522} (\bibinfo{year}{2014}), \eprint{1311.4286}.

\bibitem[{\citenamefont{Kehagias et~al.}(2014)\citenamefont{Kehagias, Perrier,
  and Riotto}}]{Kehagias:2013paa}
\bibinfo{author}{\bibfnamefont{A.}~\bibnamefont{Kehagias}},
  \bibinfo{author}{\bibfnamefont{H.}~\bibnamefont{Perrier}}, \bibnamefont{and}
  \bibinfo{author}{\bibfnamefont{A.}~\bibnamefont{Riotto}},
  \bibinfo{journal}{Mod. Phys. Lett.} \textbf{\bibinfo{volume}{A29}},
  \bibinfo{pages}{1450152} (\bibinfo{year}{2014}), \eprint{1311.5524}.

\bibitem[{\citenamefont{Schneider and Bartelmann}(1995)}]{1995MNRAS.273..475S}
\bibinfo{author}{\bibfnamefont{P.}~\bibnamefont{Schneider}} \bibnamefont{and}
  \bibinfo{author}{\bibfnamefont{M.}~\bibnamefont{Bartelmann}},
  \bibinfo{journal}{Mon. Not. Roy. Astron. Soc.}
  \textbf{\bibinfo{volume}{273}}, \bibinfo{pages}{475} (\bibinfo{year}{1995}).

\bibitem[{\citenamefont{Taylor and Hamilton}(1996)}]{Taylor:1996ne}
\bibinfo{author}{\bibfnamefont{A.~N.} \bibnamefont{Taylor}} \bibnamefont{and}
  \bibinfo{author}{\bibfnamefont{A.~J.~S.} \bibnamefont{Hamilton}},
  \bibinfo{journal}{Mon. Not. Roy. Astron. Soc.}
  \textbf{\bibinfo{volume}{282}}, \bibinfo{pages}{767} (\bibinfo{year}{1996}),
  \eprint{astro-ph/9604020}.

\bibitem[{\citenamefont{Blas et~al.}(2016{\natexlab{b}})\citenamefont{Blas,
  Garny, Ivanov, and Sibiryakov}}]{Blas:2016sfa}
\bibinfo{author}{\bibfnamefont{D.}~\bibnamefont{Blas}},
  \bibinfo{author}{\bibfnamefont{M.}~\bibnamefont{Garny}},
  \bibinfo{author}{\bibfnamefont{M.~M.} \bibnamefont{Ivanov}},
  \bibnamefont{and}
  \bibinfo{author}{\bibfnamefont{S.}~\bibnamefont{Sibiryakov}},
  \bibinfo{journal}{JCAP} \textbf{\bibinfo{volume}{1607}}, \bibinfo{pages}{028}
  (\bibinfo{year}{2016}{\natexlab{b}}), \eprint{1605.02149}.

\bibitem[{\citenamefont{Peloso and Pietroni}(2017)}]{Peloso:2016qdr}
\bibinfo{author}{\bibfnamefont{M.}~\bibnamefont{Peloso}} \bibnamefont{and}
  \bibinfo{author}{\bibfnamefont{M.}~\bibnamefont{Pietroni}},
  \bibinfo{journal}{JCAP} \textbf{\bibinfo{volume}{1701}}, \bibinfo{pages}{056}
  (\bibinfo{year}{2017}), \eprint{1609.06624}.

\bibitem[{\citenamefont{Senatore and Trevisan}(2018)}]{Senatore:2017pbn}
\bibinfo{author}{\bibfnamefont{L.}~\bibnamefont{Senatore}} \bibnamefont{and}
  \bibinfo{author}{\bibfnamefont{G.}~\bibnamefont{Trevisan}},
  \bibinfo{journal}{JCAP} \textbf{\bibinfo{volume}{1805}}, \bibinfo{pages}{019}
  (\bibinfo{year}{2018}), \eprint{1710.02178}.

\bibitem[{\citenamefont{Noda et~al.}(2017)\citenamefont{Noda, Peloso, and
  Pietroni}}]{Noda:2017tfh}
\bibinfo{author}{\bibfnamefont{E.}~\bibnamefont{Noda}},
  \bibinfo{author}{\bibfnamefont{M.}~\bibnamefont{Peloso}}, \bibnamefont{and}
  \bibinfo{author}{\bibfnamefont{M.}~\bibnamefont{Pietroni}},
  \bibinfo{journal}{JCAP} \textbf{\bibinfo{volume}{1708}}, \bibinfo{pages}{007}
  (\bibinfo{year}{2017}), \eprint{1705.01475}.

\bibitem[{\citenamefont{Valageas}(2007{\natexlab{b}})}]{Valageas:2007ge}
\bibinfo{author}{\bibfnamefont{P.}~\bibnamefont{Valageas}},
  \bibinfo{journal}{Astron. Astrophys.}  (\bibinfo{year}{2007}{\natexlab{b}}),
  \bibinfo{note}{[Astron. Astrophys.476,31(2007)]}, \eprint{0706.2593}.

\bibitem[{\citenamefont{Valageas et~al.}(2013)\citenamefont{Valageas,
  Nishimichi, and Taruya}}]{Valageas:2013gba}
\bibinfo{author}{\bibfnamefont{P.}~\bibnamefont{Valageas}},
  \bibinfo{author}{\bibfnamefont{T.}~\bibnamefont{Nishimichi}},
  \bibnamefont{and} \bibinfo{author}{\bibfnamefont{A.}~\bibnamefont{Taruya}},
  \bibinfo{journal}{Phys. Rev.} \textbf{\bibinfo{volume}{D87}},
  \bibinfo{pages}{083522} (\bibinfo{year}{2013}), \eprint{1302.4533}.

\bibitem[{\citenamefont{Cooray and Sheth}(2002)}]{Cooray:2002dia}
\bibinfo{author}{\bibfnamefont{A.}~\bibnamefont{Cooray}} \bibnamefont{and}
  \bibinfo{author}{\bibfnamefont{R.~K.} \bibnamefont{Sheth}},
  \bibinfo{journal}{Phys. Rept.} \textbf{\bibinfo{volume}{372}},
  \bibinfo{pages}{1} (\bibinfo{year}{2002}), \eprint{astro-ph/0206508}.

\bibitem[{\citenamefont{Coles et~al.}(1992)\citenamefont{Coles, Melott, and
  Shandarin}}]{Coles:1992vr}
\bibinfo{author}{\bibfnamefont{P.}~\bibnamefont{Coles}},
  \bibinfo{author}{\bibfnamefont{A.~L.} \bibnamefont{Melott}},
  \bibnamefont{and} \bibinfo{author}{\bibfnamefont{S.~F.}
  \bibnamefont{Shandarin}} (\bibinfo{year}{1992}).

\bibitem[{\citenamefont{Peebles}(1980)}]{Peebles1980}
\bibinfo{author}{\bibfnamefont{P.~J.~E.} \bibnamefont{Peebles}},
  \emph{\bibinfo{title}{The large-scale structure of the universe}}
  (\bibinfo{publisher}{Princeton: Princeton Univ. Press},
  \bibinfo{year}{1980}).

\bibitem[{\citenamefont{Polyanin and Zaitsev}(2017)}]{Polyanin2017}
\bibinfo{author}{\bibfnamefont{A.}~\bibnamefont{Polyanin}} \bibnamefont{and}
  \bibinfo{author}{\bibfnamefont{V.}~\bibnamefont{Zaitsev}},
  \emph{\bibinfo{title}{Handbook of Ordinary Differential Equations: Exact
  Solutions, Methods, and Problems}} (\bibinfo{publisher}{Chapman and
  Hall/CRC}, \bibinfo{year}{2017}).

\bibitem[{\citenamefont{Makino et~al.}(1992)\citenamefont{Makino, Sasaki, and
  Suto}}]{Makino:1991rp}
\bibinfo{author}{\bibfnamefont{N.}~\bibnamefont{Makino}},
  \bibinfo{author}{\bibfnamefont{M.}~\bibnamefont{Sasaki}}, \bibnamefont{and}
  \bibinfo{author}{\bibfnamefont{Y.}~\bibnamefont{Suto}},
  \bibinfo{journal}{Phys. Rev.} \textbf{\bibinfo{volume}{D46}},
  \bibinfo{pages}{585} (\bibinfo{year}{1992}).

\bibitem[{\citenamefont{Scoccimarro and Frieman}(1996)}]{Scoccimarro:1996se}
\bibinfo{author}{\bibfnamefont{R.}~\bibnamefont{Scoccimarro}} \bibnamefont{and}
  \bibinfo{author}{\bibfnamefont{J.}~\bibnamefont{Frieman}},
  \bibinfo{journal}{Astrophys. J.} \textbf{\bibinfo{volume}{473}},
  \bibinfo{pages}{620} (\bibinfo{year}{1996}), \eprint{astro-ph/9602070}.

\bibitem[{\citenamefont{Scoccimarro}(1997)}]{Scoccimarro:1996jy}
\bibinfo{author}{\bibfnamefont{R.}~\bibnamefont{Scoccimarro}},
  \bibinfo{journal}{Astrophys. J.} \textbf{\bibinfo{volume}{487}},
  \bibinfo{pages}{1} (\bibinfo{year}{1997}), \eprint{astro-ph/9612207}.

\bibitem[{\citenamefont{Valageas and Nishimichi}(2011)}]{Valageas:2010yw}
\bibinfo{author}{\bibfnamefont{P.}~\bibnamefont{Valageas}} \bibnamefont{and}
  \bibinfo{author}{\bibfnamefont{T.}~\bibnamefont{Nishimichi}},
  \bibinfo{journal}{Astron. Astrophys.} \textbf{\bibinfo{volume}{527}},
  \bibinfo{pages}{A87} (\bibinfo{year}{2011}), \eprint{1009.0597}.

\bibitem[{\citenamefont{Noda et~al.}(2019)\citenamefont{Noda, Peloso, and
  Pietroni}}]{Noda:2019pzd}
\bibinfo{author}{\bibfnamefont{E.}~\bibnamefont{Noda}},
  \bibinfo{author}{\bibfnamefont{M.}~\bibnamefont{Peloso}}, \bibnamefont{and}
  \bibinfo{author}{\bibfnamefont{M.}~\bibnamefont{Pietroni}}
  (\bibinfo{year}{2019}), \eprint{1901.06854}.

\bibitem[{\citenamefont{Okamura et~al.}(2011)\citenamefont{Okamura, Taruya, and
  Matsubara}}]{Okamura:2011nu}
\bibinfo{author}{\bibfnamefont{T.}~\bibnamefont{Okamura}},
  \bibinfo{author}{\bibfnamefont{A.}~\bibnamefont{Taruya}}, \bibnamefont{and}
  \bibinfo{author}{\bibfnamefont{T.}~\bibnamefont{Matsubara}},
  \bibinfo{journal}{JCAP} \textbf{\bibinfo{volume}{1108}}, \bibinfo{pages}{012}
  (\bibinfo{year}{2011}), \eprint{1105.1491}.

\bibitem[{\citenamefont{Baldauf et~al.}(2015)\citenamefont{Baldauf, Mirbabayi,
  Simonovi{\'c}, and Zaldarriaga}}]{Baldauf:2015xfa}
\bibinfo{author}{\bibfnamefont{T.}~\bibnamefont{Baldauf}},
  \bibinfo{author}{\bibfnamefont{M.}~\bibnamefont{Mirbabayi}},
  \bibinfo{author}{\bibfnamefont{M.}~\bibnamefont{Simonovi{\'c}}},
  \bibnamefont{and}
  \bibinfo{author}{\bibfnamefont{M.}~\bibnamefont{Zaldarriaga}},
  \bibinfo{journal}{Phys. Rev.} \textbf{\bibinfo{volume}{D92}},
  \bibinfo{pages}{043514} (\bibinfo{year}{2015}), \eprint{1504.04366}.

\bibitem[{\citenamefont{Tassev}(2014)}]{Tassev:2013rta}
\bibinfo{author}{\bibfnamefont{S.}~\bibnamefont{Tassev}},
  \bibinfo{journal}{JCAP} \textbf{\bibinfo{volume}{1406}}, \bibinfo{pages}{008}
  (\bibinfo{year}{2014}), \eprint{1311.4884}.

\bibitem[{\citenamefont{Gradshteyn and Ryzhik}(1965)}]{Gradshteyn1965}
\bibinfo{author}{\bibfnamefont{I.~S.} \bibnamefont{Gradshteyn}}
  \bibnamefont{and} \bibinfo{author}{\bibfnamefont{I.~M.}
  \bibnamefont{Ryzhik}}, \emph{\bibinfo{title}{Table of integrals, series, and
  products}} (\bibinfo{publisher}{New York Academic Press},
  \bibinfo{year}{1965}), \bibinfo{edition}{4th} ed.,
  \urlprefix\url{http://openlibrary.org/books/OL5955048M}.

\bibitem[{\citenamefont{Crocce and
  Scoccimarro}(2006{\natexlab{b}})}]{Crocce:2005xz}
\bibinfo{author}{\bibfnamefont{M.}~\bibnamefont{Crocce}} \bibnamefont{and}
  \bibinfo{author}{\bibfnamefont{R.}~\bibnamefont{Scoccimarro}},
  \bibinfo{journal}{Phys. Rev.} \textbf{\bibinfo{volume}{D73}},
  \bibinfo{pages}{063520} (\bibinfo{year}{2006}{\natexlab{b}}),
  \eprint{astro-ph/0509419}.

\end{thebibliography}

%\bibliography{Biblio_scalar}

\end{document}